\documentclass[twocolumn,reprint,superscriptaddress,aps,prb,floatfix]{revtex4-2}

\usepackage{amsmath,amssymb,mathrsfs,amsthm}

\usepackage{graphicx}% Include figure files
\usepackage{dcolumn}% Align table columns on decimal point
\usepackage{bm}% bold math
\usepackage{subcaption}
\usepackage[justification=Justified]{caption}
\usepackage{braket}
\usepackage{xcolor}

\usepackage[normalem]{ulem}
\usepackage[pdftex, colorlinks=true, linkcolor=blue, citecolor=blue, urlcolor=blue]{hyperref}
\definecolor{DarkGreen}{RGB}{0,142,0}

\newcommand{\acd}{\textcolor{black}}
\newcommand{\super}{\textcolor{black}}

\begin{document}

\preprint{APS/123-QED}

\title{Interface States in Space-Time Photonic Crystals: Topological Origin, Propagation and Amplification}% Force line breaks with \\
%\thanks{A footnote to the article title}%

 \author{Alejandro Caballero}
 \email{alejandro.caballero@uam.es}
 \author{Thomas F. Allard}
 \author{Paloma A. Huidobro}
 \email{p.arroyo-huidobro@uam.es}
 
 \affiliation{Departamento de Física Teórica de la Materia Condensada, Universidad Autónoma de Madrid, E28049 Madrid, Spain}
 
 \affiliation{Condensed Matter Physics Center (IFIMAC), Universidad Autónoma de Madrid, E28049 Madrid, Spain}

\begin{abstract}
  Studying the topology of spatiotemporal media poses a fundamental challenge: their remarkable properties stem from breaking spatial and temporal symmetries, yet this same breaking obscures their topological characterization. Here, we show that space-time symmetries persist in crystals with travelling-wave modulations whose velocities can be either lower (subluminal) or higher (superluminal) than the speed of light, enabling the study of their topological properties and the prediction of spatiotemporal interface states. For each modulation regime, we use a Lorentz transformation to a frame in which the modulation depends on only one of the transformed variables. Then, we identify a conserved joint parity-time-reversal symmetry in the new variables that enforces the quantization of a spatiotemporal Zak phase, elevating it to a $\mathbb{Z}_2$ topological invariant. Finally, we calculate the associated interface states and uncover unique features arising from time-varying effects, including selective directional amplification, propagation along subluminal and superluminal boundaries, frequency- and momentum-converted replicas, and broadband amplification even in the absence of momentum gaps. Our framework holds for spatiotemporal modulations of any velocity, unifying a wide class of systems that includes photonic time crystals, and clarifying their topological origin.
\end{abstract}

\maketitle

\section{Introduction}

In time-varying media, the temporal modulation of a material's properties lifts the usual constrains of passive systems by breaking fundamental symmetries such as continuous time translation \cite{Galiffi2022,PTCTut}. As a result, energy conservation no longer holds, enabling light amplification and frequency conversion. When temporal and spatial modulations are combined, additional constrains are lifted, and further wave manipulation can be achieved through frequency-momentum transitions \cite{IntBandTrans, IBTchip,FreqMomTransAcous}. In particular, periodic space-time modulations of the travelling-wave form $f(x,t) = f(x-c_gt)$, where $c_g$ corresponds to the modulation speed, have attracted significant attention since early studies \cite{OldTWref,CassedyTW1,CassedyTW2}. \super{Specifically, since the material itself is stationary, the grating speed can take any value up to infinity while respecting special relativity. As a result, two qualitatively distinct regimes emerge, determined by the modulation speed relative to the speed of light in the medium: a subluminal (space-like) regime characterized by conventional band gaps where momentum becomes a complex magnitude, and a superluminal (time-like) regime featuring momentum gaps where frequency is complex, known as $k$-gaps \cite{CassedyTW2,UnVelSTcrys}. Furthermore,} such modulations introduce a linear-momentum bias that enables magnet-free non-reciprocity, manifested as asymmetric band gaps supporting unidirectional propagation \cite{TWMetasurf,ElasTW,FreqConvCaloz,NonRecPho,3DTWFilipa}, as well as synthetic Fresnel drag effects in the long-wavelength limit \cite{Fresnel}.  Interestingly, such spatiotemporal (ST) phenomena originate fundamentally from wave interference effects, allowing their experimental realization across diverse platforms: from mechanical waves in water \cite{WaterTimeRef,FloquetWater}, elastic \cite{LemoultSTInt,STPiezoelectric} and acoustic systems \cite{NonRecipSound,STAcousticMeta,AcousMetaSurf}, to electromagnetic (EM) waves spanning a wide range of frequencies in metasurfaces \cite{STMetaSurf}, transmission lines \cite{TempRefinTL,ExpTransLines} and ENZ materials \cite{TimeDiffENZ,STDiffENZ}.

Symmetries also play a central role in topological physics, where the Altland–Zirnbauer classification groups topological insulators into ten classes according to the presence or absence of three key symmetries: time-reversal, particle-hole, and chiral \cite{TopClass, Tenfold}. Depending on the dimensionality of the system, each class may host topological phases identified by a characteristic bulk topological invariant. A quantized invariant predicts the emergence of robust boundary states between materials of different phases, a principle known as bulk–boundary correspondence. These concepts can be applied to photonic crystals (PhCs), as topological invariants can also be attributed to their band structures \cite{QHEPhCs,ObsUniTopStates,MarioPhTopInv, MarioPhBulkEdge}. However, PhCs typically lack chiral and particle-hole symmetries which, together with the modified parity of the time-reversal operator ($\mathcal{T}^2 = +1$) due to their bosonic nature, limit the number of possible non-trivial phases. Crucially, richer topological effects can emerge by invoking additional symmetries such as parity-time-duality \cite{PTDMario,SymmProCTChan}, which restores a fermionic-like time-reversal operator and enables the photonic analogue of the quantum spin Hall effect in photonics \cite{PhQSHE,TopPhIns}.
Moreover, crystalline symmetries reveal finer topological phases within classes that would otherwise appear trivial \cite{TopCrysIns,StaCrysWannCent,QChePhCs,CornerTopPh,2DPhTop}. For instance, one-dimensional (1D) PhCs present topological states thanks to the presence of parity (inversion) symmetry \cite{SurfImp}, with their topological invariant being the Zak phase \cite{ZakPhase}. 

Recently, it has been shown that the band structures of photonic time crystals \super{(PTCs)}, where the optical properties of an homogeneous material periodically vary in time instead of space as in PhCs, are also classified by a quantized Zak phase when the modulation preserves temporal inversion symmetry. In turn, interface states localized in time emerge \cite{Lustig2018}, which can be robust against disorder for chiral-symmetric modulations \cite{ChiralTIS}. Localization in time has also been proposed in different time-varying systems, as in experiments with Floquet lattices \cite{SynthLattKgapTop,KTopExp}, or two-level systems with parity-time symmetry \cite{TwoLevelLoc}.

Beyond purely temporal modulations, combining space and time variations produces ST systems with a remarkably rich topological landscape. Two broad regimes can be distinguished. First, in decoupled modulations of the form $f(x,t) = g(x)h(t)$, space and time act as independent degrees of freedom, enabling higher-dimensional topological phases such as D+1 Floquet insulators \cite{TopSTCrystal,STsymSTgropu}, synthetic dimensions in frequency space \cite{FreqDimSFan,MechQuantHall,SFanTopRes}, and systems with both momentum and frequency band gaps that allow for localization in space and time \cite{TopSTEvents}. Second, in coupled ST modulations, exemplified by travelling-wave type modulations, time is no longer independent and the system remains effectively 1D, which forbids the well-established Chern classification applicable to decoupled modulations, although a recent extension has been proposed through a synthetic vector potential approach \cite{TWTop_Joao}. Previous studies in acoustic systems have investigated the topology of \super{subluminal} modulations in experiment by invoking an adiabatic approximation, which leads to a decoupled 1+1D formalism with its associated Chern number, but overlooks the effects introduced by the dynamical modulation \cite{AcousPump}. Other works, avoiding this approximation, revealed interface states accompanied by frequency-converted replicas at frequencies far from the band gap \cite{STPhonIS}. However, in that case the topological characterization relied on a time-dependent Zak phase, leaving open the question of whether a symmetry-protected invariant underlies the existence of these states. \super{Furthermore, the extension of any topological characterization into the superluminal regime has so far remained unexplored in the literature.}

In this work, we answer this question affirmatively by considering a PhC whose permittivity is modulated in a travelling-wave form and analyzing the ST symmetries it supports. \super{First, we develop the formalism for subluminal modulations, whose spatial nature enables a direct comparison with well-established spatial PhCs, and then extend it to the superluminal regime.} For both types of modulations, we perform a Lorentz transformation to \super{a frame in which the modulation depends on only one transformed variable, allowing us to} identify the conserved symmetries and \super{to} show that a combined parity–time-reversal transformation remains invariant. This symmetry enforces the quantization of a \textit{spatiotemporal} Zak phase defined along the Brillouin Zone (BZ) \super{in the new frame}, thereby establishing a $\mathbb{Z}_2$ topological invariant. Furthermore, by calculating the band energy density, we distinguish the two resulting phases given by such an invariant as a trivial phase and an obstructed atomic limit. This allows us to predict and unveil the interface states that arise between slabs of different ST Zak phases, considering \super{different types of boundaries depending on the modulation regime, including spatial, temporal and spatiotemporal ones}. Importantly, these states persist regardless of the modulation speed and \super{they are pinned at the bulk band-crossing position}, confirming their topological origin. Owing to their ST nature, the interface states present unique features such as selective directional \super{amplification}, frequency- \super{and momentum-}converted replicas, \super{as well as} propagation along \super{subluminal and superluminal} boundaries, \super{and} broadband amplification even in the absence of momentum gaps. Finally, we analyze the effect of relevant perturbations on the modulation, demonstrating the robustness of the interface states and clarifying the conditions under which they remain topological.

\section{Methods}

\subsection{Bloch-Floquet theory of travelling-wave PhCs}

\begin{figure}[h]
    \centering
    \includegraphics[width=240pt]{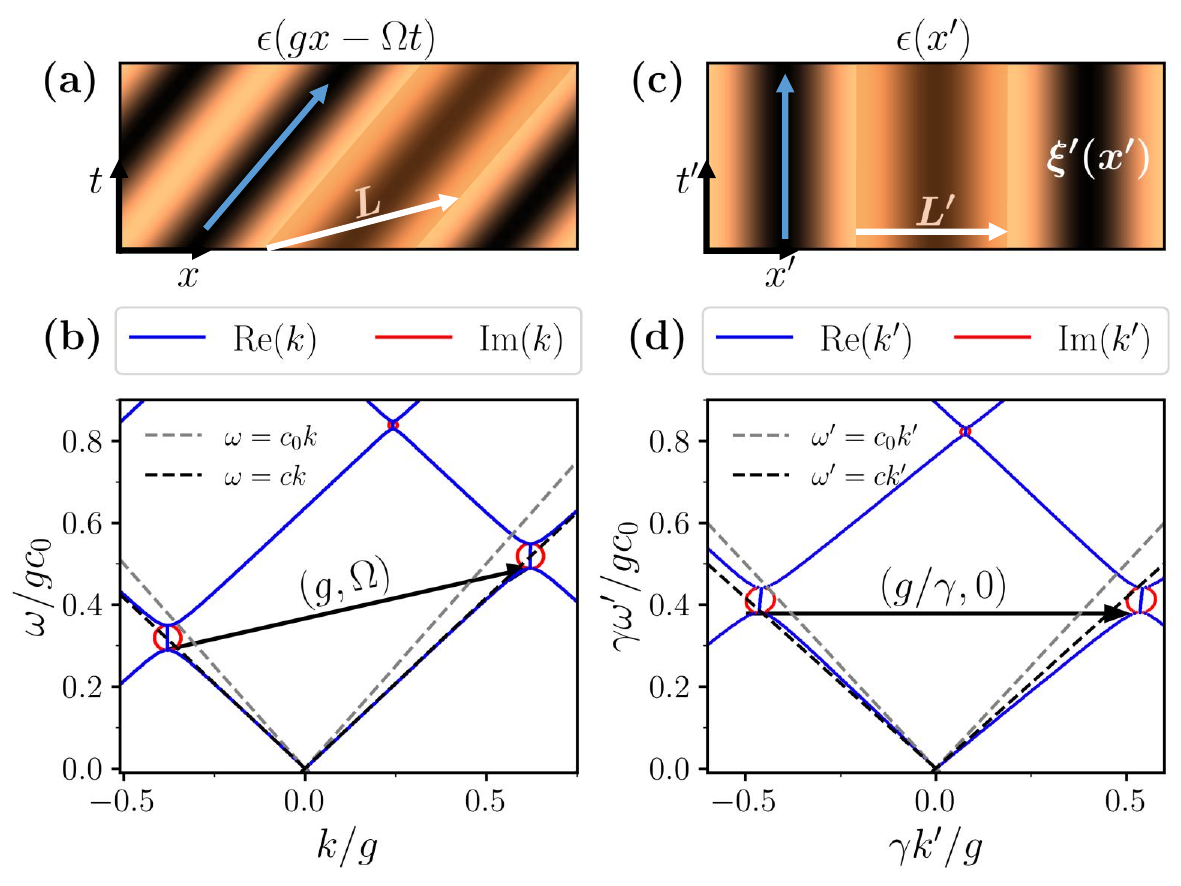}
    \caption{Spatiotemporal photonic crystal in the laboratory and comoving frames \acd{for subluminal modulation}.
    (a) Travelling-wave modulated permittivity in the lab-frame. A blue arrow signals the direction of the continuous space-time translation symmetry that defines the unit cell highlighted in the orange shaded region. The white arrow corresponds to the lattice vector.
    (b) Band structure of a STPhC for a modulation speed $c_g=\Omega/g=0.2c_0$, modulation strength $\alpha=0.3$ and $\epsilon_m=\mu_m=1.2$. Grey and black dashed lines correspond to the dispersion relation in free space and in the unmodulated material, respectively. Black arrow represents the reciprocal lattice vector $\textbf{p}$.
    (c) Modulated permittivity in the comoving frame. The previous magnitudes are now spatial-like, but a new magneto-electric coupling $\xi'(x')$ appears. (d) Band structure of the same STPhC in the comoving frame, with the black arrow now representing the transformed reciprocal vector $\textbf{p}'$.}
    \label{fig:band_eps_dist}
\end{figure}

We consider a medium whose permittivity is modulated in space and time following the travelling form
\begin{equation}\label{eq:eps}
    \epsilon(x,t) = \epsilon_0 \epsilon_m [1 + \alpha \cos(g x - \Omega t)],
\end{equation}
with $\epsilon_m$ the background relative permittivity of the medium, $g$ and $\Omega$ the spatial and temporal modulation frequencies with periods $a = 2 \pi/g$ and $T = 2 \pi/\Omega$, and $\alpha$ the modulation strength. The permeability of the medium $\mu = \mu_0\mu_m$ is considered as constant, where $\mu_m$ is the background permeability, and $c_0 = 1/\sqrt{\epsilon_0 \mu_0}$ is the speed of light in vacuum. As sketched in Fig.~\ref{fig:band_eps_dist}(a), this specific modulation where space and time are coupled creates a moving grating with phase velocity  $c_g = \Omega/g$ along the $x$ direction. Since the material itself is not moving, this grating speed can take any value up to infinity while respecting special relativity. This allows for the same model described by Eq.~\eqref{eq:eps} to host different regimes depending on whether the grating velocity is lower (subluminal) or higher (superluminal) than the speed of light in the medium $c = c_0/\sqrt{\epsilon_m \mu_m}$ \cite{Fresnel,Hom}, leading to drastically different behaviours. \super{First, we focus on the subluminal regime, defined for our travelling-wave modulation as $0 \leq c_g \leq c/\sqrt{1+\alpha}$ \cite{Hom}, with the upper bound marking the onset of the luminal regime, \super{which we will not consider here due to the absence of the notion of band structure.} In a later section, we study the superluminal regime by considering modulation velocities ranging in $c/\sqrt{1-\alpha} \leq c_g \leq \infty$.}

We calculate the band structure of the spatiotemporal PhC (STPhC) by solving Maxwell's equations, $\nabla \times \textbf{E} = - \partial_t \textbf{B}$ and $\nabla \times \textbf{H} = \partial_t\textbf{D}$. Considering s-polarization and normal incidence, the latter equations can be written in matrix form as
\begin{equation}\label{eq:Maxwell_lab}
    \begin{pmatrix}
        0 &  \partial_{x}\\
         \partial_{x} & 0
    \end{pmatrix}
    \begin{bmatrix}
        E_z\\
        H_y
    \end{bmatrix}
    =
       \partial_{t}  \left(\hat{\textbf{M}}(x,t)  
    \begin{bmatrix}
        E_z\\
        H_y
    \end{bmatrix} \right),
\end{equation}
where $\hat{\textbf{M}}(x,t)$ corresponds to the constitutive matrix of the relevant components of the EM field
\begin{equation}\label{eq:mat_mat_lab}
    \hat{\textbf{M}}(x,t) = 
    \begin{pmatrix}
        \epsilon(x,t) & 0\\
        0 & \mu
    \end{pmatrix},
\end{equation}
 such that
\begin{equation}\label{eq:const_trans}
    \begin{bmatrix}
        D_z\\
        B_y
    \end{bmatrix}
    = \hat{\textbf{M}}(x,t) \,
    \begin{bmatrix}
        E_z\\
        H_y
    \end{bmatrix}.
\end{equation}
Then, we use the Bloch-Floquet ansatz
\begin{equation}
    \begin{bmatrix}
        E_z (x,t)\\
        H_y (x,t)
    \end{bmatrix} 
    = e^{i(k x - \omega t)} \; \sum_n
    \begin{bmatrix}
        E_n \\
        H_n
    \end{bmatrix}
    \, e^{i\, n (g x - \Omega t)} 
\end{equation}
to derive an eigenvalue problem for the Bloch-Floquet amplitudes $E_n$ and $H_n$. This allows us to compute the dispersion relation as $k(\omega)$ \super{for the subluminal regime, where the band gaps present complex momenta, and $\omega(k)$ for the superluminal case, whose momentum gaps host complex frequencies. Finally, we also obtain} the mode decomposition of the eigenfunctions of the STPhC (see Supporting Information (SI) for detailed derivations). 

We represent the obtained band diagram of the STPhC \super{in the subluminal regime} in Fig.~\ref{fig:band_eps_dist}(b), together with the dispersion relation in free space and in the unmodulated medium \acd{(dashed lines)}. \super{In this regime}, the band structure displays frequency band gaps akin to spatial PhCs, in contrast to the momentum band gaps \super{present in} PTCs \cite{Galiffi2022} \super{or in the superluminal regime, as we will see in a later section}. Interestingly, however, the gaps are asymmetric and appear at different frequencies for forward ($k>0$) and backward ($k<0$) waves, revealing the mechanism by which STPhCs enable one-way propagation \cite{IntBandTrans, IBTchip,FreqConvCaloz}

The asymmetric band structure stems from the coupling between spatial and temporal modulations induced by the travelling wave form of Eq.~\eqref{eq:eps}. Since the STPhC is effectively a 1D system, it is described by a single lattice vector. While the temporal modulation breaks continuous time translation symmetry, a travelling wave modulation exhibits a continuous ST translational symmetry that conserves a linear superposition of energy and momentum \cite{STsymm}. This is highlighted by the blue arrow in Fig.~\ref{fig:band_eps_dist}(a). Such continuous symmetry of the system defines the unit cell, sketched as a shaded trapezoid, and a ST lattice vector $\textbf{\text{L}}$ with both spatial and temporal components. The reciprocal lattice vector is also a ST one, $\textbf{p} = (g,\Omega)$, see Fig.~\ref{fig:band_eps_dist}(b), and its non-zero frequency component tilts the entire band structure and explains the asymmetry of the band gaps.

\subsection{Reference frame transformation}

To facilitate the description of the crystal's underlying ST symmetries, we reformulate our problem in a \super{reference} frame \super{where the modulation depends on a single transformed variable, either space or time, depending on the modulation regime. We begin with the subluminal case, for which we choose a frame comoving with the modulation} \cite{Hom,UnVelSTcrys,IsoRefSTCrystal,STGratings,QEDSimonPendry,NonRecMink,TWAmpSimonPendry}. \acd{To do so, we use Lorentz transformations to ensure that Maxwell's equations are conserved, as we are interested in studying the different symmetries of the system.} After a Lorentz boost in the $x$ direction, the comoving coordinates read
\begin{equation}\label{eq:Lor_trans_sp1}
        x' = \gamma \, (x - c_g \, t), \; y' = y, \; z' = z,
\end{equation}
and
\begin{equation}\label{eq:Lor_trans_sp2}
    t' = \gamma \,  (t - \frac{c_g}{c_0^2} \, x),
\end{equation}
where the Lorentz factor $\gamma = 1/{\sqrt{1-(c_g/c_0)^2}}$. By transforming the EM fields accordingly \cite{Kong}, we then obtain Maxwell's equations in the comoving frame
\begin{equation}\label{eq:Maxwell_com}
    \begin{pmatrix}
        0 &  \partial_{x'}\\
         \partial_{x'} & 0
    \end{pmatrix}
    \begin{bmatrix}
        E_z'\\
        H_y'
    \end{bmatrix}
    =
    \hat{\textbf{M}}'(x') \; \partial_{t'}
    \begin{bmatrix}
        E_z'\\
        H_y'
    \end{bmatrix} ,
\end{equation}
with the transformed constitutive matrix
\begin{equation}\label{eq:mat_mat_com}
    \hat{\textbf{M}}'(x') = 
    \begin{pmatrix}
        \epsilon'_{\perp}(x') & \xi'(x')\\
        \xi'(x') & \mu'_{\perp}(x')
    \end{pmatrix},
\end{equation}
whose elements in the new frame read
\begin{equation}
    \epsilon'_{\perp}(x') = \frac{\epsilon(x')}{\gamma^2 (1-\epsilon(x') \mu \, c_g^2)},
\end{equation}
\begin{equation}
    \mu'_{\perp}(x') = \frac{\mu}{\gamma^2 (1-\epsilon(x') \mu \, c_g^2)},
\end{equation}
and
\begin{equation}\label{eq:xi_com}
    \xi'(x') = c_g \,\frac{\epsilon(x') \mu \, - c_0^{-2}}{1-\epsilon(x')\mu \, c_g^2}.
\end{equation}
In the comoving frame, the constitutive parameters depend on the spatial coordinate $x'$ only, and the modulation loses the explicit temporal dependence. More importantly, \acd{moving between} frames leads to the material acquiring a bianisotropic coupling $\xi'(x')$ proportional to the grating velocity, resulting in a moving-medium type coupling between the electric and magnetic fields \cite{Hom}. \acd{This can be understood by noting that, in the lab-frame, only the modulation propagates while the material remains stationary. Therefore, when changing to the moving frame, the modulation becomes static, but the material appears to be moving in the opposite direction, giving rise to the bianisotropic response $\xi'(x')$ akin to a moving medium.}

We sketch the permittivity distribution in the comoving frame in Fig.~\ref{fig:band_eps_dist}(c). In this frame, there is a continuous ST translation symmetry along $t'$, such that the unit cell is defined with the spatial coordinate only as $x' \in [-\gamma a/2, \gamma a/2]$, with lattice vector $\textbf{\text{L}}' = (\gamma a, 0)$. Using the reciprocal coordinates in the comoving frame
\begin{equation}\label{eq:k_Lorent}
    k' = \gamma \, (k - \frac{c_g}{c_0^2} \, \omega), \; k'_y = k_y, \; k'_z = k_z, 
\end{equation}
and
\begin{equation}\label{eq:w_Lorentz}
    \omega' = \gamma \, (\omega - c_g \, k),
\end{equation}
we obtain the reciprocal lattice vector $\textbf{p}' = (g/\gamma,0)$ as well as the BZ defined as $k' \in [-g/2\gamma, g/2\gamma]$. Finally, transforming back to the laboratory frame, the corresponding lattice vector can be calculated as $\textbf{L} = \gamma^2 a (1, c_g/c_0^2)$. \acd{This derivation highlights the importance of employing a Lorentz transformation rather than a Galilean one for our purposes, as the mixed nature of the space–time dimension cannot be taken into account without transforming the temporal variable.}

With the lattice vector now established in the comoving frame, the periodicity of the crystal becomes explicit, namely, $\hat{\textbf{M}}'(x' + \gamma a) = \hat{\textbf{M}}'(x')$. This allows us to apply Bloch’s theorem to Eq.~\eqref{eq:Maxwell_com} and obtain the dispersion relation $\omega'(k')$ together with the comoving-frame eigenfunctions
\begin{equation}\label{eq:com_eigenfun}
    \begin{bmatrix}
        E'(x')\\
        H'(x')
    \end{bmatrix}_{k',m}
    = e^{i k' x'} \, 
    \begin{bmatrix}
        u{'}^E(x')\\
        u{'}^H(x')
    \end{bmatrix}_{k',m} ,
\end{equation}
where the subscripts of the electric and magnetic fields have been dropped for clarity,  $u{'}^{E/H}_{k',m}(x'+\gamma a) = u{'}^{E/H}_{k',m}(x')$ correspond to the periodic part of the eigenfields, and $m$ represents the band index (see SI).

Figure \ref{fig:band_eps_dist}(d) shows the dispersion relation of the STPhC in the comoving frame. The absence of frequency component in $\textbf{p}'$ removes the tilted nature of the band structure observed in the laboratory frame [cf. Fig.~\ref{fig:band_eps_dist}(b)], rendering the band gaps symmetric. However, the system still presents non-reciprocal features, as visible from the fact that $\omega'(-k') \neq \omega'(k')$. Indeed, an increased group velocity is observed for backward propagating waves, while forward propagating waves present a reduced one. This is a result of the non-zero magneto-electric coupling that appears in the comoving frame, which breaks time-reversal symmetry $\mathcal{T'} (t' \rightarrow -t')$ and, consequently, reciprocity \cite{EMnonRec}.

Therefore, although the reciprocal lattice vector is spatial-like in the comoving frame, the effective magneto-electric coupling induced by the \acd{frame transformation} still underlies the breaking of fundamental symmetries as well as the emergence of non-reciprocal features. In the following, we analyze the impact of this bianisotropic coupling on the ST symmetries present in the comoving frame, as well as its consequences in the topological characterization.

\section{Results and discussion}

\subsection{Topological characterization of STPhCs}

In spatial non-magnetic PhCs, only time-reversal $\mathcal{T}$ is present out of the three fundamental symmetries that define the ten-fold way \cite{Tenfold,TopClass}, which for general 1D systems classify them as trivial. However, crystalline symmetries can enrich PhCs with nontrivial topology \cite{StaCrysWannCent}. In 1D, this role is played by parity (inversion) symmetry $\mathcal{P}$ ($x \rightarrow -x$), which quantizes the Zak phase \cite{ZakPhase}, defining it as a $\mathbb{Z}_2$ topological invariant. This symmetry also establishes a direct connection between the parity of the Bloch functions and the existence of surface states \cite{SymmCritZak,SurfImp}, i.e., it establishes a bulk-interface correspondence.

Motivated by this context, we now discuss the topological characterization of a \super{subluminal} STPhC by studying the space–time counterpart of parity symmetry, defined in the comoving frame as $\mathcal{P'}$ ($x' \rightarrow -x'$). Applying this transformation to Eq.~\eqref{eq:Maxwell_com} reveals that, although each entry of the material matrix is symmetric under $\mathcal{P'}$, the presence of the magneto-electric coupling $\xi'$ breaks the invariance of Maxwell’s equations. As a result, the direct generalization of the symmetry that enables non-trivial topology in static crystals is not possible, raising the question of whether STPhCs can still be topologically classified. Crucially, however, while $\mathcal{T'}$ and $\mathcal{P'}$ are individually broken, their combined operation $\mathcal{P'T'}$ [$(x',t') \rightarrow -(x',t')$] is preserved. \acd{Physically, $\mathcal{P'T'}$-symmetry has the same relevance as conventional $\mathcal{PT}$-symmetry since both transformations are equivalent, as can be seen in Eqs.\eqref{eq:Lor_trans_sp1}-\eqref{eq:Lor_trans_sp2}.} In what follows, we examine whether this joint symmetry can play a similar role as parity symmetry does in 1D spatial PhCs.

To do so, we study the Zak phase defined with the eigenfunctions of Eq.~\eqref{eq:com_eigenfun} and along the comoving-frame BZ
\begin{equation}\label{eq:STZak}
    \theta^{\mathrm{ST}}_m =  i \int_{-g/2\gamma}^{+g/2\gamma} dk'\braket{\textbf{u}{'}_{k',m} | \partial_{k'} \textbf{u}'_{k',m}}_{\acd{\mathbf{\hat{M}}'}},
\end{equation}
where the integrand corresponds to the Berry connection and $\textbf{u}'_{k',m} = [u{'}^E_{k',m}, u{'}^H_{k',m}]^T$. Importantly, the Hermiticity of the constitutive matrix in Eq.~\eqref{eq:mat_mat_com}, together with $\mathcal{P'T'}$ symmetry, allows us to relate the left and right eigenvectors of the system (see SI) \cite{NonBlochTheoSTcrys}. As a result, the conventional Berry connection can be used in place of the biorthogonal one, which would otherwise be required by the non-Hermitian nature of the differential operator in Eq.~\eqref{eq:Maxwell_com}. This Berry connection is defined with a weighted scalar product over the unit cell given by
\begin{equation}\label{eq:scalar_prod}
    \braket{\mathbf{\Psi}'_1 | \mathbf{\Psi}'_2}_{\acd{\mathbf{\hat{M}}'}} = \int_{- \gamma a /2}^{+\gamma a/2} dx' \,  \, \mathbf{\Psi}_1'^\dagger(x') \, \mathbf{\hat{M}}'(x') \mathbf{\Psi}'_2(x').
\end{equation}
The scalar product ensures the normalization of the comoving-frame eigenfunctions by establishing $\braket{\mathbf{\Psi}'_{k',m}|\mathbf{\Psi}'_{k',m}}_{\acd{\mathbf{\hat{M}}'}} = 1$, with $\mathbf{\Psi}'_{k',m} = [E'_{k',m}, H'_{k',m}]^T$. 

Furthermore, building upon the results of Ref.~\cite{BerrConnProof} on the Hermitian Berry connection, one can show that $\mathcal{P'T'}$-symmetry alone allows the quantization of the ST Zak phase. The quantization defines $\theta^{\mathrm{ST}}_m$ as a $\mathbb{Z}_2$ topological invariant, which means that one can distinguish between two different phases, delimited by a topological transition where the band gap must close. Therefore, solely from this property we can predict the existence of interface states between two STPhCs of different topological phase. 

To understand the physical meaning behind these two phases, we use Eq.~\eqref{eq:scalar_prod} to establish a direct relation between Wannier centers and the ST Zak phase, as is done in the Hermitian case \cite{vanderbilt}. We can then write
\begin{equation}\label{eq:wannier_center}
    \bar{x}'_m = \braket{\textbf{w}'_{L',m}|x'|\textbf{w}'_{L',m}}_{\acd{\mathbf{\hat{M}}'}} = \gamma \theta^{\mathrm{ST}}_m/g,
\end{equation}
with $\textbf{w}'_{L',m}(x') = \gamma a/2\pi \int dk' \; e^{ik' L'} \mathbf{\Psi}'_{k',m}$ the Wannier function associated to the eigenfields $\mathbf{\Psi}'_{k',m}$ and $L' = n \gamma a$, with $n \in \mathbb{Z}$, labeling the position of the unit cell. Therefore, the quantization of $\theta^{\mathrm{ST}}_m$ enforced by $\mathcal{P'T'}$-symmetry translates directly into the quantization of the Wannier centers. In particular, since the Zak phase can only take the values $0, \pi$ (mod $2\pi$), the Wannier functions can only be localized at the positions $\bar{x}' = 0, \gamma a/2$, which correspond to the inversion centers of the unit cell. 

In electronic insulators, Wannier centers mark the positions of electronic charge within the unit cell, effectively defining atomic orbitals \cite{vanderbilt}. The quantization of $\bar{x}_m'$ thus distinguishes two phases, with orbitals localized at the cell center or edges. Within the band representation framework \cite{TopQChem,StaCrysWannCent,WannCentSpec}, these correspond in one dimension to trivial and obstructed (topological) phases, respectively. This formalism extends to PhCs, where EM energy density acts as an analogue of electronic charge \cite{EnergyDens}, enabling the same physical interpretation of the two phases defined by the ST Zak phase in our photonic system. Indeed, this can be derived directly from the time-averaged energy density of the comoving-frame eigenfields. Integrating over the whole BZ, the latter density reads \cite{Kong}
\begin{equation}\label{eq:en_dens}
    n_m(x') = \int_{-g/2\gamma}^{+g/2\gamma} \frac{E{'}_{k',m} \, D'^*_{k',m} + H{'}_{k',m} \, B'^*_{k',m}}{2} dk',
\end{equation}
where $D'_{k',m}$ and $B'_{k',m}$ correspond to the displacement and magnetic flux eigenfields, respectively, obtained through the constitutive matrix in Eq.~\eqref{eq:mat_mat_com}. The total EM energy density is then $n_\text{T}(x') = \sum_m n_m(x')$.

\begin{figure}[t]
    \includegraphics[clip,width =220pt]{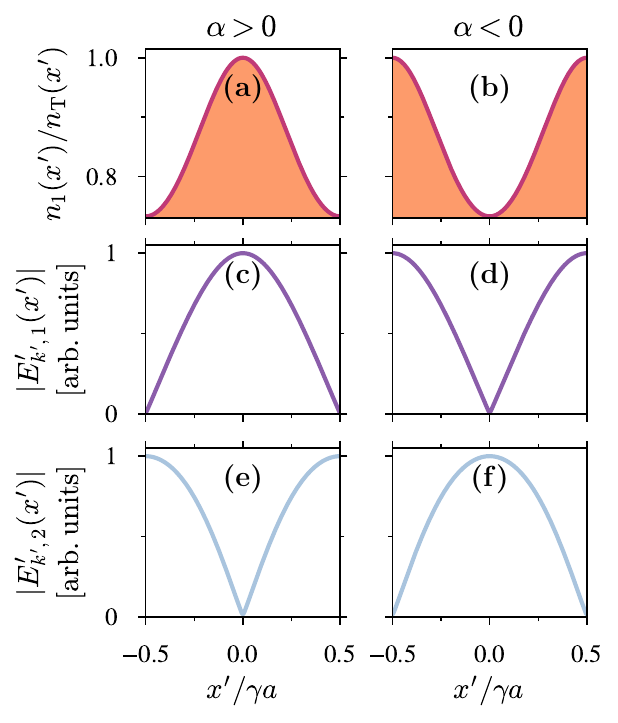}
    \centering
    \caption{Distinct topological phases in STPhCs. (a)-(b) Energy density distribution over the unit cell of the first band $m = 1$, for $\alpha > 0$ and $\alpha < 0$. (c)-(d) Absolute value of the electric field eigenfunction $|E'_{k'_{\text{gap}},m = 1}(x')|$ located at the first band gap shown in Fig.~\ref{fig:band_eps_dist}(d) for $k' < 0$, considering $\alpha>0$ and $\alpha<0$. (e)-(f) $|E'_{k'_{\text{gap}},m = 2}(x')|$ at the same gap for the second band $m=2$, considering $\alpha > 0$ and $\alpha < 0$. All the parameters are the same as in Fig.~\ref{fig:band_eps_dist}.}
    \label{fig:band_inv}
\end{figure}

In order to corroborate the existence of the two topologically distinct phases in our model, we study the transition between positive and negative modulation strength, with $\alpha = 0$ in Eq.~\eqref{eq:eps} marking the topological transition point. This is shown in Fig.~\ref{fig:band_inv}(a)–(b), where we represent the distribution of the energy density of the first band $m=1$ along the unit cell for both phases. Indeed, we find that for $\alpha > 0$ the energy density of Eq.~\eqref{eq:en_dens} is localized at the center of the unit cell, corresponding to the trivial phase, whereas for $\alpha < 0$ it shifts to the edges, which marks the obstructed (topological) phase [see Fig.~\ref{fig:band_inv}(b)]. We note that a subtlety arises in the calculation of $\theta^{\mathrm{ST}}_m$ for the first band $m = 1$ near the singular point $k' = \omega' = 0$, where the parity of the electric and magnetic fields becomes ill-defined. Even though this does not affect the quantization of the Zak phase, it obscures the direct correspondence with the Wannier centers (see SI). Crucially, only the relative difference of Zak phases between crystals matters for the appearance of interface states \cite{ExpGeomPhase,ZakPhMeas}, so this subtlety does not affect the topological characterization of STPhCs.

In Fig.~\ref{fig:band_inv}(c)–(f) we further represent the electric eigenfields at the gap between the first and second bands for both signs of $\alpha$. The eigenfunctions clearly interchange across the $\alpha = 0$ transition, a hallmark of band inversion that, as established in Ref.~\cite{PTexactTop}, characterizes a topological phase transition even in non-Hermitian systems as long as they remain in the $\mathcal{P'T'}$-exact phase. Taken together, these results confirm that $\mathcal{P'T'}$-symmetry alone suffices to topologically classify \super{subluminal} STPhCs ensuring the existence of two distinct phases even in the absence of parity symmetry.

\subsection{Interface state}

\begin{figure*}[h!]
    \includegraphics[clip,width=\textwidth]{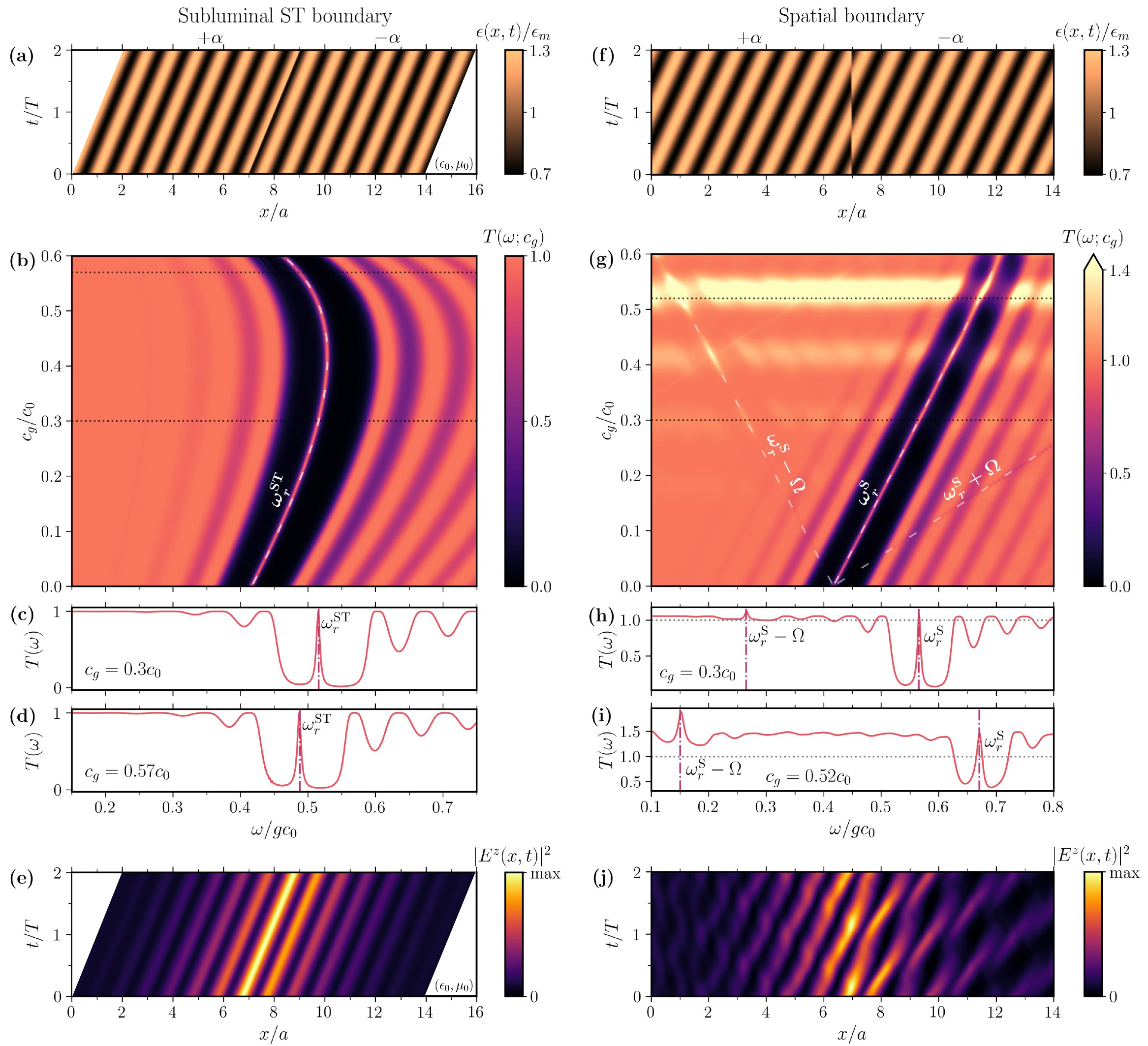}
    \caption{Interface states between a trivial ($\alpha>0$) and a topological ($\alpha<0$) spatiotemporal slab, considering (a)-(e) a spatiotemporal boundary and (f)-(j) a spatial boundary. 
    (a) Permittivity profile $\epsilon(x,t)$ of the two-slab configuration for $c_g = 0.3 c_0$. 
    (b) Transmittance spectrum of two spatiotemporal slabs whose boundary moves along with the modulation, as a function of the modulation speed $c_g$. An interface state with resonance frequency following the bulk-predicted $\omega^{\text{ST}}_r$ is observed. 
    (c) Horizontal cross section of the transmittance at $c_g = 0.3 c_0$. A clear peak inside the gap corresponding to the interface state is visible. 
    (d) Same magnitude for $c_g = 0.57 c_0$.
    (e) Intensity distribution of the interface state observed when exciting the left slab with a plane wave of frequency $\omega^{\text{ST}}_r$. A propagating interface state is observed.
    (f) Permittivity profile $\epsilon(x,t)$ of the two spatial slabs for $c_g = 0.3 c_0$.
    (g) Transmittance spectrum as a function of $c_g$ for two slabs of the STPhC with purely spatial boundaries. An interface state is also present for the whole modulation interval. Furthermore, the frequency conversion induced by the temporal modulation enables replicas of the interface states at $\omega_r^{\text{S}} \pm \Omega$.
    (h) Horizontal cross section of the transmittance for $c_g = 0.3 c_0$, revealing the interface state inside the gap as well as the replicas at $\omega^{\text{S}}_r - \Omega$.
    (i) Same magnitude for $c_g = 0.52 c_0$, where broadband amplification ($T(\omega)>1$) is observed.
    (j) Intensity distribution of the interface state for a static boundary, showing no propagation.}
    \label{fig:int_states}
\end{figure*}

Building on the prediction of two topologically distinct phases, we now employ a semi-analytical scattering matrix formalism (detailed in the SI) to investigate the presence of interface states at the boundary between two \super{subluminal} STPhCs, as well as discuss their novel properties due to the temporal modulation. To do so, we consider two types of slabs by truncating the same STPhC in different space-time directions: along the modulation (\super{subluminal ST} boundary) [see Fig.~\ref{fig:int_states}(a)] and at a fixed position in space (spatial boundary) [see Fig.~\ref{fig:int_states}(f)]. \\

\textit{\super{Subluminal} Spatiotemporal Interface} - We begin by considering two ST slabs, each composed of seven unit cells of the STPhC defined in Eq.~(\ref{eq:eps}), as represented in Fig.~\ref{fig:int_states}(a). Specifically, we set the medium's parameters to $\epsilon_m = \mu_m = 1.2$ and $\alpha = 0.3$ as in Figs.~\ref{fig:band_eps_dist} and \ref{fig:band_inv}, with different signs of the modulation strength $\alpha$ for each slab. As discussed in the previous section, positive and negative values of $\alpha$ correspond to different topological phases \acd{while presenting the same band structure. Consequently,} interface states are expected to appear. In Fig.~\ref{fig:int_states}(b), we show the transmission spectrum of this configuration as a function of the grating speed $c_g$, obtained by exciting the left slab with a plane wave and measuring the transmitted wave at the right end.  Horizontal cross sections at $c_g = 0.3 c_0$ and $c_g = 0.57 c_0$ are presented in Fig.~\ref{fig:int_states}(c)-(d), respectively. We observe that the darker regions in Fig.~\ref{fig:int_states}(a), corresponding to lower transmission, mark the band gap, while the continuous peak within the gap identifies the interface state. Its persistence across the entire modulation range confirms the prediction of the ST Zak phase and supports the conclusion that $\mathcal{P'T'}$-symmetry alone suffices for the topological characterization of the system. Indeed, despite the progressive breaking of $\mathcal{P'}$-symmetry with increasing magneto-electric coupling $\xi'(x'; c_g)$, the interface state persists. \acd{We note that all the phenomenology discussed in this section is also present when the incident wave excites the composite system in the opposite direction. More interestingly, the band gap, and therefore the interface state, are observed at lower frequencies due to the non-reciprocal nature of the band structure of STPhCs.}

Furthermore, we can predict the resonance frequency $\omega_r^{\text{ST}}$ at which interface states will be excited for each $c_g$ through the following intuition: a topological transition necessarily entails a band-gap closing, so the frequency at which this occurs identifies the precise point in the band structure where an interface state must emerge between materials at opposite sides of this transition. Therefore, the resonance frequency can be determined by finding the band-crossing position in the unmodulated case $\alpha = 0$, which marks the transition point. 

In order to find the band crossing point that we argue predicts the resonance frequency of the interface state, we use homogenization theory to obtain a linear approximation of the band dispersion of the STPhC. The dispersion is constructed by copying replicas of the $\omega = \pm v_{\text{eff}} k$ curves, where $v_{\text{eff}} = 1/\sqrt{\epsilon_{\text{eff}} \mu_{\text{eff}}}$ corresponds to the effective velocity defined by the homogenized material parameters (see Ref.~\cite{Hom}). These replicas are translated by the reciprocal lattice vector $\textbf{p} = (g,\Omega)$ an integer number of times $n$, yielding $\omega + n \Omega = \pm v_{\text{eff}} (k + n g)$. Then, calculating the crossing point between the fundamental ($n=0$) forward mode, and the first replica of the backward mode ($n = -1$), we obtain the crossing point corresponding to the first gap in the forward direction (see sketch in SI):
\begin{equation}\label{eq:freq_gap}
    \frac{\omega_{\text{gap}}}{gc_0} =  \frac{c_g + v_{\text{eff}}(c_g)}{2c_0}.
\end{equation}
However, since we are considering a ST interface, it is essential to account for the modified conservation law: it is $\omega'$ (the comoving frequency) that is conserved, rather than $\omega$. This means that, when a monochromatic wave of frequency $\omega_0$  encounters the first interface coming from free space, it will not couple to the same frequency inside the material $\omega_{\text{mat}}$, but rather to a Doppler shifted frequency given by the expression \cite{UnVelSTcrys}
\begin{equation}\label{eq:Doppler_freq}
    \omega'/\gamma = (1-c_g/v_{\text{eff}}) \, \omega_{\text{mat}} = (1-c_g/c_0) \, \omega_0 .
\end{equation}
This explains why the band-gap position in Fig.~\ref{fig:int_states}(b) does not grow linearly with $c_g$, as might be expected from the tilt induced by the reciprocal lattice vector, but instead shifts to lower frequencies once the modulation speed becomes sufficiently large. This red shift effect can be clearly seen by comparing the transmission spectra for $c_g = 0.3c_0$ [Fig.~\ref{fig:int_states}(c)] and $c_g = 0.57c_0$ [Fig.~\ref{fig:int_states}(d)]. As expressed in Eq.~\eqref{eq:Doppler_freq}, when $c_0 \neq v_{\text{eff}}$ the mismatch between $\omega_0$ and $\omega_{\text{mat}}$ grows with increasing $c_g$, leading to the more intricate dependence observed in Fig.~\ref{fig:int_states}(b). Interestingly, this effect makes the gap felt by the incoming wave larger than the width of the actual band gap. Furthermore, from Eq.~\eqref{eq:Doppler_freq} we can directly determine the incident frequency necessary to excite $\omega_{\text{gap}}$ inside the material, which corresponds to the resonance frequency of the topological states at a ST interface:
\begin{equation}\label{eq:ST_res_freq}
    \omega^{\mathrm{ST}}_r =  \frac{1-c_g/v_{\text{eff}}}{1-c_g/c_0} \, \;\omega_{\text{gap}}.
\end{equation}
This frequency is shown as a function of $c_g$ in Fig.~\ref{fig:int_states}(b) with a dashed line and, as we can see, it predicts with high accuracy the numerically calculated position of the interface state within the gap, showing the connection of these states to the properties of the bulk. This connection, together with the persistence of the interface states, establishes a ST analogue of the bulk-interface correspondence of 1D spatial PhCs. 

Finally, we present in Fig.~\ref{fig:int_states}(e) the intensity distribution of the interface state excited by a plane wave at its resonance frequency $\omega^{\mathrm{ST}}_r$ considering a grating velocity $c_g=0.3c_0$. As observed, the topological state remains localized, decaying evanescently away from the moving boundary. Interestingly, this creates a 1D propagating interface state, in contrast to the 0D static state typically found in 1D spatial PhCs. \acd{We note that such a propagation of the interface state at a subluminal spatiotemporal interface can be understood from the tilted nature of the bandstructure discussed in Fig.~\ref{fig:band_eps_dist}(b). Indeed, for a spatial PhC and stationary boundaries, the interface state appears in the band structure of the composite system as a flat band connecting the two edges of the BZ \cite{STPhonIS}. Therefore, in a STPhC with ST boundaries, the topological state would also appear joining the two edges of the BZ which, given the tilted character of the band structure, would grant the state a nonzero group velocity that exactly matches the boundary speed $c_g$.}\\

\textit{Spatial interface} - We now turn to the case of two spatial slabs of the same STPhC, as illustrated in Fig.~\ref{fig:int_states}(f), in order to examine how changing the type of boundary affects the interface state. The static interface implies that the slabs are constructed with a purely spatial unit cell, defined with the same spatial period as the modulation $a = 2\pi/g$. Since this does not correspond to the ST unit cell used to characterize the infinite crystal, the derivation of the scattering matrix for this case needs a tailored approach (see SI). As for the ST interface, we plot in Fig.~\ref{fig:int_states}(g) the transmittance spectrum of the slabs in this configuration as a function of the modulation velocity, together with two horizontal cross sections at $c_g = 0.3 c_0$ and $c_g = 0.52 c_0$. Crucially, we find that an interface state also appears inside the gap regardless of the value of the grating speed. This demonstrates the robustness and predictive power of the ST Zak phase, which correctly anticipates the existence of such states for two very different types of boundaries. Furthermore, since the spatial interface conserves the lab-frame frequency $\omega$, we can directly predict the resonance frequency of this state through Eq.~\eqref{eq:freq_gap} as $\omega^\mathrm{S}_r = \omega_{\text{gap}}$, without the need of any Doppler shift correction as for the ST boundary case.

From Fig.~\ref{fig:int_states}(g) we observe two key differences from the ST boundary case: (i) the appearance of replicas of the interface state at frequencies that lie outside the band gap, highlighted by dashed lines, and (ii) the amplification of the transmitted wave with $T(\omega) > 1$. 

First, as reported in Ref.~\cite{STPhonIS} for a ST phononic crystal, the temporal modulation enables the excitation of the interface state from frequencies shifted in multiples of $\Omega$ thanks to the frequency conversion processes occurring at each boundary. This is confirmed by the appearance of a second-order transmission peak at $\omega^{\text{S}}_r - \Omega$ in Fig.~\ref{fig:int_states}(h)-(i) highlighted by a vertical line. To understand why this frequency conversion only happens for spatial slabs, we again put our attention on the conservation law for each boundary: conservation of the incident lab-frame frequency $\omega$ implies that an incoming plane wave excites all the modes of the band structure that lay in the horizontal line defined by a given $\omega_0$, whereas conservation of $\omega'$ implies exciting all the modes laying in a diagonal line with slope equal to $c_g$ \cite{UnVelSTcrys}. As a consequence, once we fold each mode into the BZ by shifting them an integer number of reciprocal vectors $\mathbf{p}$, we find that the spatial boundary excites an infinite number of modes inside the BZ with frequencies that differ an integer number of $\Omega$ with respect to $\omega_0$ \cite{FreqConvCaloz}. However, for the ST boundary, each mode along the diagonal line folds back into the same mode inside the BZ, such that we can only excite one truly distinct mode inside the material, explaining why we do not observe frequency conversion in this configuration. We have verified this by means of our semi-analytical scattering matrix formalism, which allows us to study the mode decomposition of the scattered fields, as well as the fields inside the slabs (see SI).

\begin{figure*}[t!]
    \includegraphics[width=\linewidth]{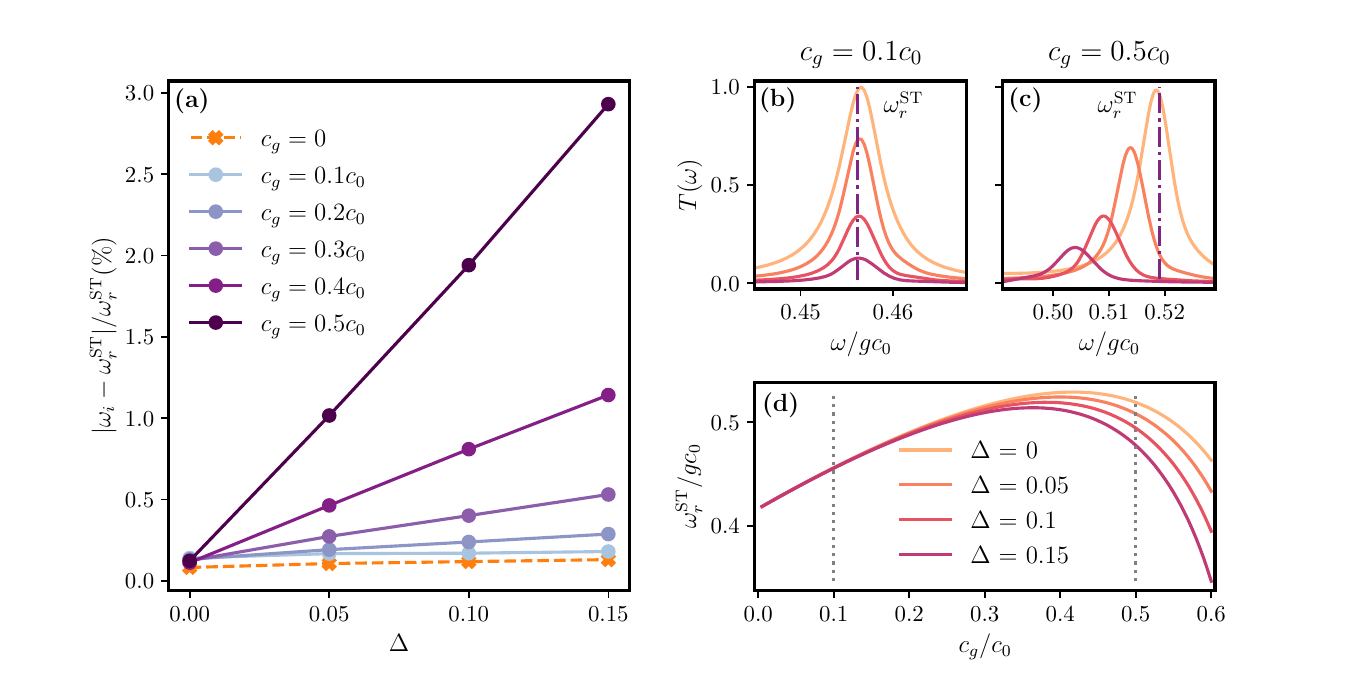}
    \caption{Robustness of interface states studied in the two-slab configuration of Fig.~\ref{fig:int_states}(a) by perturbing the system with an increased modulation strength in the second slab by $(\alpha+\Delta)$, with $\Delta$ the perturbation parameter.
    (a) Deviation of the numerically obtained resonance frequency $\omega_i$ from the bulk-predicted $\omega^{\mathrm{ST}}_r$ as we increase $\Delta$ and $c_g$.
    (b) Zoom in at the peaks of transmittance observed inside the gap for $c_g = 0.1 c_0$, with a vertical line showing $\omega_r^{\text{ST}}$ as a reference. 
    (c) Same magnitude for $c_g = 0.5 c_0$.
    (d) Analytical resonance frequency $\omega^{\mathrm{ST}}_r$ for different modulation strengths ($\alpha + \Delta$) as a function of the modulation speed $c_g$.
    }
    \label{fig:symm_pro}
\end{figure*}

Second, the amplification of the transmitted wave can be clearly observed by the horizontal bands of transmittance larger than one in Fig.~\ref{fig:int_states}(g), together with Fig.~\ref{fig:int_states}(i) where most part of the transmission spectrum present $T(\omega) > 1$, highlighting the broadband nature of this effect. Within our framework, we can identify frequency conversion processes as the underlying mechanism behind the amplification. Indeed, as shown in the SI, the permittivity encountered by the wave at the second and third interfaces determines whether higher-frequency modes are created in the scattering processes. For a given size of the slab, this condition is satisfied depending on the modulation velocity, explaining the periodic character of the amplification. Moreover, the amplification increases with $c_g$, since the wave experiences more modulation cycles while traversing the finite slab\acd{, allowing for a larger energy transfer from the temporally modulated medium to the wave}. In contrast, this effect is absent at ST boundaries, where frequency conversion does not occur. Crucially, this mechanism enables amplification even in the subluminal regime, without the need for the opening of a momentum gap as in PTC\acd{s} \cite{PhotonLoc}\acd{, facilitating potential experimental realizations with a reduced modulation speed.}

Finally, we plot in Fig.~\ref{fig:int_states}(j) the interface state excited by an incoming plane wave of frequency $\omega^\mathrm{S}_r$ considering a modulation velocity $c_g = 0.3 c_0$. As we can see, although the overall intensity profile presents a similar spatial structure to Fig.~\ref{fig:int_states}(e), the peak of the field's intensity remains pinned at the static interface, and the state does not propagate without perturbation along the boundary, in contrast to the ST boundary case.

\subsection{Robustness against perturbations}

An important aspect of interface states of topological origin is their inherent robustness against certain perturbations. In the previous section, we showed that the interface state presents robustness against an increase of the symmetry-breaking parameter $\xi'$. Indeed, not only did the states persist, but their resonance frequency remained pinned to the one predicted by the band-crossing position $\omega_r^{\text{ST/S}}$. Here, we extend this analysis by modifying the properties of one of the slabs, such that the two are no longer related by a simple spatial shift. Moreover, we also discuss the effect of truncating the crystal at arbitrary points rather than at the inversion centers. These configurations represent more general and realistic scenarios, allowing us to test whether the robustness of the interface states extends beyond idealized conditions that might not be met in experimental realizations.

Specifically, we revisit the configuration of ST slabs of Fig.~\ref{fig:int_states}(a), introducing a perturbation $\Delta$ that increases the modulation strength of the second slab to $\alpha+\Delta$. In Fig.~\ref{fig:symm_pro}(a), we evaluate numerically the resonance frequency $\omega_i$ of the interface state and plot it as a function of $\Delta$ and $c_g$ to study its deviation from the bulk prediction $\omega_r^{\text{ST}}$. Importantly, while this frequency plays a role similar to the mid-gap position in chiral-symmetric systems, here a deviation does not always signal a loss of bulk connection, and thus the source of the deviation needs to be carefully studied. In the standard case of a spatial 1D PhC, $c_g = 0$, we observe no deviation with $\Delta$, highlighting the robustness of the interface states, since they are completely insensitive to the perturbation. 
As we turn on and increase the modulation speed, however, the asymmetry in the properties of each slab becomes more relevant, as the deviation grows with increasing $\Delta$ and $c_g$. This effect is clearly observed when comparing the position of the transmission peaks for $c_g = 0.1c_0$ and $c_g = 0.5c_0$ in Fig.~\ref{fig:symm_pro}(b)-(c), respectively.

The deviation can be understood by noting that the effective parameters ($\epsilon_{\text{eff}}, \mu_{\text{eff}}$) used to derive the band-crossing position depend on the modulation strength when $c_g \neq 0$ \cite{Hom}. Indeed, Fig.~\ref{fig:symm_pro}(d) shows how the predicted resonance frequency $\omega_r^{\text{ST}}$ for a STPhC with modulation strength $\alpha+\Delta$ splits into distinct values for different $\Delta$ as the modulation speed increases. Consequently, in a finite system composed of two slabs with different modulation strengths, each slab predicts a distinct $\omega_r^{\text{ST}}$, explaining the deviation and the apparent loss of predictive power. Crucially, however, this does not imply a breakdown of bulk-interface correspondence, but rather highlights the need for a more refined bulk theory to determine the exact band-crossing position in such asymmetric slab configurations. To emphasize this point, we note that the deviation in frequency observed in Fig.~\ref{fig:symm_pro} is therefore different in nature from the one that would induce a phase perturbation $\phi$ in the modulation Eq.~\eqref{eq:eps}. Indeed, such a shift truncates the slabs at points different from the inversion centers of the unit cells, which obscures the distinction between trivial and topological phases by lifting the quantization of the Zak phase. While interface states may still appear in that configuration, as reported for the static case in Refs.~\cite{TopTruncStates,MarioTopPump}, their resonance frequency cannot be related anymore to the bulk-predicted value, signaling in that case a loss of topological nature.

\subsection{Superluminal regime}

Once we have characterized the topology of STPhCs in the subluminal regime, we now extend our formalism to investigate the interface states of superluminal STPhCs. To do so, we briefly discuss the main differences between regimes, and then establish the new reference frame in which we study the relevant symmetries and define the corresponding topological invariant.

\begin{figure}[h]
    \centering
    \includegraphics[width=240pt]{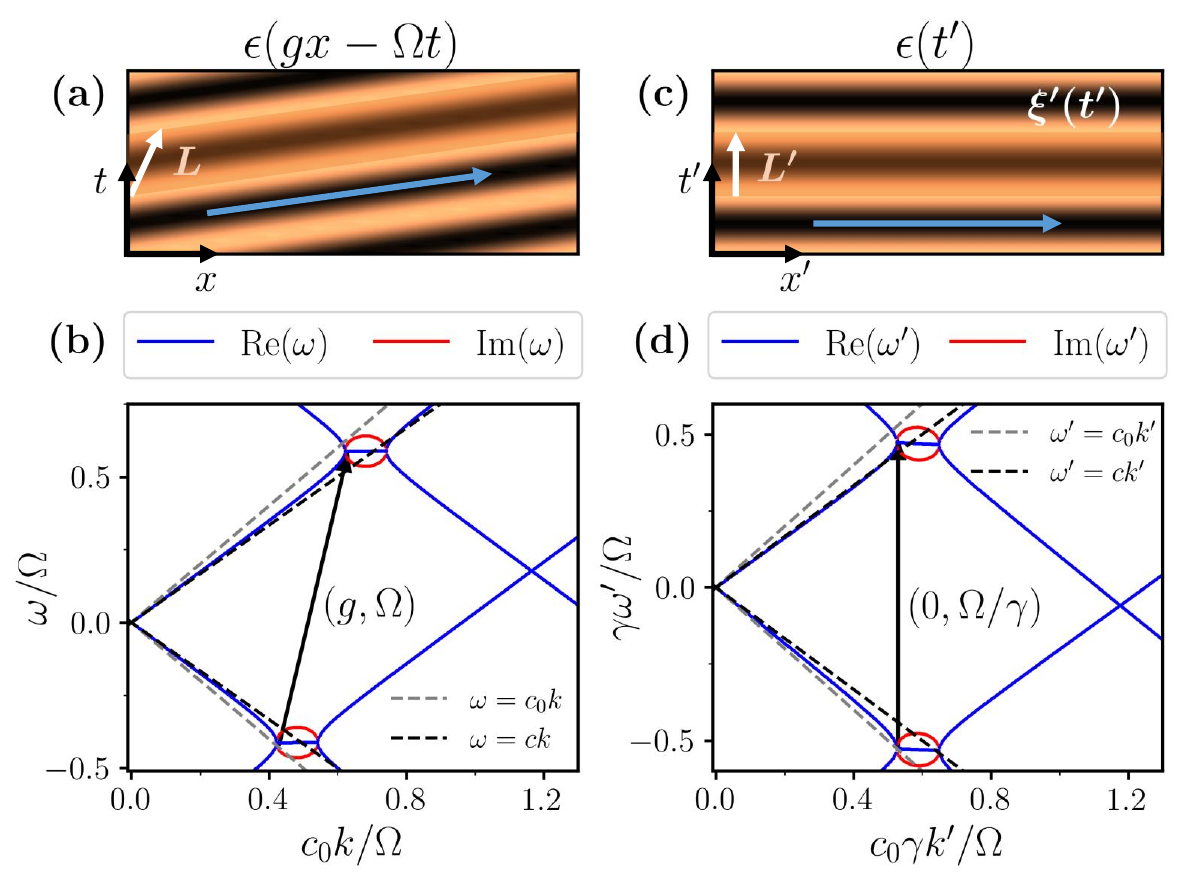}
    \caption{Spatiotemporal photonic crystal in the laboratory and time-like frames for superluminal modulation.
    (a) Travelling-wave modulated permittivity in the lab-frame. A blue arrow signals the direction of the continuous space-time translation symmetry that defines the unit cell highlighted in the orange shaded region. The white arrow corresponds to the lattice vector.
    (b) Band structure of a STPhC for a modulation speed $c_g=5c_0$, modulation strength $\alpha=0.4$ and $\epsilon_m=\mu_m=1.2$. Grey and black dashed lines correspond to the dispersion relation in free space and in the unmodulated material, respectively. Black arrow represents the reciprocal lattice vector $\textbf{p}$.
    (c) Modulated permittivity in the time-like frame. The previous magnitudes are now temporal-like and a magneto-electric coupling $\xi'(x')$ appears. (d) Band structure of the same STPhC in the comoving frame, with the black arrow now representing the transformed reciprocal vector $\textbf{p}'$.}
    \label{fig:sup_band_eps_dist}
\end{figure}

Figure ~\ref{fig:sup_band_eps_dist}(a) shows the permittivity distribution for a modulation speed of $c_g = 5 c_0$. As in the subluminal case, the system remains effectively 1D, exhibiting a moving grating profile and a continuous symmetry along the space–time direction. A key difference, however, is that although the reciprocal lattice vector is still given by $\textbf{p} = (g,\Omega)$, the lattice vector $\textbf{L}$ now has a larger temporal than spatial component, hinting at the time-like nature of the system. This feature is confirmed when we study its band structure, shown in Fig.~\ref{fig:sup_band_eps_dist}(b), where momentum gaps ($k$-gaps) with complex frequency appear, analogous to those found in PTCs. Recently, these $k$-gaps have attracted special attention due to their ability to amplify an incoming wave in time, as the usual energy conservation constraint is lifted \cite{Galiffi2022}. Furthermore, contrary to PTCs, the $k$-gaps found in superluminal STPhCs are asymmetric, which further enriches this phenomenology by enabling unidirectional amplification.

In order to apply the same methodology as in the subluminal case, we need to make a change of reference frame such that its magnitudes and symmetries are purely temporal or spatial. In this case, however, we cannot make the same transformation to a frame that comoves with the modulation, since the Lorentz factor $\gamma$ would give unphysical imaginary values due to the modulation speed being higher than the speed of light. Instead, given the temporal character of the band structure, we want to turn the modulation in the transformed frame purely temporal. A transformation into such a time-like frame is achieved by setting the frame velocity in the Lorentz transformation from Eqs.~\eqref{eq:Lor_trans_sp1}-\eqref{eq:Lor_trans_sp2} equal to \cite{UnVelSTcrys}
\begin{equation}\label{eq:frame_vel}
    c_f = c_0^2/c_g .
\end{equation}
Crucially, as it remains a Lorentz transformation, the structure of Maxwell's equations, as well as the expressions for the constitutive parameters, are the same as in the subluminal regime [Eqs.~\eqref{eq:Maxwell_com}-\eqref{eq:xi_com}] with the modulation speed $c_g$ changed to $c_f$.

Figure \ref{fig:sup_band_eps_dist}(c) shows the permittivity distribution as seen in the time-like frame, only periodic in $t'$ and presenting a moving-medium type magneto-electric coupling $\xi'(t')$. In this frame, the unit cell is defined in the temporal variable as $t' \in [-\gamma T/2, \gamma T/2]$, with a lattice vector $\textbf{L}' = [0,\gamma T]$, where $T = 2\pi/\Omega$ is the temporal period. Furthermore, we can use Eqs.~\eqref{eq:k_Lorent}-\eqref{eq:w_Lorentz} with $c_f$ to transform the reciprocal variables in the new frame $(k',\omega')$ and obtain the transformed reciprocal vector $\textbf{p}' = [0, \Omega/\gamma]$, as well as the time-like-frame BZ $\omega' \in [-\Omega/2\gamma,\Omega/2\gamma]$. Finally, making the inverse transformation, we obtain the lattice vector in the lab-frame as $\textbf{L} = \gamma^2 T \; [c_f, 1] = \gamma^2 T \; [c_0^2/c_g, 1]$. 

As we can see, an important consequence of the periodicty being present in time is that the BZ is now defined in the frequency dimension, similarly to Floquet systems \cite{TopSTCrystal,STsymSTgropu}. Therefore, to obtain the eigenfunctions in the time-like frame, we derive an eigenvalue equation for $k'$. Due to the temporal dependence of the constitutive matrix, Eq.~\eqref{eq:Maxwell_com} takes the form
\begin{equation}
    \begin{pmatrix}
        0 & \partial_{x'}\\
        \partial_{x'} & 0
    \end{pmatrix}
    \begin{bmatrix}
        E_z'\\
        H_y'
    \end{bmatrix} =
    \partial_{t'}\left( \hat{\textbf{M}}'(t') \begin{bmatrix}
        E_z'\\
        H_y'
    \end{bmatrix} \right).
\end{equation}
Using Eq.~\eqref{eq:const_trans}, we arrive at the following expression
\begin{equation}\label{eq:sup_Maxwell_com}
\hat{\textbf{M}}'(t')
    \begin{pmatrix}
        0 & \partial_{t'}\\
        \partial_{t'} & 0
    \end{pmatrix} 
    \begin{bmatrix}
        D_z'\\
        B_y'
    \end{bmatrix} =
    \partial_{x'} \begin{bmatrix}
        D_z'\\
        B_y'
    \end{bmatrix}, 
\end{equation}
where we clearly see that $[D'_z,B'_y]^T$ corresponds to the natural choice of eigenfunctions in the time-like frame. Finally, making use of the temporal periodicity, we apply a Floquet's ansatz to Eq.~\eqref{eq:sup_Maxwell_com} to obtain the dispersion relation $k'(\omega')$, together with the time-like frame eigenfunctions
\begin{equation}\label{eq:com_eigenfun_sup}
    \begin{bmatrix}
        D'(t')\\
        B'(t')
    \end{bmatrix}_{\omega',m}
    = e^{-i \omega' t'} \, 
    \begin{bmatrix}
        u{'}^D(t')\\
        u{'}^B(t')
    \end{bmatrix}_{\omega',m} ,
\end{equation}
where  $u{'}^{D/B}_{\omega',m}(t'+\gamma T) = u{'}^{D/B}_{\omega',m}(t')$ correspond to the periodic part of the eigenfields, and $m$ represents the band index. 

The corresponding band dispersion is shown in Fig.~\ref{fig:sup_band_eps_dist}(d). As before, the change of reference frame removes the asymmetry in the band gaps, owing to the absence of a momentum component in $\textbf{p}'$. However, the system still exhibits nonreciprocal behavior, as $k'(-\omega') \neq k'(\omega')$, highlighting the role of the bianisotropic term $\xi'(t')$. Indeed, since Maxwell’s equations in the time-like frame retain the same structure as in the comoving frame of the subluminal case, the system inherits the same symmetry properties in the new frame. Consequently, even though the constitutive parameters are symmetric under $\mathcal{T'}$, the magneto-electric coupling breaks parity and time-reversal symmetry individually, which leads to nonreciprocity. 

In this case, however, increasing the modulation speed $c_g$ does not necessarily lead to stronger symmetry breaking, since the bianisotropic coupling $\xi'(t')$ now depends on $c_f$. In the limit $c_g \rightarrow \infty$, corresponding to a purely temporal modulation where the system reduces to a PTC, one has $c_f \rightarrow 0$. In this regime, the magneto-electric coupling vanishes, and both parity and time-reversal symmetries are restored. This shows that, in either reference frame, it is the proximity to the luminal regime that induces the symmetry breaking, rather than an increase in the modulation speed.\\

\textit{Topological Characterization} - Importantly, $\mathcal{P'T'}$-symmetry is always preserved. Building on the topological framework developed for subluminal STPhCs, we examine whether a symmetry-protected topological invariant can also be defined through this symmetry in a system with a predominantly temporal character. To this end, we evaluate the Zak phase using the eigenfunctions of Eq.~\eqref{eq:com_eigenfun_sup} and integrate it over the BZ defined in the time-like frame
\begin{equation}\label{eq:STZak_sup}
    \theta^{\mathrm{ST}}_m =  i \int_{-\Omega/2\gamma}^{+\Omega/2\gamma} d\omega'\braket{\textbf{u}{'}_{\omega',m} | \partial_{\omega'} \textbf{u}'_{\omega',m}}_{{(\mathbf{\hat{M}}')^{-1}}},
\end{equation}
where the integrand corresponds to the Berry connection and $\textbf{u}'_{\omega',m} = [u{'}^D_{\omega',m}, u{'}^B_{\omega',m}]^T$. Crucially, {even if the system is clearly non-Hermitian due to its temporal modulation}, we can also relate the left and right eigenvectors of the system in the time-like frame thanks to the Hermiticity of the constitutive matrix and $\mathcal{P'T'}$ symmetry (see SI). As a result, we can use the conventional Berry connection defined with a weighted scalar product over the unit cell given by
\begin{equation}\label{eq:scalar_prod_sup}
    \braket{\mathbf{\Psi}'_1 | \mathbf{\Psi}'_2}_{{(\mathbf{\hat{M}}')^{-1}}} = \int_{- \gamma T /2}^{+\gamma T/2} dt' \,  \, \mathbf{\Psi}_1'^\dagger(t') \, \left( \mathbf{\hat{M}}'(t')\right)^{-1} \mathbf{\Psi}'_2(t'),
\end{equation}
with $\mathbf{\Psi}'_{\omega',m} = [D'_{\omega',m}, B'_{\omega',m}]^T$ the time-like-frame eigenfunctions, normalized by establishing $\braket{\mathbf{\Psi}'_{\omega',m}|\mathbf{\Psi}'_{\omega',m}}_{{(\mathbf{\hat{M}}')^{-1}}} = 1$. Therefore, since we are able to write the ST Zak phase with a Hermitian Berry connection, we can make use of the results of Ref.~\cite{BerrConnProof} to conclude that $\mathcal{P'T'}$-symmetry alone allows the quantization of the ST Zak phase defined in Eq.~\eqref{eq:STZak_sup}. With this, we have derived a $\mathbb{Z}_2$ topological invariant for superluminal STPhCs, extending our topological characterization of STPhCs to its whole modulation regime. {This result unifies a wide class of systems, ranging from spatial PhCs to PTCs, and brings further light into why the Hermitian definition of the Zak phase was able to characterize the topology of PTC's band structures in previous works \cite{Lustig2018}.}\\

Building on this result, we now investigate the existence of interface states between two superluminal STPhC slabs characterized by different ST Zak phases. To do so, we consider two types of truncations: a spatiotemporal boundary that follows the space–time symmetry highlighted in Fig.~\ref{fig:sup_band_eps_dist}(a), and a purely temporal boundary.

These configurations correspond to the temporal counterparts of the cases studied in the subluminal regime. However, the temporal nature of the boundaries introduces important differences. First, frequency is no longer conserved, but rather momentum is: $k'$ for spatiotemporal boundaries and $k$ for purely temporal ones. Second, temporal reflection is fundamentally different from conventional spatial reflection. Due to causality, a temporally reflected wave cannot propagate within the same space-time region as the incident wave, as this would imply backward propagation in time. Instead, both reflected and transmitted waves remain in the same medium after the boundary, but each one propagates in a different spatial direction. Taking these considerations into account, we generalize our scattering-matrix formalism to compute the interface states in superluminal STPhCs and explore their distinctive features (see SI).\\

\begin{figure*}[h!]
    \centering
    \includegraphics[width=\linewidth]{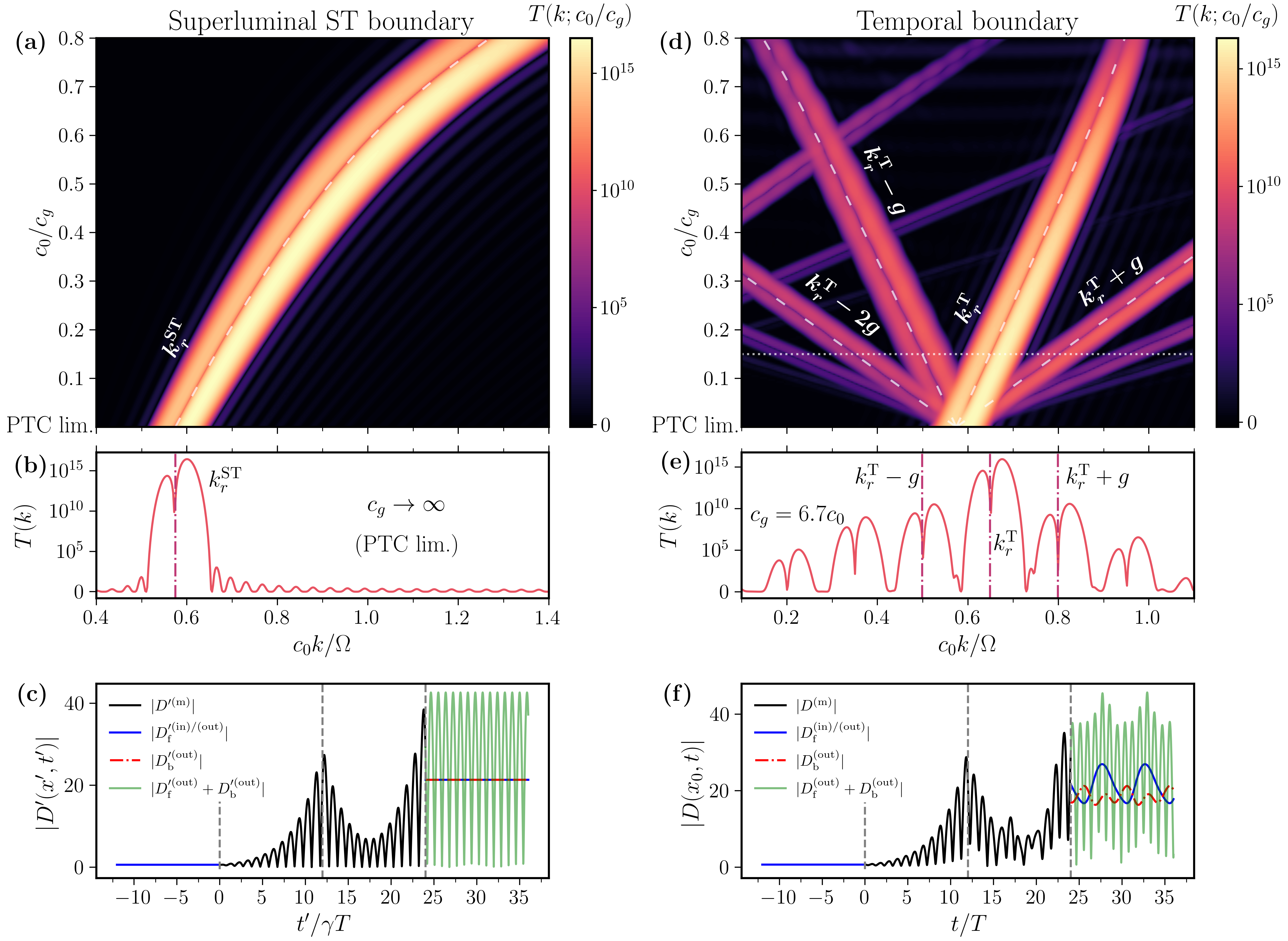}
    \caption{Interface states in the superluminal regime between a trivial ($\alpha>0$) and a topological ($\alpha<0$) spatiotemporal slab, considering (a)-(c) a spatiotemporal boundary and (d)-(f) a temporal boundary.
    (a) Transmittance spectrum of two spatiotemporal slabs whose boundary moves along with the modulation, as a function of $c_0/c_g$. An interface state with resonance frequency following the bulk-predicted $k^{\text{ST}}_r$ is observed. 
    (b) Horizontal cross section of the transmittance at the PTC limit $c_g \rightarrow \infty$. A clear dip inside the amplifying gap corresponding to the interface state is visible. 
    (c) Absolute value of the displacement field {for $c_g = 5c_0$ and twelve unit cells per slab} as a function of time at every region of space-time delimited by grey dashed lines: vacuum before the slabs ($D{'}_{\mathrm{f}}^{(\mathrm{in})}$), inside the two STPhC slabs ($D{'}^{(\mathrm{m})}$), and vacuum after the slabs ($D{'}_{\mathrm{f/b}}^{(\mathrm{out})}$). Localization at the temporal interface is observed.
    (d) Transmittance spectrum as a function of $c_0/c_g$ for two slabs of the STPhC with purely temporal boundaries. An interface state, as well as its replicas at $k_r^{\text{S}} \pm m g$ due to momentum conversion, are observed similar to the subluminal case.
    (e) Horizontal cross section of the transmittance for $c_g = 6.7 c_0$, revealing the interface state inside the gap as well as the many replicas.
    (j) Absolute value of the fields at different regions of space-time. Temporal dependence of the transmitted wave shows the presence of different harmonic modes, excited due to momentum conversion.}
    \label{fig:sup_trans}
\end{figure*}

\textit{Superluminal Spatiotemporal Interface} - We first consider two slabs, each composed of fifteen ST unit cells, with the same parameters as in Fig.~\ref{fig:sup_band_eps_dist}. To visualize the composite system in a space-time diagram analogous to the subluminal case [Fig.~\ref{fig:int_states}(a)], one may invoke the space–time duality between modulation regimes. In particular, subluminal and superluminal systems can be related by reflecting the slabs along the light line \cite{UnVelSTcrys}. Under this transformation, the permittivity distribution of the two superluminal slabs can be visualized by interchanging the spatial and temporal axes in Fig.~\ref{fig:int_states}(a). A subtlety arises, however, when considering the wave excitation of the composite system. Since the velocity of the slab exceeds the wave velocity, the usual picture of a wave impinging on a boundary no longer applies. Instead, in order to excite the finite system, it is the slab that sweeps over the wave.

Figure \ref{fig:sup_trans}(a) shows the transmittance spectrum of the two ST slabs as a function of the inverse modulation speed $c_0/c_g = c_f/c_0$, obtained by exciting the composite system with plane waves of momentum $k$ at normal incidence. An horizontal cross section at the PTC limit is shown in Fig.~\ref{fig:sup_trans}(b), presenting remarkable amplification of the transmitted wave due to the complex frequency modes in the $k$-gaps, as well as a dip in transmission. {Interestingly, we note that this dip has been predicted and observed experimentally in PTCs, as well as its associated interface states localized in time \cite{PTacousticFloq,
ExpTopPTC}. In our case, we observe such a dip in transmission throughout the whole modulation regime, therefore confirming} the existence of robust interface states also in superluminal STPhCs. 

Furthermore, we can relate the position of the boundary state through the band-crossing point derived for the subluminal regime in Eq.~\eqref{eq:freq_gap} with 
\begin{equation} \label{mom_gap}
k_{\mathrm{gap}} = \omega_{\mathrm{gap}}/v_{\mathrm{eff}}.
\end{equation}
Then, since we need to account for the conservation of $k'$, we know that the incident wave with $k_0$ will couple to a mode inside the material with momentum $k_{\mathrm{mat}}$ given by
\begin{equation}\label{eq:Doppler_mom}
    k'/\gamma = (1-c_f/v_{\text{eff}}) \, k_{\text{mat}} = (1-c_f/c_0) \, k_0 .
\end{equation}
Again, the enforcement of this modified conservation law explains why the position of the band gap does not grow linearly with $c_f$, but rather presents a more intricate dependence. Taking this into account, we can derive an analytical expression for the resonant momentum of the interface state as
\begin{equation}\label{eq:ST_res_mom}
    k^{\mathrm{ST}}_r =  \frac{1-c_f/v_{\text{eff}}}{1-c_f/c_0} \, \;k_{\text{gap}}.
\end{equation}
This magnitude is represented as a function of $c_0/c_g$ in Fig.~\ref{fig:sup_trans}(a) with a dashed line. As we can see, it perfectly follows the dip in transmission, which confirms the topological nature of the state in the superluminal regime.

Finally, in Fig.~\ref{fig:sup_trans}(c) we plot the field distribution of the interface state $|D_z'(x',t')|$ in the time-like frame across each region of the composite system, delimited by grey dashed lines: vacuum before the slabs ($D{'}_{\mathrm{f}}^{(\mathrm{in})}$), inside the positive and negative modulation strength $\alpha$ slabs ($D{'}_{\mathrm{f}}^{(\mathrm{in})}$), and vacuum after the slabs ($D{'}_{\mathrm{f/b}}^{(\mathrm{out})}$).  As shown, the state is pinned to the temporal interface $t' = 12 \, \gamma T$, where it presents a sharp decay within an otherwise amplifying medium. As a result, the field amplitude grows again until it reaches the third boundary, which leads to a reduced amplification of the transmitted wave. This behavior explains why the interface state manifests as a dip in the transmittance spectrum. In the time-like frame, the localization in {the transformed time $t'$ } is analogous to the interface states observed in PTCs. However, upon transforming back to the lab frame, we see the state propagating {at a superluminal speed} together with the moving boundary. Finally, beyond the slabs, two plane waves with constant absolute value are observed, propagating in opposite directions with an amplified amplitude.\\

\textit{Temporal Interface} - We now turn to the case of two temporal slabs of the same superluminal STPhC with $\alpha$ of opposite signs, composed of purely temporal unit cells of period $T$, whose permittivity distribution can be obtained by interchanging the spatial and temporal axes of Fig.~\ref{fig:int_states}(f). For this configuration, Fig.~\ref{fig:sup_trans}(d) shows the transmittance spectrum of a forward-propagating incident wave, analogous to the ST-boundary case. A transmission dip is again observed within the $k$-gap throughout the superluminal modulation regime, confirming the presence of interface states. Moreover, since the conserved quantity at a temporal boundary is the standard momentum $k$, the resonant momentum can be accurately predicted from Eq.~\eqref{mom_gap} as $k_r^{\mathrm{T}} = k_{\mathrm{gap}}$.

More interestingly, the conservation of $k$, together with the tilted band structure shown in Fig.~\ref{fig:sup_band_eps_dist}(b), leads to momentum-conversion at each scattering processes analogous to the frequency-conversion observed in the subluminal regime. Through the same mechanism, replicas of the $k$-gaps, and therefore of the interface states, are created. In this regime, these replicas are more pronounced due to the amplifying nature of the gaps, which allows the observation of higher-order excitations. Indeed, Fig.~\ref{fig:sup_trans}(e) presents a horizontal cross section of the transmittance spectrum at $c_g = 6.7c_0$, where up to fourth-order replicas can be observed at lower momentum values. Furthermore, since the spacing between replicas depends on the modulation spatial frequency $g$, the gaps can overlap at sufficiently small $g$ (high modulation speeds), resulting in broadband amplification that is significantly stronger than that observed in subluminal STPhCs.

Finally, Fig.~\ref{fig:sup_trans}(f) shows the field distribution $|D_z(x_0,t)|$ of the interface state at a fixed spatial position $x_0$ in the lab frame. Although the temporal dependence is more intricate, the state remains localized in time at the interface $t = 12 \, T$. Beyond the slabs, forward- ($D_{\mathrm{f}}^{(\mathrm{out})}$) and backward-propagating waves ($D_{\mathrm{b}}^{(\mathrm{out})}$) are observed whose absolute value vary in time. This indicates that both transmitted and reflected fields contain multiple harmonic components, which is a direct consequence of momentum conversion.

\section{Conclusions}

In this work, we investigate the topological origin of interface states in spatiotemporal photonic crystals with travelling-wave modulation and uncover their unique properties. Through the appropriate consideration of the symmetries present in the system, we provide a framework to topologically classify time-dependent modulations of travelling-wave type \super{for both subluminal and superluminal regimes}. Notably, we show that computing topological invariants in a frame \super{where} the modulation \super{depends on only one transformed variable, either space for the subluminal regime or time for the superluminal one,} leads to a well-defined and meaningful topological classification. \super{Using Lorentz transformations,} we show that the spatiotemporal counterpart of parity-time-reversal symmetry is conserved. This symmetry enforces the quantization of the spatiotemporal Zak phase defined along the \super{transformed} Brillouin Zone, yielding a $\mathbb{Z}_2$ topological invariant. By calculating the electromagnetic band energy density, we distinguish the two resulting phases as a trivial phase and an obstructed atomic limit, which clarifies the topological nature of interface states in spatiotemporal photonic crystals without further symmetries. \super{Our formalism holds for arbitrary modulation velocities, bridging a wide range of systems from conventional photonic crystals to photonic time crystals, and providing new insights into the topological characterization of the latter.}

We prove this point by calculating semi-analytically the interface states that arise between slabs of different spatiotemporal Zak phase for \super{different} types of boundaries \super{depending on the modulation regime: spatial, temporal and spatiotemporal.} From the transmittance spectra, we find that\super{, for each boundary,} an in-gap state exists \super{in both subluminal and superluminal regimes}, and that its resonance frequency\super{, or momentum,} can be accurately predicted \acd{solely} from bulk properties. The interface states also present novel features stemming from the non-reciprocal nature of travelling-wave media and the lack of energy conservation. For a spatiotemporal interface, the topological state propagates along with the moving boundary \super{even at superluminal speeds}, a remarkable feature in an effectively one-dimensional system. In contrast, \super{both} static \super{and temporal} interfaces introduces frequency \super{or momentum} conversion at each scattering event, producing replicas of the interface states and broadband amplification of the transmitted wave. Futhermore, non-reciprocity in this system also enables selective excitation\super{, which in the superluminal regime gives rise to directional amplification due to the presence of momentum gaps.}

Our results highlight the potential of temporal modulation to enrich the properties of interface states in spatiotemporal crystals. This paves the way to explore time-varying effects in higher-dimensional systems with additional symmetries, where richer topological phases already exist in the static limit. Although this work focuses on photonic systems, our predictions are general and extend to other wave platforms where spatiotemporal interfaces and travelling-wave modulations have already been experimentally demonstrated: elastic materials such as elastic strips \cite{LemoultSTInt} or piezoelectric crystals \cite{STPiezoelectric}, acoustic systems \cite{AcousPump}, and transmission lines \cite{ExpTransLines}. This underscores the feasibility of realizing these effects experimentally and broadens their potential applications beyond those of their static counterparts \cite{ExpFanoResTopState,Exp1DTopState,TopNanoCavLas}. Altogether, these directions point to a broader landscape in which spatiotemporal modulation becomes a key ingredient for engineering novel topological phases for wave control.

\section*{Acknowledgements}

We acknowledge financial support from the EU (ERC  grant TIMELIGHT, GA 101115792) and MCIUN/AEI (PID2022-141036NA-I00 through MCIUN/AEI/10.1303 9/501100011033, FSE+ and FPI grant PREP2022-000455; RYC2021-031568-I; Programme for Units of Excellence in R\&D CEX2023-001316-M).

\bibliography{ref}

@article{Galiffi2022,
    author = {Emanuele Galiffi and Romain Tirole and Shixiong Yin and Huanan Li and Stefano Vezzoli and Paloma A. Huidobro and M{\'a}rio G. Silveirinha and Riccardo Sapienza and Andrea Al{\`u} and J. B. Pendry},
    title = {{Photonics of time-varying media}},
    volume = {4},
    journal = {Adv. Photonics},
    number = {1},
    publisher = {SPIE},
    pages = {014002},
    year = {2022},
    doi = {10.1117/1.AP.4.1.014002},
    URL = {https://doi.org/10.1117/1.AP.4.1.014002}
}

@article{PTCTut,
author = {Mohammad M. Asgari and Puneet Garg and Xuchen Wang and Mohammad S. Mirmoosa and Carsten Rockstuhl and Viktar Asadchy},
journal = {Adv. Opt. Photon.},
number = {4},
pages = {958--1063},
publisher = {Optica Publishing Group},
title = {Theory and applications of photonic time crystals: a tutorial},
volume = {16},
month = {Dec},
year = {2024},
url = {https://opg.optica.org/aop/abstract.cfm?URI=aop-16-4-958},
doi = {10.1364/AOP.525163},
}

@article{FreqMomTransAcous,
author = {Zhaoxian Chen  and Yugui Peng  and Haoxiang Li  and Jingjing Liu  and Yujiang Ding  and Bin Liang  and Xue-Feng Zhu  and Yanqing Lu  and Jianchun Cheng  and Andrea Alù },
title = {Efficient nonreciprocal mode transitions in spatiotemporally modulated acoustic metamaterials},
journal = {Science Advances},
volume = {7},
number = {45},
pages = {eabj1198},
year = {2021},
doi = {10.1126/sciadv.abj1198},
URL = {https://www.science.org/doi/abs/10.1126/sciadv.abj1198}
}

@article{QHEPhCs,
  title = {Possible Realization of Directional Optical Waveguides in Photonic Crystals with Broken Time-Reversal Symmetry},
  author = {Haldane, F. D. M. and Raghu, S.},
  journal = {Phys. Rev. Lett.},
  volume = {100},
  issue = {1},
  pages = {013904},
  numpages = {4},
  year = {2008},
  month = {Jan},
  publisher = {American Physical Society},
  doi = {10.1103/PhysRevLett.100.013904},
  url = {https://link.aps.org/doi/10.1103/PhysRevLett.100.013904}
}

@article{ObsUniTopStates,
  author       = {Wang, Zheng and Chong, Yidong and Joannopoulos, J. D. and Soljačić, Marin},
  title        = {Observation of unidirectional backscattering-immune topological electromagnetic states},
  journal      = {Nature},
  year         = {2009},
  volume       = {461},
  number       = {7265},
  pages        = {772--775},
  month        = {October},
  doi          = {10.1038/nature08293},
  url          = {https://doi.org/10.1038/nature08293},
}

@article{Lustig2018,
    author = {Eran Lustig and Yonatan Sharabi and Mordechai Segev},
    journal = {Optica},
    number = {11},
    pages = {1390--1395},
    publisher = {Optica Publishing Group},
    title = {Topological aspects of photonic time crystals},
    volume = {5},
    month = {Nov},
    year = {2018},
    url = {https://opg.optica.org/optica/abstract.cfm?URI=optica-5-11-1390},
    doi = {10.1364/OPTICA.5.001390},
}

@article{SurfImp,
  title = {Surface Impedance and Bulk Band Geometric Phases in One-Dimensional Systems},
  author = {Xiao, Meng and Zhang, Z. Q. and Chan, C. T.},
  journal = {Phys. Rev. X},
  volume = {4},
  issue = {2},
  pages = {021017},
  numpages = {12},
  year = {2014},
  month = {Apr},
  publisher = {American Physical Society},
  doi = {10.1103/PhysRevX.4.021017},
  url = {https://link.aps.org/doi/10.1103/PhysRevX.4.021017}
}

@article{ZakPhase,
  title = {Berry's phase for energy bands in solids},
  author = {Zak, J.},
  journal = {Phys. Rev. Lett.},
  volume = {62},
  issue = {23},
  pages = {2747--2750},
  numpages = {0},
  year = {1989},
  month = {Jun},
  publisher = {American Physical Society},
  doi = {10.1103/PhysRevLett.62.2747},
  url = {https://link.aps.org/doi/10.1103/PhysRevLett.62.2747}
}

@article{AcousPump,
  title = {Physical Observation of a Robust Acoustic Pumping in Waveguides with Dynamic Boundary},
  author = {Xu, Xianchen and Wu, Qian and Chen, Hui and Nassar, Hussein and Chen, Yangyang and Norris, Andrew and Haberman, Michael R. and Huang, Guoliang},
  journal = {Phys. Rev. Lett.},
  volume = {125},
  issue = {25},
  pages = {253901},
  numpages = {6},
  year = {2020},
  month = {Dec},
  publisher = {American Physical Society},
  doi = {10.1103/PhysRevLett.125.253901},
  url = {https://link.aps.org/doi/10.1103/PhysRevLett.125.253901}
}

@article{STsymSTgropu,
  title = {Space-Time Crystal and Space-Time Group},
  author = {Xu, Shenglong and Wu, Congjun},
  journal = {Phys. Rev. Lett.},
  volume = {120},
  issue = {9},
  pages = {096401},
  numpages = {6},
  year = {2018},
  month = {Feb},
  publisher = {American Physical Society},
  doi = {10.1103/PhysRevLett.120.096401},
  url = {https://link.aps.org/doi/10.1103/PhysRevLett.120.096401}
}

@article{TopSTCrystal,
  title = {Topological Space-Time Crystal},
  author = {Peng, Yang},
  journal = {Phys. Rev. Lett.},
  volume = {128},
  issue = {18},
  pages = {186802},
  numpages = {6},
  year = {2022},
  month = {May},
  publisher = {American Physical Society},
  doi = {10.1103/PhysRevLett.128.186802},
  url = {https://link.aps.org/doi/10.1103/PhysRevLett.128.186802}
}

@article{ChiralTIS,
author = {Yang, Yukun and Hu, Hao and Liu, Liangliang and Yang, Yihao and Yu, Youxiu and Long, Yang and Zheng, Xuezhi and Luo, Yu and Li, Zhuo and Garcia-Vidal, Francisco J.},
title = {Topologically Protected Edge States in Time Photonic Crystals with Chiral Symmetry},
journal = {ACS Photonics},
volume = {12},
number = {5},
pages = {2389-2396},
year = {2025},
doi = {10.1021/acsphotonics.4c01785},
URL = {https://doi.org/10.1021/acsphotonics.4c01785}}

@article{StaCrysWannCent,
  title = {Topological phases of photonic crystals under crystalline symmetries},
  author = {Vaidya, Sachin and Ghorashi, Ali and Christensen, Thomas and Rechtsman, Mikael C. and Benalcazar, Wladimir A.},
  journal = {Phys. Rev. B},
  volume = {108},
  issue = {8},
  pages = {085116},
  numpages = {25},
  year = {2023},
  month = {Aug},
  publisher = {American Physical Society},
  doi = {10.1103/PhysRevB.108.085116},
  url = {https://link.aps.org/doi/10.1103/PhysRevB.108.085116}
}

@Article{TopQChem,
    author={Bradlyn, Barry
    and Elcoro, L.
    and Cano, Jennifer
    and Vergniory, M. G.
    and Wang, Zhijun
    and Felser, C.
    and Aroyo, M. I.
    and Bernevig, B. Andrei},
    title={Topological quantum chemistry},
    journal={Nature},
    year={2017},
    month={Jul},
    day={01},
    volume={547},
    number={7663},
    pages={298-305},
    issn={1476-4687},
    doi={10.1038/nature23268},
    url={https://doi.org/10.1038/nature23268}
}

@article{WannCentSpec,
  title = {Wannier center spectroscopy to identify boundary-obstructed topological insulators},
  author = {Ligthart, R. A. M. and Herrera, M. A. J. and Visser, A. C. H. and Vlasblom, A. and Bercioux, D. and Swart, I.},
  journal = {Phys. Rev. Res.},
  volume = {7},
  issue = {1},
  pages = {012076},
  numpages = {6},
  year = {2025},
  month = {Mar},
  publisher = {American Physical Society},
  doi = {10.1103/PhysRevResearch.7.012076},
  url = {https://link.aps.org/doi/10.1103/PhysRevResearch.7.012076}
}

@article{Fresnel,
author = {Paloma A. Huidobro  and Emanuele Galiffi  and Sébastien Guenneau  and Richard V. Craster  and J. B. Pendry },
title = {Fresnel drag in space–time-modulated metamaterials},
journal = {Proc. Natl. Acad. Sci. U.S.A.},
volume = {116},
number = {50},
pages = {24943-24948},
year = {2019},
doi = {10.1073/pnas.1915027116},
URL = {https://www.pnas.org/doi/abs/10.1073/pnas.1915027116}}

@article{Hom,
  title = {Homogenization Theory of Space-Time Metamaterials},
  author = {Huidobro, P.A. and Silveirinha, M.G. and Galiffi, E. and Pendry, J.B.},
  journal = {Phys. Rev. Appl.},
  volume = {16},
  issue = {1},
  pages = {014044},
  numpages = {13},
  year = {2021},
  month = {Jul},
  publisher = {American Physical Society},
  doi = {10.1103/PhysRevApplied.16.014044},
  url = {https://link.aps.org/doi/10.1103/PhysRevApplied.16.014044}
}

@article{EMnonRec,
  title = {Electromagnetic Nonreciprocity},
  author = {Caloz, Christophe and Al\`u, Andrea and Tretyakov, Sergei and Sounas, Dimitrios and Achouri, Karim and Deck-L\'eger, Zo\'e-Lise},
  journal = {Phys. Rev. Appl.},
  volume = {10},
  issue = {4},
  pages = {047001},
  numpages = {26},
  year = {2018},
  month = {Oct},
  publisher = {American Physical Society},
  doi = {10.1103/PhysRevApplied.10.047001},
  url = {https://link.aps.org/doi/10.1103/PhysRevApplied.10.047001}
}

@article{PTexactTop,
  title = {Coalescence of exceptional points and phase diagrams for one-dimensional $\mathcal{P}\mathcal{T}$-symmetric photonic crystals},
  author = {Ding, Kun and Zhang, Z. Q. and Chan, C. T.},
  journal = {Phys. Rev. B},
  volume = {92},
  issue = {23},
  pages = {235310},
  numpages = {7},
  year = {2015},
  month = {Dec},
  publisher = {American Physical Society},
  doi = {10.1103/PhysRevB.92.235310},
  url = {https://link.aps.org/doi/10.1103/PhysRevB.92.235310}
}

@article{NonBlochTheoSTcrys,
  title = {Non-Bloch theory for spatiotemporal photonic crystals assisted by continuum effective medium},
  author = {Ding, Haozhi and Ding, Kun},
  journal = {Phys. Rev. Res.},
  volume = {6},
  issue = {3},
  pages = {033167},
  numpages = {12},
  year = {2024},
  month = {Aug},
  publisher = {American Physical Society},
  doi = {10.1103/PhysRevResearch.6.033167},
  url = {https://link.aps.org/doi/10.1103/PhysRevResearch.6.033167}
}

@article{STsymm,
author = {Liberal, I{\~n}igo and Ganfornina-Andrades, Antonio and Vázquez-Lozano, J. Enrique},
title = {Spatiotemporal Symmetries and Energy-Momentum Conservation in Uniform Spacetime Metamaterials},
journal = {ACS Photonics},
volume = {11},
number = {12},
pages = {5273-5280},
year = {2024},
doi = {10.1021/acsphotonics.4c01496},
URL = {https://doi.org/10.1021/acsphotonics.4c01496}}

@article{STPhonIS,
  title = {Space-time phononic crystals with anomalous topological edge states},
  author = {Oudich, Mourad and Deng, Yuanchen and Tao, Molei and Jing, Yun},
  journal = {Phys. Rev. Res.},
  volume = {1},
  issue = {3},
  pages = {033069},
  numpages = {9},
  year = {2019},
  month = {Nov},
  publisher = {American Physical Society},
  doi = {10.1103/PhysRevResearch.1.033069},
  url = {https://link.aps.org/doi/10.1103/PhysRevResearch.1.033069}
}

@book{Kong,
  title={Electromagnetic Wave Theory},
  author={Kong, J.A.},
  isbn={9780471828235},
  lccn={lc85009554},
  series={A Wiley-Interscience publication},
  url={https://books.google.es/books?id=OposAAAAYAAJ},
  year={1986},
  publisher={Wiley}
}

@book{vanderbilt,
  title={Berry Phases in Electronic Structure Theory: Electric Polarization, Orbital Magnetization and Topological Insulators},
  author={Vanderbilt, D.},
  isbn={9781107157651},
  lccn={2018018455},
  series={Titolo collana},
  url={https://books.google.es/books?id=485FtgEACAAJ},
  year={2018},
  publisher={Cambridge University Press}
}

@article{EnergyDens,
doi = {10.1088/1361-648X/ac73cf},
url = {https://dx.doi.org/10.1088/1361-648X/ac73cf},
year = {2022},
month = {jun},
publisher = {IOP Publishing},
volume = {34},
number = {31},
pages = {314002},
author = {Blanco de Paz, M and Herrera, M A J and Arroyo Huidobro, P and Alaeian, H and Vergniory, M G and Bradlyn, B and Giedke, G and García-Etxarri, A and Bercioux, D},
title = {Energy density as a probe of band representations in photonic crystals},
journal = {J. Phys. Condens. Matter},
}

@article{UnVelSTcrys,
author = {Zo{\'e}-Lise Deck-L{\'e}ger and Nima Chamanara and Maksim Skorobogatiy and M{\'a}rio G. Silveirinha and Christophe Caloz},
title = {{Uniform-velocity spacetime crystals}},
volume = {1},
journal = {Adv. Photonics},
number = {5},
publisher = {SPIE},
pages = {056002},
year = {2019},
doi = {10.1117/1.AP.1.5.056002},
URL = {https://doi.org/10.1117/1.AP.1.5.056002}
}

@article{FreqConvCaloz,
  title = {Optical isolation based on space-time engineered asymmetric photonic band gaps},
  author = {Chamanara, Nima and Taravati, Sajjad and Deck-L\'eger, Zo\'e-Lise and Caloz, Christophe},
  journal = {Phys. Rev. B},
  volume = {96},
  issue = {15},
  pages = {155409},
  numpages = {12},
  year = {2017},
  month = {Oct},
  publisher = {American Physical Society},
  doi = {10.1103/PhysRevB.96.155409},
  url = {https://link.aps.org/doi/10.1103/PhysRevB.96.155409}
}

@article{Tenfold,
doi = {10.1088/1367-2630/12/6/065010},
url = {https://dx.doi.org/10.1088/1367-2630/12/6/065010},
year = {2010},
month = {jun},
publisher = {},
volume = {12},
number = {6},
pages = {065010},
author = {Ryu, Shinsei and Schnyder, Andreas P and Furusaki, Akira and Ludwig, Andreas W W},
title = {Topological insulators and superconductors: tenfold way and dimensional hierarchy},
journal = {New J. Phys.},
}

@article{IsoRefSTCrystal,
  title        = {Replicating Physical Motion with {Minkowskian} Isorefractive Spacetime Crystals},
  author       = {Prud\^encio, Filipa R. and Silveirinha, Mário G.},
  journal      = {Nanophotonics},
  volume       = {12},
  number       = {14},
  pages        = {3007--3017},
  year         = {2023},
  doi          = {10.1515/nanoph-2023-0144},
  issn         = {2192-8606 (print), 2192-8614 (online)},
  url          = {https://doi.org/10.1515/nanoph-2023-0144},
}

@article{NonRecMink,
  title = {Engineering nonreciprocal responses in traveling-wave spacetime crystals},
  author = {Prud\^{e}ncio, F. R. and Silveirinha, M. G.},
  journal = {Phys. Rev. Appl.},
  volume = {22},
  number = {5},
  pages = {054020},
  year = {2024},
  doi = {10.1103/PhysRevApplied.22.054020}
}

@article{STGratings,
  title = {Analysis of Metallic Space-Time Gratings Using {Lorentz} Transformations},
  author = {Alex-Amor, Antonio and Molero, Carlos and Silveirinha, M\'ario G.},
  journal = {Phys. Rev. Appl.},
  volume = {20},
  issue = {1},
  pages = {014063},
  numpages = {15},
  year = {2023},
  month = {Jul},
  publisher = {American Physical Society},
  doi = {10.1103/PhysRevApplied.20.014063},
  url = {https://link.aps.org/doi/10.1103/PhysRevApplied.20.014063}
}

@article{BerrConnProof,
doi = {10.1088/2053-1583/4/1/015023},
url = {https://dx.doi.org/10.1088/2053-1583/4/1/015023},
year = {2016},
month = {nov},
publisher = {IOP Publishing},
volume = {4},
number = {1},
pages = {015023},
author = {van Miert, Guido and Ortix, Carmine and Smith, Cristiane Morais},
title = {Topological origin of edge states in two-dimensional inversion-symmetric insulators and semimetals},
journal = {2D Mater.},
}

@article{ExpGeomPhase,
  title        = {Geometric phase and band inversion in periodic acoustic systems},
  author       = {Xiao, Meng and Ma, Guancong and Yang, Zhiyu and Sheng, Ping and Zhang, Z. Q. and Chan, C. T.},
  journal      = {Nat. Phys.},
  volume       = {11},
  number       = {3},
  pages        = {240--244},
  year         = {2015},
  doi          = {10.1038/nphys3228},
}

@article{TopTruncStates,
  title = {Topological Boundary Modes from Translational Deformations},
  author = {Nakata, Yosuke and Ito, Yoshitaka and Nakamura, Yasunobu and Shindou, Ryuichi},
  journal = {Phys. Rev. Lett.},
  volume = {124},
  issue = {7},
  pages = {073901},
  numpages = {6},
  year = {2020},
  month = {Feb},
  publisher = {American Physical Society},
  doi = {10.1103/PhysRevLett.124.073901},
  url = {https://link.aps.org/doi/10.1103/PhysRevLett.124.073901}
}

@article{ZakPhMeas,
  author       = {Atala, Marcos and Aidelsburger, Monika and Barreiro, Julio T. and Abanin, Dmitry and Kitagawa, Takuya and Demler, Eugene and Bloch, Immanuel},
  title        = {Direct measurement of the {Zak} phase in topological {Bloch} bands},
  journal      = {Nat. Phys.},
  volume       = {9},
  number       = {12},
  pages        = {795--800},
  year         = {2013},
  doi          = {10.1038/nphys2790},
}

@article{MarioTopPump,
  title = {Topological pumping and {Tamm} states in photonic systems},
  author = {Silva, Solange V. and Fernandes, David E. and Morgado, Tiago A. and Silveirinha, M\'ario G.},
  journal = {Phys. Rev. B},
  volume = {105},
  issue = {15},
  pages = {155133},
  numpages = {11},
  year = {2022},
  month = {Apr},
  publisher = {American Physical Society},
  doi = {10.1103/PhysRevB.105.155133},
  url = {https://link.aps.org/doi/10.1103/PhysRevB.105.155133}
}

@article{FreqDimSFan,
author = {Luqi Yuan and Qian Lin and Meng Xiao and Shanhui Fan},
journal = {Optica},
number = {11},
pages = {1396--1405},
publisher = {Optica Publishing Group},
title = {Synthetic dimension in photonics},
volume = {5},
month = {Nov},
year = {2018},
url = {https://opg.optica.org/optica/abstract.cfm?URI=optica-5-11-1396},
doi = {10.1364/OPTICA.5.001396},
}

@article{MechQuantHall,
  title = {Mechanical Quantum {Hall} Effect in Time-Modulated Elastic Materials},
  author = {Chen, H. and Yao, L.Y. and Nassar, H. and Huang, G.L.},
  journal = {Phys. Rev. Appl.},
  volume = {11},
  issue = {4},
  pages = {044029},
  numpages = {15},
  year = {2019},
  month = {Apr},
  publisher = {American Physical Society},
  doi = {10.1103/PhysRevApplied.11.044029},
  url = {https://link.aps.org/doi/10.1103/PhysRevApplied.11.044029}
}

@misc{KTopExp,
      title={Observation of Momentum-Band Topology in PT-Symmetric acoustic Floquet Lattices}, 
      author={Shuaishuai Tong and Qicheng Zhang and Gaohan Li and Kun Zhang and Chun Xie and Chunyin Qiu},
      eprint={2507.04068},
      archivePrefix={arXiv}
}

@article{SynthLattKgapTop,
  author       = {Yudong Ren and Kangpeng Ye and Qiaolu Chen and Fujia Chen and Li Zhang and Yuang Pan and Wenhao Li and Xinrui Li and Lu Zhang and Hongsheng Chen and Yihao Yang},
  title        = {Observation of momentum-gap topology of light at temporal interfaces in a time-synthetic lattice},
  journal      = {Nat. Commun.},
  year         = {2025},
  volume       = {16},
  pages      = {707},
  doi          = {10.1038/s41467-025-56021-7},
  url          = {https://www.nature.com/articles/s41467-025-56021-7},
}

@misc{TwoLevelLoc,
      title={Topological Localisation in Time from {PT} Symmetry}, 
      author={Tom Sheppard and C. B. B. Camacho and Sebastian Weidemann and Alexander Szameit and Joshua Feis and Frank Schindler and Hannah M. Price},
      eprint={2509.06679},
      archivePrefix={arXiv}
}

@article{SFanTopRes,
  author       = {Lin, Qian and Xiao, Meng and Yuan, Luqi and Fan, Shanhui},
  title        = {Photonic {Weyl} point in a two-dimensional resonator lattice with a synthetic frequency dimension},
  journal      = {Nat. Commun.},
  year         = {2016},
  volume       = {7},
  pages        = {13731},
  month        = {December},
  doi          = {10.1038/ncomms13731},
  url          = {https://doi.org/10.1038/ncomms13731},
}

@article{TopSTEvents,
  author       = {Feis, Joshua and Weidemann, Sebastian and Sheppard, Tom and Price, Hannah M. and Szameit, Alexander},
  title        = {Space-time-topological events in photonic quantum walks},
  journal      = {Nat. Photonics},
  year         = {2025},
  volume       = {19},
  number       = {5},
  pages        = {518--525},
  month        = {April},
  doi          = {10.1038/s41566-025-01653-w},
  url          = {https://www.nature.com/articles/s41566-025-01653-w},
}

@article{TopClass,
  title = {Classification of topological insulators and superconductors in three spatial dimensions},
  author = {Schnyder, Andreas P. and Ryu, Shinsei and Furusaki, Akira and Ludwig, Andreas W. W.},
  journal = {Phys. Rev. B},
  volume = {78},
  issue = {19},
  pages = {195125},
  numpages = {22},
  year = {2008},
  month = {Nov},
  publisher = {American Physical Society},
  doi = {10.1103/PhysRevB.78.195125},
  url = {https://link.aps.org/doi/10.1103/PhysRevB.78.195125}
}

@article{WaterTimeRef,
  title = {Time Reversal of Water Waves},
  author = {Przadka, A. and Feat, S. and Petitjeans, P. and Pagneux, V. and Maurel, A. and Fink, M.},
  journal = {Phys. Rev. Lett.},
  volume = {109},
  issue = {6},
  pages = {064501},
  numpages = {5},
  year = {2012},
  month = {Aug},
  publisher = {American Physical Society},
  doi = {10.1103/PhysRevLett.109.064501},
  url = {https://link.aps.org/doi/10.1103/PhysRevLett.109.064501}
}

@article{LemoultSTInt,
  title = {Elastic Wave Packets Crossing a Space-Time Interface},
  author = {Delory, Alexandre and Prada, Claire and Lanoy, Maxime and Eddi, Antonin and Fink, Mathias and Lemoult, Fabrice},
  journal = {Phys. Rev. Lett.},
  volume = {133},
  issue = {26},
  pages = {267201},
  numpages = {6},
  year = {2024},
  month = {Dec},
  publisher = {American Physical Society},
  doi = {10.1103/PhysRevLett.133.267201},
  url = {https://link.aps.org/doi/10.1103/PhysRevLett.133.267201}
}

@article{FloquetWater,
title = {Floquet scattering of shallow water waves by a vertically oscillating plate},
journal = {Wave Motion},
volume = {136},
pages = {103530},
year = {2025},
issn = {0165-2125},
doi = {https://doi.org/10.1016/j.wavemoti.2025.103530},
url = {https://www.sciencedirect.com/science/article/pii/S0165212525000411},
author = {Magdalini Koukouraki and Philippe Petitjeans and Agnès Maurel and Vincent Pagneux}
}

@article{STPiezoelectric,
    author = {Tessier Brothelande, S. and Croënne, C. and Allein, F. and Vasseur, J. O. and Amberg, M. and Giraud, F. and Dubus, B.},
    title = {Experimental evidence of nonreciprocal propagation in space-time modulated piezoelectric phononic crystals},
    journal = {Appl. Phys. Lett.},
    volume = {123},
    number = {20},
    pages = {201701},
    year = {2023},
    month = {11},
    issn = {0003-6951},
    doi = {10.1063/5.0169265},
    url = {https://doi.org/10.1063/5.0169265},
}

@article{NonRecipSound,
  title = {Nonreciprocal Sound Propagation via Cascaded Time-Modulated Slab Resonators},
  author = {Wan, Sheng and Cao, Liyun and Zhu, Yifan and Oudich, Mourad and Assouar, Badreddine},
  journal = {Phys. Rev. Appl.},
  volume = {16},
  issue = {6},
  pages = {064061},
  numpages = {11},
  year = {2021},
  month = {Dec},
  publisher = {American Physical Society},
  doi = {10.1103/PhysRevApplied.16.064061},
  url = {https://link.aps.org/doi/10.1103/PhysRevApplied.16.064061}
}

@article{STAcousticMeta,
  title = {Realizing spatiotemporal effective media for acoustic metamaterials},
  author = {Wen, Xinhua and Zhu, Xinghong and Wu, Hong Wei and Li, Jensen},
  journal = {Phys. Rev. B},
  volume = {104},
  issue = {6},
  pages = {L060304},
  numpages = {8},
  year = {2021},
  month = {Aug},
  publisher = {American Physical Society},
  doi = {10.1103/PhysRevB.104.L060304},
  url = {https://link.aps.org/doi/10.1103/PhysRevB.104.L060304}
}

@article{AcousMetaSurf,
author = {Yang, Yunhan and Jia, Han and Lu, Jiuyang and Yang, Yuzhen and Liu, Tuo and Yang, Jun and Liu, Zhengyou},
title = {Acoustic Metasurface for Space-time Reflection Manipulation},
journal = {Adv. Sci.},
volume = {12},
number = {36},
pages = {e06308},
doi = {https://doi.org/10.1002/advs.202506308},
url = {https://advanced.onlinelibrary.wiley.com/doi/abs/10.1002/advs.202506308},
year = {2025}
}

@article{STMetaSurf,
  author       = {Guo, Xuexue and Ding, Yimin and Duan, Yao and Ni, Xingjie},
  title        = {Nonreciprocal metasurface with space–time phase modulation},
  journal      = {Light Sci. Appl.},
  year         = {2019},
  volume       = {8},
  number       = {1},
  pages        = {123},
  doi          = {10.1038/s41377-019-0225-z},
  url          = {https://doi.org/10.1038/s41377-019-0225-z},
}

@article{TempRefinTL,
  author       = {Hady Moussa and Gengyu Xu and Shixiong Yin and Emanuele Galiffi and Younes Radi and Andrea Alù},
  title        = {Observation of temporal reflection and broadband frequency translation at photonic time interfaces},
  journal      = {Nat. Phys.},
  year         = {2023},
  volume       = {19},
  number       = {7},
  pages        = {863--868},
  month        = {July},
  doi          = {10.1038/s41567-023-01975-y},
  url          = {https://www.nature.com/articles/s41567-023-01975-y},
}

@article{TimeDiffENZ,
  author       = {Romain Tirole and Stefano Vezzoli and Emanuele Galiffi and Iain Robertson and Dries Maurice and Benjamin Tilmann and Stefan A. Maier and John B. Pendry and Riccardo Sapienza},
  title        = {Double-slit time diffraction at optical frequencies},
  journal      = {Nat. Phys.},
  year         = {2023},
  volume       = {19},
  number       = {7},
  pages        = {999--1002},
  month        = {July},
  doi          = {10.1038/s41567-023-01993-w},
  url          = {https://doi.org/10.1038/s41567-023-01993-w},
}

@article{STDiffENZ,
  author       = {Harwood, Anthony C. and Vezzoli, Stefano and Raziman, T. V. and Hooper, Calvin and Tirole, Romain and Wu, Fanyi and Maier, Stefan A. and Pendry, John B. and Horsley, Simon A. R. and Sapienza, Riccardo},
  title        = {Space-time optical diffraction from synthetic motion},
  journal      = {Nat. Commun.},
  year         = {2025},
  volume       = {16},
  pages        = {5147},
  month        = {June},
  doi          = {10.1038/s41467-025-60159-9},
  url          = {https://www.nature.com/articles/s41467-025-60159-9},
}

@article{ExpFanoResTopState,
author = {Wei Gao and Xiaoyong Hu and Chong Li and Jinghuan Yang and Zhen Chai and Jingya Xie and Qihuang Gong},
journal = {Opt. Express},
number = {7},
pages = {8634--8644},
publisher = {Optica Publishing Group},
title = {Fano-resonance in one-dimensional topological photonic crystal heterostructure},
volume = {26},
month = {Apr},
year = {2018},
url = {https://opg.optica.org/oe/abstract.cfm?URI=oe-26-7-8634},
doi = {10.1364/OE.26.008634},
}

@article{Exp1DTopState,
author = {Li, Chong and Hu, Xiaoyong and Gao, Wei and Ao, Yutian and Chu, Saisai and Yang, Hong and Gong, Qihuang},
title = {Thermo-optical Tunable Ultracompact Chip-Integrated {1D} Photonic Topological Insulator},
journal = {Adv. Opt. Mater},
volume = {6},
number = {4},
pages = {1701071},
doi = {https://doi.org/10.1002/adom.201701071},
url = {https://advanced.onlinelibrary.wiley.com/doi/abs/10.1002/adom.201701071},
year = {2018}
}

@article{TopNanoCavLas,
  author       = {Yasutomo Ota and Ryota Katsumi and Katsuyuki Watanabe and Satoshi Iwamoto and Yasuhiko Arakawa},
  title        = {Topological photonic crystal nanocavity laser},
  journal      = {Commun. Phys.},
  year         = {2018},
  volume       = {1},
  pages      = {86},
  doi          = {10.1038/s42005-018-0083-7}
}

@article{SymmCritZak,
  title = {Symmetry criterion for surface states in solids},
  author = {Zak, J.},
  journal = {Phys. Rev. B},
  volume = {32},
  issue = {4},
  pages = {2218--2226},
  numpages = {0},
  year = {1985},
  month = {Aug},
  publisher = {American Physical Society},
  doi = {10.1103/PhysRevB.32.2218},
  url = {https://link.aps.org/doi/10.1103/PhysRevB.32.2218}
}

@article{IntBandTrans,
  title = {Interband transitions in photonic crystals},
  author = {Winn, Joshua N. and Fan, Shanhui and Joannopoulos, John D. and Ippen, Erich P.},
  journal = {Phys. Rev. B},
  volume = {59},
  issue = {3},
  pages = {1551--1554},
  numpages = {0},
  year = {1999},
  month = {Jan},
  publisher = {American Physical Society},
  doi = {10.1103/PhysRevB.59.1551},
  url = {https://link.aps.org/doi/10.1103/PhysRevB.59.1551}
}

@article{IBTchip,
  title = {Electrically Driven Nonreciprocity Induced by Interband Photonic Transition on a Silicon Chip},
  author = {Lira, Hugo and Yu, Zongfu and Fan, Shanhui and Lipson, Michal},
  journal = {Phys. Rev. Lett.},
  volume = {109},
  issue = {3},
  pages = {033901},
  numpages = {5},
  year = {2012},
  month = {Jul},
  publisher = {American Physical Society},
  doi = {10.1103/PhysRevLett.109.033901},
  url = {https://link.aps.org/doi/10.1103/PhysRevLett.109.033901}
}

@article{OldTWref,
  author={Oliner, A.A. and Hessel, A.},
  journal={IRE Trans. Microw. Theory Tech.}, 
  title={Wave Propagation in a Medium with a Progressive Sinusoidal Disturbance}, 
  year={1961},
  volume={9},
  number={4},
  pages={337-343},
  doi={10.1109/TMTT.1961.1125340}}

@article{CassedyTW1,
  author={Cassedy, E.S. and Oliner, A.A.},
  journal={Proc. IEEE}, 
  title={Dispersion relations in time-space periodic media: Part {I}—{Stable} interactions}, 
  year={1963},
  volume={51},
  number={10},
  pages={1342-1359},
  doi={10.1109/PROC.1963.2566}}

@article{CassedyTW2,
  author={Cassedy, E.S.},
  journal={Proc. IEEE}, 
  title={Dispersion relations in time-space periodic media: {Part II}—{Unstable} interactions}, 
  year={1967},
  volume={55},
  number={7},
  pages={1154-1168},
  doi={10.1109/PROC.1967.5775}}

@article{SymmProCTChan,
  author       = {Wen-Jie Chen and Zhao-Qing Zhang and Jian-Wen Dong and C.~T. Chan},
  title        = {Symmetry-protected transport in a pseudospin-polarized waveguide},
  journal      = {Nat. Commun.},
  year         = {2015},
  volume       = {6},
  pages      = {8183},
  doi          = {10.1038/ncomms9183},
}

@article{PhQSHE,
author = {Cheng He  and Xiao-Chen Sun  and Xiao-Ping Liu  and Ming-Hui Lu  and Yulin Chen  and Liang Feng  and Yan-Feng Chen },
title = {Photonic topological insulator with broken time-reversal symmetry},
journal = {Proc. Natl. Acad. Sci. U.S.A.},
volume = {113},
number = {18},
pages = {4924-4928},
year = {2016},
doi = {10.1073/pnas.1525502113},
URL = {https://www.pnas.org/doi/abs/10.1073/pnas.1525502113},
}

@article{PTDMario,
  title = {$\mathcal{P}\ifmmode\cdot\else\textperiodcentered\fi{}\mathcal{T}\ifmmode\cdot\else\textperiodcentered\fi{}\mathcal{D}$ symmetry-protected scattering anomaly in optics},
  author = {Silveirinha, M\'ario G.},
  journal = {Phys. Rev. B},
  volume = {95},
  issue = {3},
  pages = {035153},
  numpages = {14},
  year = {2017},
  month = {Jan},
  publisher = {American Physical Society},
  doi = {10.1103/PhysRevB.95.035153},
  url = {https://link.aps.org/doi/10.1103/PhysRevB.95.035153}
}

@article{MarioPhTopInv,
  title = {Chern invariants for continuous media},
  author = {Silveirinha, M\'ario G.},
  journal = {Phys. Rev. B},
  volume = {92},
  issue = {12},
  pages = {125153},
  numpages = {16},
  year = {2015},
  month = {Sep},
  publisher = {American Physical Society},
  doi = {10.1103/PhysRevB.92.125153},
  url = {https://link.aps.org/doi/10.1103/PhysRevB.92.125153}
}

@article{MarioPhBulkEdge,
  title = {Proof of the Bulk-Edge Correspondence through a Link between Topological Photonics and Fluctuation-Electrodynamics},
  author = {Silveirinha, M\'ario G.},
  journal = {Phys. Rev. X},
  volume = {9},
  issue = {1},
  pages = {011037},
  numpages = {18},
  year = {2019},
  month = {Feb},
  publisher = {American Physical Society},
  doi = {10.1103/PhysRevX.9.011037},
  url = {https://link.aps.org/doi/10.1103/PhysRevX.9.011037}
}

@article{NonRecPho,
  author       = {Sounas, Dimitrios L. and Alù, Andrea},
  title        = {Non-reciprocal photonics based on time modulation},
  journal      = {Nat. Photonics},
  year         = {2017},
  volume       = {11},
  number       = {12},
  pages        = {774--783},
  month        = {December},
  doi          = {10.1038/s41566-017-0051-x},
  url          = {https://doi.org/10.1038/s41566-017-0051-x},
}

@article{QEDSimonPendry,
author = {Simon A. R. Horsley  and John B. Pendry },
title = {Quantum electrodynamics of time-varying gratings},
journal = {Proc. Natl. Acad. Sci. U.S.A.},
volume = {120},
number = {36},
pages = {e2302652120},
year = {2023},
doi = {10.1073/pnas.2302652120},
URL = {https://www.pnas.org/doi/abs/10.1073/pnas.2302652120}}

@article{TWAmpSimonPendry,
  title = {Traveling Wave Amplification in Stationary Gratings},
  author = {Horsley, S. A. R. and Pendry, J. B.},
  journal = {Phys. Rev. Lett.},
  volume = {133},
  issue = {15},
  pages = {156903},
  numpages = {6},
  year = {2024},
  month = {Oct},
  publisher = {American Physical Society},
  doi = {10.1103/PhysRevLett.133.156903},
  url = {https://link.aps.org/doi/10.1103/PhysRevLett.133.156903}
}

@article{3DTWFilipa,
  title = {Engineering nonreciprocal responses in traveling-wave spacetime crystals via {Clausius-Mossotti} homogenization},
  author = {Prud\^encio, Filipa R. and Silveirinha, M\'ario G.},
  journal = {Phys. Rev. Appl.},
  volume = {22},
  issue = {5},
  pages = {054080},
  numpages = {16},
  year = {2024},
  month = {Nov},
  publisher = {American Physical Society},
  doi = {10.1103/PhysRevApplied.22.054080},
  url = {https://link.aps.org/doi/10.1103/PhysRevApplied.22.054080}
}

@article{ExpTransLines,
  author={Qin, Shihan and Xu, Qiang and Wang, Yuanxun Ethan},
  journal={IEEE Trans. Microw. Theory Tech.}, 
  title={Nonreciprocal Components With Distributedly Modulated Capacitors}, 
  year={2014},
  volume={62},
  number={10},
  pages={2260-2272},
  doi={10.1109/TMTT.2014.2347935}}

@article{TWMetasurf,
  title = {Space-time gradient metasurfaces},
  author = {Hadad, Y. and Sounas, D. L. and Alu, A.},
  journal = {Phys. Rev. B},
  volume = {92},
  issue = {10},
  pages = {100304},
  numpages = {6},
  year = {2015},
  month = {Sep},
  publisher = {American Physical Society},
  doi = {10.1103/PhysRevB.92.100304},
  url = {https://link.aps.org/doi/10.1103/PhysRevB.92.100304}
}

@article{ElasTW,
doi = {10.1088/1367-2630/18/8/083047},
url = {https://doi.org/10.1088/1367-2630/18/8/083047},
year = {2016},
month = {aug},
publisher = {IOP Publishing},
volume = {18},
number = {8},
pages = {083047},
author = {Trainiti, G and Ruzzene, M},
title = {Non-reciprocal elastic wave propagation in spatiotemporal periodic structures},
journal = {New J. Phys.}
}

@article{TopCrysIns,
  title = {Topological Crystalline Insulators},
  author = {Fu, Liang},
  journal = {Phys. Rev. Lett.},
  volume = {106},
  issue = {10},
  pages = {106802},
  numpages = {4},
  year = {2011},
  month = {Mar},
  publisher = {American Physical Society},
  doi = {10.1103/PhysRevLett.106.106802},
  url = {https://link.aps.org/doi/10.1103/PhysRevLett.106.106802}
}

@article{TopPhIns,
  author       = {Lu, Ling and Joannopoulos, John D. and Soljačić, Marin},
  title        = {Topological photonics},
  journal      = {Nat. Photonics},
  volume       = {8},
  number       = {11},
  pages        = {821--829},
  year         = {2014},
  doi          = {10.1038/nphoton.2014.248},
}

@article{QChePhCs,
  title = {Engineering fragile topology in photonic crystals: Topological quantum chemistry of light},
  author = {de Paz, Mar\'{\i}a Blanco and Vergniory, Maia G. and Bercioux, Dario and Garc\'{\i}a-Etxarri, Aitzol and Bradlyn, Barry},
  journal = {Phys. Rev. Res.},
  volume = {1},
  issue = {3},
  pages = {032005},
  numpages = {6},
  year = {2019},
  month = {Oct},
  publisher = {American Physical Society},
  doi = {10.1103/PhysRevResearch.1.032005},
  url = {https://link.aps.org/doi/10.1103/PhysRevResearch.1.032005}
}

@article{CornerTopPh,
  title = {Robustness of topological corner modes in photonic crystals},
  author = {Proctor, Matthew and Huidobro, Paloma Arroyo and Bradlyn, Barry and de Paz, Mar\'{\i}a Blanco and Vergniory, Maia G. and Bercioux, Dario and Garc\'{\i}a-Etxarri, Aitzol},
  journal = {Phys. Rev. Res.},
  volume = {2},
  issue = {4},
  pages = {042038},
  numpages = {7},
  year = {2020},
  month = {Dec},
  publisher = {American Physical Society},
  doi = {10.1103/PhysRevResearch.2.042038},
  url = {https://link.aps.org/doi/10.1103/PhysRevResearch.2.042038}
}

@article{2DPhTop,
  title = {Prevalence of Two-Dimensional Photonic Topology},
  author = {Ghorashi, Ali and Vaidya, Sachin and Rechtsman, Mikael C. and Benalcazar, Wladimir A. and Solja\ifmmode \check{c}\else \v{c}\fi{}i\ifmmode \acute{c}\else \'{c}\fi{}, Marin and Christensen, Thomas},
  journal = {Phys. Rev. Lett.},
  volume = {133},
  issue = {5},
  pages = {056602},
  numpages = {8},
  year = {2024},
  month = {Aug},
  publisher = {American Physical Society},
  doi = {10.1103/PhysRevLett.133.056602},
  url = {https://link.aps.org/doi/10.1103/PhysRevLett.133.056602}
}

@article{PhotonLoc,
  title = {Photon localization and {Bloch} symmetry breaking in luminal gratings},
  author = {Galiffi, E. and Silveirinha, M. G. and Huidobro, P. A. and Pendry, J. B.},
  journal = {Phys. Rev. B},
  volume = {104},
  issue = {1},
  pages = {014302},
  numpages = {6},
  year = {2021},
  month = {Jul},
  publisher = {American Physical Society},
  doi = {10.1103/PhysRevB.104.014302},
  url = {https://link.aps.org/doi/10.1103/PhysRevB.104.014302}
}

@article{PTacousticFloq,
  author       = {Tong, Shuaishuai and Zhang, Qicheng and Li, Gaohan and Zhang, Kun and Xie, Chun and Qiu, Chunyin},
  title        = {Observation of momentum-band topology in PT-symmetric acoustic Floquet lattices},
  journal      = {Nature Communications},
  year         = {2025},
  volume       = {16},
  article      = {9975},
  doi          = {10.1038/s41467-025-64915-9},
  url          = {https://www.nature.com/articles/s41467-025-64915-9},
}

@article{ExpTopPTC,
  author       = {Xiong, Jiang and Zhang, Xudong and Duan, Longji and Wang, Jiarui and Long, Yang and Hou, Haonan and Yu, Letian and Zou, Linyang and Zhang, Baile},
  title        = {Observation of wave amplification and temporal topological state in a non-synthetic photonic time crystal},
  journal      = {Nature Communications},
  year         = {2025},
  volume       = {16},
  article      = {11182},
  doi          = {10.1038/s41467-025-66154-4},
  url          = {https://www.nature.com/articles/s41467-025-66154-4},
}

@article{TWTop_Joao,
author = {Serra, João C. and Silveirinha, Mário G.},
title = {Engineering Topological Phases with a Traveling-Wave Spacetime Modulation},
journal = {Laser \& Photonics Reviews},
volume = {n/a},
number = {n/a},
pages = {e00560},
year = {2025},
doi = {https://doi.org/10.1002/lpor.202500560},
url = {https://onlinelibrary.wiley.com/doi/abs/10.1002/lpor.202500560},
}

\end{document}

\title{Supporting Information - Interface States in Space-Time Photonic Crystals: Topological Origin, Propagation and Amplification}

\author{Alejandro Caballero$^{1,2}$}
\email{alejandro.caballero@uam.es}

\author{Thomas F. Allard$^{1,2}$}

\author{Paloma A. Huidobro$^{1,2}$}
\email{p.arroyo-huidobro@uam.es}

\affiliation{${}^{1}$ Departamento de F\'isica Te\'orica de la Materia Condensada, Universidad Aut\'onoma de Madrid, E28049 Madrid, Spain}
\affiliation{${}^{2}$ Condensed Matter Physics Center (IFIMAC), Universidad Aut\'onoma de Madrid, E28049 Madrid, Spain}

\maketitle

\tableofcontents

\section{Maxwell's equations in spatiotemporal photonic crystals}

In this section we detail our approach to solve Maxwell's equations for spatiotemporal photonic crystals (STPhCs). Let us first consider a general scenario where both the permittivity and permeability are modulated in space and time following a travelling-wave:
\begin{equation}
    \epsilon(x,t) = \epsilon_0 \epsilon_m (1 + 2 \alpha_e \cos(gx-\Omega t)),
\end{equation}
\begin{equation}
    \mu(x,t) = \mu_0 \mu_m (1 + 2 \alpha_m \cos(gx-\Omega t)).
\end{equation}
Then, we can write Maxwell's equations in the lab-frame as
\begin{equation}\label{eq_pre_eigen}
    \hat{\textbf{L}} \mathbf{\Psi} = \partial_t \left( \hat{\textbf{M}} \mathbf{\Psi} \right),
\end{equation}
with
\begin{equation}
    \hat{\textbf{L}} = 
    \begin{pmatrix}
        \mathbb{0}_{3\times 3} & \nabla\times \\
        -\nabla\times & \mathbb{0}_{3\times 3}
    \end{pmatrix}, \; \; \; 
    \hat{\textbf{M}}(x,t) =
    \begin{pmatrix}
        \epsilon(x,t) \mathds{1}_{3\times 3} & \mathbb{0}_{3\times 3}\\
        \mathbb{0}_{3\times 3} & \mu(x,t) \mathds{1}_{3\times 3}
    \end{pmatrix}
    , \; \; \;
    \mathbf{\Psi} = 
    \begin{bmatrix}
        \textbf{E} \\
        \textbf{H}
    \end{bmatrix}.
\end{equation}
On the other hand, the spatiotemporal (ST) nature of the system complicates the definition of the relevant symmetries and magnitudes compared to regular photonic crystals. As discussed in the main text, we can overcome this difficulty by performing a Lorentz transformation to the frame comoving with the modulation. Specifically, we apply the Lorentz boost defined in Eqs.~(6–7) of the main text, which preserves the form of Maxwell’s equations but, crucially, modifies the constitutive relations of the transformed fields. After this transformation, we obtain
\begin{equation}\label{eq_pre_eigen}
    \hat{\textbf{L}}' \mathbf{\Psi}' = \hat{\textbf{M}}' \partial_{t'} \mathbf{\Psi}',
\end{equation}
with
\begin{equation}\label{eq:gen_com_Maxwell}
    \hat{\textbf{L}}' = 
    \begin{pmatrix}
        \mathbb{0}_{3\times 3} & \nabla' \times \\
        -\nabla' \times & \mathbb{0}_{3\times 3}
    \end{pmatrix}, \; \; \; 
    \hat{\textbf{M}}'(x') = 
    \begin{pmatrix}
        \epsilon'_\parallel(x') & 0 & 0 & 0 & 0 & 0\\
        0 & \epsilon'_\perp(x') & 0 & 0 & 0 & -\xi'(x')\\
        0 & 0 & \epsilon'_\perp(x') & 0 & \xi'(x') & 0 \\
        0 & 0 & 0 & \mu'_\parallel(x') & 0 & 0 \\
        0 & 0 & \xi'(x') & 0 & \mu'_\perp(x') & 0 \\
        0 & -\xi'(x') & 0 & 0 & 0 & \mu'_\perp(x')
    \end{pmatrix}, \; \; \;
    \mathbf{\Psi}' = 
    \begin{bmatrix}
        \textbf{E}' \\
        \textbf{H}'
    \end{bmatrix}.
\end{equation}
The expressions for the components of the constitutive matrix are given in the main text, and $\epsilon'_\parallel(x') = \epsilon(x'/\gamma)$, $\mu'_\parallel(x') = \mu(x'/\gamma)$. As we can see, a moving medium type bianisotropic coupling between electric and magnetic fields appears.

\subsection{Eigenvalue problem for the lab-frame}\label{sec:eigen_lab}

As discussed in the main text, Maxwell's equations can be solved by expanding the fields in a Bloch-Floquet form. Considering an s-polarized wave travelling along with the modulation \cite{CassedyTW1,Fresnel},
\begin{equation}\label{eq:BFansatz}
    \begin{bmatrix}
        E_z (x,t)\\
        H_y (x,t)
    \end{bmatrix} 
    = e^{i(k x - \omega t)} \; \sum_{n}
    \begin{bmatrix}
        E_n \\
        H_n
    \end{bmatrix}
    \, e^{i\, n (g x - \Omega t)},
\end{equation}
where $-N_{\mathrm{F}} \le n \le N_{\mathrm{F}}$, and $2N_{\mathrm{F}} + 1$ is the total number of Bloch-Floquet harmonics included in the truncated field expansion. Substituting Eq.~\eqref{eq:BFansatz} into Maxwell's equations yields the following eigenvalue problem for the field Floquet amplitudes:
\begin{equation}\label{eigen_lab}
\begin{pmatrix}
    \mathbf{\mathds{M}}^{\text{EE}}(\omega) & \mathbf{\mathds{M}}^{\text{EH}}(\omega) \\
    \mathbf{\mathds{M}}^{\text{HE}}(\omega) & \mathbf{\mathds{M}}^{\text{HH}}(\omega)
\end{pmatrix}
\,
\begin{bmatrix}
    \textbf{E}_{\omega} \\
    \textbf{H}_{\omega}
\end{bmatrix} 
=
k_{\omega} \,
\begin{bmatrix}
    \textbf{E}_{\omega} \\
    \textbf{H}_{\omega}
\end{bmatrix},
\end{equation}
where each component of the $2 \times 2$ block matrix is another square matrix of dimension $(2N_{\text{F}} + 1) \times (2N_{\text{F}} + 1)$, giving a total of $2  (2 N_{\mathrm{F}} + 1)$ eigenvalues. Since only s-polarization is considered throughout this work, the subscripts $y$ and $z$ are omitted for simplicity. Furthermore, we make the dependence with $\omega$ explicit in order to distinguish between solutions in the laboratory and comoving frames, since the latter is defined for a Lorentz transformed frequency $\omega'$.

The row ($n$) and columns ($n'$) entries of the matrices read:
\begin{equation}\label{eq:1smatrix_lab}
    \mathbf{\mathds{M}}^{\text{EE}}_{n,n'} =  \mathbf{\mathds{M}}^{\text{HH}}_{n,n'} = -n g \delta_{n,n'},
\end{equation}
\begin{equation}
    \begin{split}
     \mathbf{\mathds{M}}^{\text{EH}}_{n,n'} = -\mu_0 \mu_m (\omega + n \Omega) [ & \delta_{n,n'} + \alpha_m (\delta_{n,n'+1}+ \delta_{n,n'-1}) ] ,
    \end{split}
\end{equation}
\begin{equation}\label{eq:last_matrix_lab}
    \begin{split}
     \mathbf{\mathds{M}}^{\text{HE}}_{n,n'} = -\epsilon_0 \epsilon_m (\omega + n \Omega)  [ & \delta_{n,n'} + \alpha_e ( \delta_{n,n'+1} +  \delta_{n,n'-1}) ] .
    \end{split}
\end{equation}
Once the eigenproblem is solved, we organize the eigenvectors as follows
\begin{equation}\label{eq:eigen_matrix}
    \mathcal{M}_{\omega}^{(\text{m})} = \left( \mathbf{\Psi}_{\omega,\tau=1,p=-N_{\text{F}}},\dots, \mathbf{\Psi}_{\omega,\tau=1,p=N_{\text{F}}}, \mathbf{\Psi}_{\omega,\tau=2,p=-N_{\text{F}}}, \dots , \mathbf{\Psi}_{\omega,\tau=2,p=N_{\text{F}}}  \right),
\end{equation}
defining a $2 \, (2 N_\text{F}+1) \times 2\, (2 N_\text{F}+1)$ matrix whose columns correspond to the eigenvectors, where
\begin{equation}
    \mathbf{\Psi}_{\omega,\tau,p} = \left[E_{\omega,\tau,p,n=-N_{\text{F}}}, \dots, E_{\omega,\tau,p,n=N_{\text{F}}}, H_{\omega,\tau,p,n=-N_{\text{F}}}, \dots, H_{\omega,\tau,p,n=N_{\text{F}}} \right]^T,
\end{equation}
with $-N_{\mathrm{F}} \le p \le N_{\mathrm{F}}$ and $\tau = 1,2$ spanning all $2 \, (2 N_{\mathrm{F}} + 1)$ eigenvalues. Therefore, the eigenfunctions can be written as
\begin{equation}\label{eq:Ef_eigen_lan}
    E_{\omega,\tau,p} (x,t) = e^{i k_{\tau,p} x - \omega t} \sum_n E_{\omega,\tau,p,n} \, e^{i\, n (g x - \Omega t)}
\end{equation}
and 
\begin{equation}\label{eq:Hf_eigen_lan}
    H_{\omega,\tau,p} (x,t) = e^{i k_{\tau,p} x - \omega t} \sum_n H_{\omega,\tau,p,n} \, e^{i\, n (g x - \Omega t)}.
\end{equation}

From the above, we can derive analytical expressions for the vacuum in the laboratory frame. To do so, we evaluate $\alpha_e = \alpha_m = 0$, as well as $\epsilon_m = \mu_m = 1$ in Eqs.~(\ref{eq:1smatrix_lab}-\ref{eq:last_matrix_lab}) to obtain:
\begin{equation}
    \mathbf{\mathds{M}}^{\text{(v)}} = 
    \begin{pmatrix}
        -n g \, \delta_{n,n'} & - \mu_0 (\omega + n\Omega) \delta_{n,n'}\\
        - \epsilon_0 (\omega + n\Omega) \delta_{n,n'} & -n g \, \delta_{n,n'}
    \end{pmatrix}.
\end{equation}
Its characteristic equation is $(k + ng)^2 - c_0^{-2} (\omega + n\Omega)^2 = 0$, such that its eigenvalues are:
\begin{equation}
    k_{n}^{\pm} = -ng \pm c_0^{-1} (\omega + n\Omega).
\end{equation}
In contrast to the eigenfunctions inside the STPhC, in vacuum we can distinguish between forward and backward modes and, furthermore, each mode will only contain one Bloch-Floquet harmonic. As such, $p$ and $n$ indices coincides, and the general $\tau = 1,2$ label now identifies forward and backward waves.
In order to distinguish between forward and backward propagating modes, we calculate the inverse group velocity $\frac{\text{d} k_n^{\pm}}{\text{d}\omega} = \pm 1/c_0$ to find that it is positive for $k_n^{+}$ (forward: $k_n^{+} \equiv k_{\omega,\mathrm{f},n}$) and negative for $k_n^{-}$ (backward: $k_n^{-} \equiv k_{\omega,\mathrm{b},n}$). Therefore, the backward and forward eigenvectors are defined as
\begin{equation}
    \begin{bmatrix}
        E_{\omega,\mathrm{f},n} \\
        H_{\omega,\mathrm{f},n}
    \end{bmatrix}
    =
    \begin{bmatrix}
        1 \\
        \frac{ -(k_{\omega,\mathrm{f},n} + ng)}{\mu_0 (\omega+n\Omega)}
    \end{bmatrix};
    \; \; 
    \begin{bmatrix}
        E_{\omega,\mathrm{b},n} \\
        H_{\omega,\mathrm{b},n}
    \end{bmatrix}
    =
    \begin{bmatrix}
        1 \\
        \frac{ -(k_{\omega,\mathrm{b},n} + ng)}{\mu_0 (\omega+n\Omega)}
    \end{bmatrix}.
\end{equation}

\acd{We can also obtain an eigenvalue problem for $\omega$, which will be needed for the scattering calculation in the superluminal regime:
\begin{equation}\label{eigen_labk}
\begin{pmatrix}
    \mathbf{\mathds{M}}^{\text{EE,L}}(k) & \mathbf{\mathds{M}}^{\text{EH,L}}(k) \\
    \mathbf{\mathds{M}}^{\text{HE,L}}(k) & \mathbf{\mathds{M}}^{\text{HH,L}}(k)
\end{pmatrix}
\,
\begin{bmatrix}
    \textbf{E}_{k} \\
    \textbf{H}_{k}
\end{bmatrix} 
=
\omega_{k} \,
\begin{pmatrix}
    \mathbf{\mathds{M}}^{\text{EE,R}}(k) & \mathbf{\mathds{M}}^{\text{EH,R}}(k) \\
    \mathbf{\mathds{M}}^{\text{HE,R}}(k) & \mathbf{\mathds{M}}^{\text{HH,R}}(k)
\end{pmatrix}
\,
\begin{bmatrix}
    \textbf{E}_{k} \\
    \textbf{H}_{k}
\end{bmatrix},
\end{equation}
where the right matrix entries read
\begin{equation}
    \mathbf{\mathds{M}}^{\text{EH,R}}_{n,n'} = \mathbf{\mathds{M}}^{\text{HE,R}}_{n,n'} = 0,
\end{equation}
\begin{equation}
    \mathbf{\mathds{M}}^{\text{EE,R}}_{n,n'} = -\epsilon_0 \epsilon_m [\delta_{n,n'} + \alpha_e (\delta_{n,n'+1} + \delta_{n,n'-1})],
\end{equation}
\begin{equation}
    \mathbf{\mathds{M}}^{\text{HH,R}}_{n,n'} = -\mu_0 \mu_m [\delta_{n,n'} + \alpha_m (\delta_{n,n'+1} + \delta_{n,n'-1})],
\end{equation}
while the left ones are
\begin{equation}
    \mathbf{\mathds{M}}^{\text{EH,L}}_{n,n'} = \mathbf{\mathds{M}}^{\text{HE,L}}_{n,n'} = (k + n\, g) \, \delta_{n,n'},
\end{equation}
\begin{equation}
    \mathbf{\mathds{M}}^{\text{EE,L}}_{n,n'} = \epsilon_0 \epsilon_m (n \Omega) [\delta_{n,n'} + \alpha_e  (\delta_{n,n'+1} +\delta_{n,n'-1})],
\end{equation}
\begin{equation}
    \mathbf{\mathds{M}}^{\text{HH,L}}_{n,n'} = \mu_0 \mu_m (n \Omega) [\delta_{n,n'} + \alpha_m (\delta_{n,n'+1} +\delta_{n,n'-1})].
\end{equation}
The eigenfunctions are then recovered using the same expressions as in Eqs.\eqref{eq:Ef_eigen_lan}-\eqref{eq:Hf_eigen_lan}.
From the above, we now derive analytical expressions for the vacuum in the laboratory frame. Again, we evaluate $\alpha_e = \alpha_m = 0$, as well as $\epsilon_m = \mu_m = 1$ in Eq.~\eqref{eigen_labk} to obtain:
\begin{equation}
    \mathbf{\mathds{M}}^{\text{(v)}} = 
    \begin{pmatrix}
        -n \Omega \, \delta_{n,n'} & - \epsilon^{-1}_0 (k + n g) \delta_{n,n'}\\
        - \mu^{-1}_0 (k + n g) \delta_{n,n'} & -n \Omega \, \delta_{n,n'}
    \end{pmatrix}.
\end{equation}
Its characteristic equation is $(\omega + n \Omega)^2 - c_0^{2} (k + n g)^2 = 0$, such that its eigenvalues are:
\begin{equation}
    \omega_{n}^{\pm} = -n\Omega \pm c_0 (k + n g).
\end{equation}
Again we find the forward modes to be $\omega_{k,\mathrm{f},n} \equiv \omega^+_n$ and the backward ones $\omega_{k,\mathrm{b},n} \equiv \omega^-_n$, with the associated eigenvectors:
\begin{equation}
    \begin{bmatrix}
        E_{k,\mathrm{f},n} \\
        H_{k,\mathrm{f},n}
    \end{bmatrix}
    =
    \begin{bmatrix}
        1 \\
        \frac{ -(k + ng)}{\mu_0 (\omega_{k,\mathrm{f},n}+n\Omega)}
    \end{bmatrix};
    \; \; 
    \begin{bmatrix}
        E_{k,\mathrm{b},n} \\
        H_{k,\mathrm{b},n}
    \end{bmatrix}
    =
    \begin{bmatrix}
        1 \\
        \frac{ -(k + ng)}{\mu_0 (\omega_{k,\mathrm{b},n}+n\Omega)}
    \end{bmatrix}.
\end{equation}}

\subsection{Eigenvalue problem for the comoving frame}

The magneto-electric coupling $\xi'(x')$ appearing in the constitutive matrix of Eq.~\eqref{eq:gen_com_Maxwell} complicates the derivation of equivalent analytical expressions for the entries of the eigenvalue matrix. To circumvent this problem, we derive an eigenvalue problem for the comoving variables starting from the lab-frame ansatz:
\begin{equation}\label{eq:mix-ans}
    \begin{bmatrix}
        E (x,t)\\
        H (x,t)
    \end{bmatrix} 
    = e^{i(k x - \omega t)} \; \sum_n
    \begin{bmatrix}
        E_n \\
        H_n
    \end{bmatrix}
    \, e^{i\, n (g x - \Omega t)} = e^{i(k' x' - \omega' t')} \, \sum_n
    \begin{bmatrix}
        E_n \\
        H_n
    \end{bmatrix}
    \, e^{i\, n g/\gamma \, x'} =
    \begin{bmatrix}
        E (x',t')\\
        H (x',t')
    \end{bmatrix} ,
\end{equation}
where the second equality follows from the invariance of the phase under Lorentz transformations. As we can see, the Bloch-Floquet amplitudes still correspond to the lab-frame, but the field is expressed fully in the comoving frame. In the same sense, we can then derive Maxwell's equations for the lab-frame fields
\begin{equation}\label{eq:Max_lab}
    \begin{cases}
        \partial_{x} E = \partial_t \left[ \mu(x-c_gt) \, H \right] \\
        \partial_{x} H = \partial_t \left[ \epsilon(x-c_gt) \, E \right]
    \end{cases},
\end{equation}
and express them in the new coordinates as
\begin{equation}\label{eq:Max}
    \begin{cases}
        \partial_{x'} \left[ E + c_g \, \mu(x') \, H \right]= \partial_{t'} \left[ \mu(x') \, H + c_g/c_0^2 E\right] \\
        \partial_{x'} \left[ H + c_g \, \epsilon(x') \, E \right]= \partial_{t'} \left[ \epsilon(x') \, E + c_g/c_0^2 H\right] 
    \end{cases}.
\end{equation}
This allows us to obtain an eigenvalue problem for $(k',\omega')$, while avoiding the bianisotropic structure, by substituting Eq.~\eqref{eq:mix-ans} in Eq.~\eqref{eq:Max}:
\begin{equation}\label{eigen_com}
\begin{pmatrix}
    \mathbf{\mathds{M}}^{\text{EE,L}}(\omega') & \mathbf{\mathds{M}}^{\text{EH,L}}(\omega') \\
    \mathbf{\mathds{M}}^{\text{HE,L}}(\omega') & \mathbf{\mathds{M}}^{\text{HH,L}}(\omega')
\end{pmatrix}
\,
\begin{bmatrix}
    \textbf{E}_{\omega'} \\
    \textbf{H}_{\omega'}
\end{bmatrix} 
=
k'_{\omega'} \,
\begin{pmatrix}
    \mathbf{\mathds{M}}^{\text{EE,R}}(\omega') & \mathbf{\mathds{M}}^{\text{EH,R}}(\omega') \\
    \mathbf{\mathds{M}}^{\text{HE,R}}(\omega') & \mathbf{\mathds{M}}^{\text{HH,R}}(\omega')
\end{pmatrix}
\,
\begin{bmatrix}
    \textbf{E}_{\omega'} \\
    \textbf{H}_{\omega'}
\end{bmatrix},
\end{equation}
where the right matrix entries read
\begin{equation}\label{eq:1smatrix_com}
    \mathbf{\mathds{M}}^{\text{EE,R}}_{n,n'} = \mathbf{\mathds{M}}^{\text{HH,R}}_{n,n'} = - \delta_{n,n'},
\end{equation}
\begin{equation}
    \begin{split}
        \mathbf{\mathds{M}}^{\text{HE,R}}_{n,n'} = -c_g \, \epsilon_0 \epsilon_m [\delta_{n,n'} + \alpha_e (\delta_{n,n'+1} + \delta_{n,n'-1})],
    \end{split}
\end{equation}
\begin{equation}
    \begin{split}
        \mathbf{\mathds{M}}^{\text{EH,R}}_{n,n'} = -c_g \, \mu_0 \mu_m [\delta_{n,n'} + \alpha_m (\delta_{n,n'+1} + \delta_{n,n'-1})],
    \end{split}
\end{equation}
while the left ones are
\begin{equation}
    \mathbf{\mathds{M}}^{\text{EE,L}}_{n,n'} = \mathbf{\mathds{M}}^{\text{HH,L}}_{n,n'} = (\omega' c_g /c_0^2 + n g/\gamma) \, \delta_{n,n'},
\end{equation}
\begin{equation}
    \begin{split}
        \mathbf{\mathds{M}}^{\text{HE,L}}_{n,n'} = \epsilon_0 \epsilon_m (\omega' + n \Omega/\gamma) [ &  \delta_{n,n'} + \alpha_e ( \delta_{n,n'+1} + \delta_{n,n'-1}) ],
    \end{split}
\end{equation}
\begin{equation}\label{eq:last_matrix_com}
    \begin{split}
        \mathbf{\mathds{M}}^{\text{EH,L}}_{n,n'} = \mu_0 \mu_m (\omega' + n \Omega/\gamma)  [ & \delta_{n,n'} + \alpha_m ( \delta_{n,n'+1}  + \delta_{n,n'-1}) ].
    \end{split}
\end{equation}
Once the eigenproblem is solved, we can then write the eigenfunctions for a given comoving-frame frequency $\omega'$ as 
\begin{equation}
    E_{\omega',\tau,p} (x',t') = e^{i( k'_{\omega',\tau,p} x' - \omega' t')} \sum_n E_{\omega',\tau,p,n} \, e^{i\, n g/\gamma \, x'}
\end{equation}
and 
\begin{equation}
    H_{\omega',\tau,p} (x',t') = e^{i (k'_{\omega',\tau,p} x' - \omega' t')} \sum_n H_{\omega',\tau,p,n} \, e^{i\, n g/\gamma \, x'} .
\end{equation}
Again, $E_{\omega',\tau,p,n}$ correspond to the first $2 N_\mathrm{F}+1$ components of the eigenvector given by $(\tau,p)$, whereas $H_{\omega',\tau,p,n}$ equals the last $2N_\mathrm{F}+1$ components. However, these expressions correspond to the lab-frame fields, and we can transform them to the comoving fields with the corresponding Lorentz transformations \cite{Kong}
\begin{equation}
    \begin{bmatrix}
        E'(x',t')\\
        H'(x',t')
    \end{bmatrix}=
    \textbf{T}(x')
    \begin{bmatrix}
        E(x',t')\\
        H(x',t')
    \end{bmatrix},
\end{equation}
with 
\begin{equation}\label{eq:trans_mat}
    \textbf{T}(x') = 
    \gamma \;
    \begin{pmatrix}
        1 & c_g \, \mu (x')\\
        c_g \, \epsilon (x') & 1
    \end{pmatrix}
\end{equation}
the transformation matrix. Thus, the comoving-frame eigenfunctions are defined as
\begin{equation}\label{eq:com_eigefun_E}
    E'_{\omega',\tau,p}(x',t') = e^{i (k'_{\omega',\tau,p} x' - \omega' t')} \sum_n \gamma \, \left[ E_{\omega',\tau,p,n} + c_g \, \mu(x') H_{\omega',\tau,p,n} \right ] e^{i\, n g/\gamma \, x'}
\end{equation}
and
\begin{equation}\label{eq:com_eigefun_H}
    H'_{\omega',\tau,p}(x',t') = e^{i (k'_{\omega',\tau,p} x' - \omega' t')} \sum_n \gamma \, \left[ H_{\omega',\tau,p,n} + c_g \, \epsilon(x') E_{\omega',\tau,p,n} \right ] e^{i\, n g/\gamma \, x'}.
\end{equation}\\

In the same manner as in the lab-frame calculation, we can derive the eigenvectors in the vacuum. To do so, we evaluate $\alpha_e = \alpha_m = 0$, as well as $\epsilon_m = \mu_m = 1$, in Eqs.~(\ref{eq:1smatrix_com}-\ref{eq:last_matrix_com}) to obtain:
\begin{equation}
    \mathbf{\mathds{M}}^{\mathrm{R}} = 
    \begin{pmatrix}
        \delta_{n,n'} & c_g \mu_0 \delta_{n,n'}\\
        c_g \epsilon_0 \delta_{n,n'} & \delta_{n,n'}
    \end{pmatrix}; \; \; 
    \mathbf{\mathds{M}}^{\mathrm{L}} = 
    \begin{pmatrix}
        - (\omega' c_g /c_0^2 + n g/\gamma) \delta_{n,n'} & -\mu_0 (\omega' + n \Omega/\gamma) \delta_{n,n'}\\
        -\epsilon_0  (\omega' + n \Omega/\gamma) \delta_{n,n'} & - (\omega' c_g /c_0^2 + n g/\gamma) \delta_{n,n'}
    \end{pmatrix}.
\end{equation}
Then, rearranging Eq.~\eqref{eigen_com} to have the same structure as Eq.~\eqref{eigen_lab}, we obtain the eigenvalue matrix in the comoving frame by defining $\mathbf{\mathds{M}}^{\text{(v)}} = (\mathbf{\mathds{M}}^{\mathrm{R}})^{-1}\,\mathbf{\mathds{M}}^{\mathrm{L}}$, which corresponds to
\begin{equation}
    \mathbf{\mathds{M}}^{\text{(v)}} = 
    \begin{pmatrix}
        -n g/\gamma \, \delta_{n,n'} & - \mu_0 \omega' \delta_{n,n'}\\
        - \epsilon_0 \omega' \delta_{n,n'} & -n g/\gamma \, \delta_{n,n'}
    \end{pmatrix}.
\end{equation}
Its characteristic equation is then $(k'+ng/\gamma)^2 - (\omega'/c_0)^2 = 0$, so the comoving-frame eigenvalues of free space are:
\begin{equation}\label{eq:eigdef}
    k{'}_n^{\pm} = - ng/\gamma \pm \omega'/c_0 
\end{equation}
Since we want to distinguish between forward and backward modes, we calculate the inverse group velocity. We find the same as in the lab-frame case: it is positive for $k{'}_n^{+}$ (forward: $k{'}_n^{+} \equiv k'_{\omega',\mathrm{f},n}$) and negative for $k{'}_n^{-}$ (backward: $k{'}_n^{-} \equiv k'_{\omega',\mathrm{b},n}$). Therefore, the backward and forward eigenvectors are defined as
\begin{equation} \label{eq:com_vacummodes}
    \begin{bmatrix}
        E_{\omega',\mathrm{f},n} \\
        H_{\omega',\mathrm{f},n}
    \end{bmatrix}
    =
    \begin{bmatrix}
        1 \\
        \frac{ -(k'_{\omega',\mathrm{f},n} + ng/\gamma)}{\mu_0 \omega'}
    \end{bmatrix} 
    =
    \frac{1}{\sqrt{1+\eta^{-2}}}
    \begin{bmatrix}
        1 \\
        -\eta^{-1}
    \end{bmatrix}
    ;
    \; \; 
    \begin{bmatrix}
        E_{\omega',\mathrm{b},n} \\
        H_{\omega',\mathrm{b},n}
    \end{bmatrix}
    =
    \begin{bmatrix}
        1 \\
        \frac{ -(k'_{\omega',\mathrm{b},n} + ng/\gamma)}{\mu_0 \omega'}
    \end{bmatrix} 
    =
    \frac{1}{\sqrt{1+\eta^{-2}}}
    \begin{bmatrix}
        1 \\
        \eta^{-1}
    \end{bmatrix},
\end{equation}
where $\eta^2 = \mu_0/\epsilon_0$ is the wave impedance in vacuum.\\

On the other hand, we can also derive an eigenvalue problem with $\omega'$ as eigenvalue
\begin{equation}\label{eigen_com_k}
\begin{pmatrix}
    \mathbf{\mathds{M}}^{\text{EE,L}}(k') & \mathbf{\mathds{M}}^{\text{EH,L}}(k') \\
    \mathbf{\mathds{M}}^{\text{HE,L}}(k') & \mathbf{\mathds{M}}^{\text{HH,L}}(k')
\end{pmatrix}
\,
\begin{bmatrix}
    \textbf{E}_{k'} \\
    \textbf{H}_{k'}
\end{bmatrix} 
=
\omega'_{k'} \,
\begin{pmatrix}
    \mathbf{\mathds{M}}^{\text{EE,R}}(k') & \mathbf{\mathds{M}}^{\text{EH,R}}(k') \\
    \mathbf{\mathds{M}}^{\text{HE,R}}(k') & \mathbf{\mathds{M}}^{\text{HH,R}}(k')
\end{pmatrix}
\,
\begin{bmatrix}
    \textbf{E}_{k'} \\
    \textbf{H}_{k'}
\end{bmatrix},
\end{equation}
where the right matrix entries read
\begin{equation}
    \mathbf{\mathds{M}}^{\text{EH,R}}_{n,n'} = \mathbf{\mathds{M}}^{\text{HE,R}}_{n,n'} = -c_g/c_0^2 \, \delta_{n,n'},
\end{equation}
\begin{equation}
    \mathbf{\mathds{M}}^{\text{EE,R}}_{n,n'} = -\epsilon_0 \epsilon_m [\delta_{n,n'} + \alpha_e (\delta_{n,n'+1} + \delta_{n,n'-1})],
\end{equation}
\begin{equation}
    \mathbf{\mathds{M}}^{\text{HH,R}}_{n,n'} = -\mu_0 \mu_m [\delta_{n,n'} + \alpha_m (\delta_{n,n'+1} + \delta_{n,n'-1})],
\end{equation}
while the left ones are
\begin{equation}
    \mathbf{\mathds{M}}^{\text{EH,L}}_{n,n'} = \mathbf{\mathds{M}}^{\text{HE,L}}_{n,n'} = (k' + n\, g/\gamma) \, \delta_{n,n'},
\end{equation}
\begin{equation}
    \mathbf{\mathds{M}}^{\text{EE,L}}_{n,n'} = c_g \, \epsilon_0 \epsilon_m (k' + n\, g/\gamma) [\delta_{n,n'} + \alpha_e  (\delta_{n,n'+1} +\delta_{n,n'-1})],
\end{equation}
\begin{equation}
    \mathbf{\mathds{M}}^{\text{HH,L}}_{n,n'} = c_g \, \mu_0 \mu_m (k' + n\, g/\gamma) [\delta_{n,n'} + \alpha_m (\delta_{n,n'+1} +\delta_{n,n'-1})].
\end{equation}
Once solved, we can recover the comoving-frame eigenfunctions as 
\begin{equation}\label{eq:com_eigefun_Ek}
    E'_{k',m}(x',t') = e^{i (k' x' - \omega'_{k',m} t')} \underbrace{\sum_n \gamma \, \left[ E_{k',m,n} + c_g \, \mu(x') H_{k',m,n} \right ] e^{i\, n g/\gamma \, x'}}_{u{'}^E_{k',m}(x')}
\end{equation}
and
\begin{equation}\label{eq:com_eigefun_Hk}
    H'_{k',m}(x',t') = e^{i (k' x' - \omega'_{k',m} t')} \underbrace{\sum_n \gamma \, \left[ H_{k',m,n} + c_g \, \epsilon(x') E_{k',m,n} \right ] e^{i\, n g/\gamma \, x'}}_{u{'}^H_{k',m}(x')}.
\end{equation}
These expressions are equivalent to Eqs.~\eqref{eq:com_eigefun_E}-\eqref{eq:com_eigefun_H}, however, they are now labeled by $k'$ and the band index $m$.

\section{Scattering matrix and transmission spectrum}

Once the eigenvalue problem for the STPhC is solved, we can compute the scattering matrix of a finite system composed of two slabs separated by either spatial or ST boundaries. The boundary conditions differ depending on the type of interface: a static boundary conserves the laboratory-frame frequency $\omega$, whereas a ST boundary conserves the comoving-frame frequency $\omega'$. Consequently, for each case we must select the eigenfunctions of the STPhC that possess a well-defined $\omega$ or $\omega'$, such that we use those obtained from Eq.~\eqref{eigen_lab} for spatial boundaries and from Eq.~\eqref{eigen_com} for spatiotemporal ones. 

\subsection{Spatiotemporal Boundary}

We start by defining the scattering matrix of two STPhC slabs truncated with ST interfaces. To do so, we need to enforce the continuity of the comoving-frame $E'_z$ and $H'_y$ fields at each interface, which are easily defined as $x' = 0, N \gamma a, 2N \gamma a$, where $N$ is the number of ST unit cells (defined as $x' \in [0, \gamma a]$) used to construct each slab. Making use of the field decomposition, we can write the matching conditions using a superposition of the eigenvectors as 
\begin{equation}
    \begin{bmatrix}
        \textbf{E}'_{\omega'} \\
        \textbf{H}'_{\omega'}
    \end{bmatrix}^{(\text{m})}
    = \mathbf{\mathcal{M}}{'}_{\omega'}^{(\text{m})} \, \textbf{e}_{\omega'}^{(\text{m})}; \; \; \; 
    \begin{bmatrix}
        \textbf{E}'_{\omega'} \\
        \textbf{H}'_{\omega'}
    \end{bmatrix}_{\text{f/b}}^{(\text{in/out})}
    = \mathbf{\mathcal{M}}{'}_{\omega',\text{f/b}}^{{(\text{v})}} \, \textbf{e}_{\omega',\text{f/b}}^{(\text{in/out})},
\end{equation}
with 
\begin{equation}
    \mathbf{\mathcal{M}}{'}_{\omega'}^{(\text{m})} = \textbf{T}^{(\text{m})} \mathbf{\mathcal{M}}_{\omega'}^{(\text{m})}; \; \; \; \mathbf{\mathcal{M}}{'}_{\omega',\text{f/b}}^{(\text{v})} = \textbf{T}^{(\text{v})} \mathbf{\mathcal{M}}_{\omega',\text{f/b}}^{(\text{v})},
\end{equation}
where $\mathbf{\mathcal{M}}_{\omega'}^{(\text{m})}$ is a $2 \, (2 N_\text{F}+1) \times 2\, (2 N_\text{F}+1)$ matrix equivalent to the one defined in Eq.~\eqref{eq:eigen_matrix} for the eigenvectors containing the amplitudes given by Eq.~\eqref{eigen_com}, and $\textbf{e}_{\omega'}^{(\text{m})}$ a $2 \, (2 N_\text{F}+1)$ vector containing the scattering amplitudes of each eigenvector. $\mathbf{\mathcal{M}}{'}_{\omega',\text{f/b}}^{{(\text{v})}}$ correspond to $2 \, (2 N_\text{F}+1) \times (2 N_\text{F}+1)$ matrices containing the forward and backward vacuum modes given by Eq.~\eqref{eq:com_vacummodes}, with $\textbf{e}_{\omega',\text{f/b}}^{(\text{in/out})}$ the amplitudes of the input and output waves in both directions of propagation. Finally, the $\textbf{T}^{(\text{v})}$ and $\textbf{T}^{(\text{m})}$ matrices correspond to the transformation matrix of Eq.~\eqref{eq:trans_mat} evaluated at the interfaces for the unmodulated and modulated cases, respectively. The periodicity of the modulation ensures that at the boundaries the matrices read
\begin{equation}\label{eq:trans_mat_exp}
    \textbf{T}^{(\text{v})} = \gamma \;
    \begin{pmatrix}
        \mathds{1} & c_g \mu_0 \mathds{1}\\
        c_g \epsilon_0 \mathds{1} & \mathds{1}
    \end{pmatrix}; \; \; \;
    \textbf{T}^{(\text{m})} = \gamma \;
    \begin{pmatrix}
        \mathds{1} & c_g \mu_0 \mu_m (1+2\alpha_m) \mathds{1}\\
        c_g \epsilon_0 \epsilon_m (1+2\alpha_e) \mathds{1} & \mathds{1}
    \end{pmatrix}.
\end{equation}

We now enforce the boundary conditions at each interface:\\

\begin{itemize}
    \item \underline{1st interface}: 
\end{itemize}
    \begin{equation}\label{eq:1stint}
        \begin{bmatrix}
            \textbf{E}'_{\omega'} \\
            \textbf{H}'_{\omega'}
        \end{bmatrix}_{\text{f}}^{\text{(in)},1}
        +
        \begin{bmatrix}
            \textbf{E}'_{\omega'} \\
            \textbf{H}'_{\omega'}
        \end{bmatrix}_{\text{b}}^{\text{(out)},1}
        =
        \begin{bmatrix}
            \textbf{E}'_{\omega'} \\
            \textbf{H}'_{\omega'}
        \end{bmatrix}^{\text{(m)},1}
        \rightarrow
        \mathbf{\mathcal{M}}{'}_{\omega',\text{f}}^{(\text{v})} \, \textbf{e}_{\omega',\text{f}}^{\text{(in),1}} + \mathbf{\mathcal{M}}{'}_{\omega',\text{b}}^{(\text{v})} \, \textbf{e}_{\omega',\text{b}}^{\text{(out),1}} = \mathbf{\mathcal{M}}{'}_{\omega'}^{(\text{m})} \, \textbf{e}_{\omega'}^{\text{(m)},1} .
    \end{equation}
\begin{itemize}
    \item \underline{2nd interface}:
\end{itemize}
    \begin{equation}\label{eq:2ndint}
        \begin{bmatrix}
            \textbf{E}'_{\omega'} \\
            \textbf{H}'_{\omega'}
        \end{bmatrix}^{(\text{m}),2}
        =
        \begin{bmatrix}
            \textbf{E}'_{\omega'} \\
            \textbf{H}'_{\omega'}
        \end{bmatrix}^{(\tilde{\text{m}}),2}
        \rightarrow
        \mathbf{\mathcal{M}}{'}_{\omega'}^{(\text{m})} \, \textbf{e}_{\omega'}^{\text{(m)},2} = \mathbf{\mathcal{M}}{'}_{\omega'}^{(\text{m})} \, \textbf{P}{'} \, \textbf{e}_{\omega'}^{\text{(m)},1} = \mathbf{\mathcal{M}}{'}_{\omega'}^{(\tilde{\text{m}})} \, \textbf{e}_{\omega'}^{(\tilde{\text{m}}),2} .
    \end{equation}
\begin{itemize}
    \item \underline{3rd interface}:
\end{itemize}
    \begin{equation}\label{eq:3rdint}
        \begin{bmatrix}
            \textbf{E}'_{\omega'} \\
            \textbf{H}'_{\omega'}
        \end{bmatrix}^{(\tilde{\text{m}}),3}
        =
        \begin{bmatrix}
            \textbf{E}'_{\omega'} \\
            \textbf{H}'_{\omega'}
        \end{bmatrix}_{\text{f}}^{\text{(out)},3}
        +
        \begin{bmatrix}
            \textbf{E}'_{\omega'} \\
            \textbf{H}'_{\omega'}
        \end{bmatrix}_{\text{b}}^{\text{(in)},3}
        \rightarrow
         \mathbf{\mathcal{M}}{'}_{\omega'}^{(\tilde{\text{m}})} \, \textbf{e}_{\omega'}^{(\tilde{\text{m}}),3} = \mathbf{\mathcal{M}}{'}_{\omega'}^{(\tilde{\text{m}})} \, \Tilde{\textbf{P}}{'} \, \textbf{e}_{\omega'}^{(\tilde{\text{m}}),2} = \mathbf{\mathcal{M}}{'}_{\omega',\text{f}}^{(\text{v})} \, \textbf{e}_{\omega',\text{f}}^{\text{(out)},3} + \mathbf{\mathcal{M}}{'}_{\omega',\text{b}}^{(\text{v})} \, \textbf{e}_{\omega',\text{b}}^{\text{(in)},3} .
    \end{equation}
$\textbf{P}'$ corresponds to the diagonal matrix containing the phase acquired by each eigenvector between surfaces $P'_{jj} = \exp(i\, k'_j  D')$, where $D' = N \gamma a$ is the width of each slab and $k'_j$ the eigenvalues of the STPhC. The matrix containing the eigenvectors of the second slab is represented with $(\tilde{\text{m}})$, and its corresponding phase matrix as $\Tilde{\textbf{P}}{'}$. Now, multiplying Eq. \eqref{eq:3rdint} by $\mathbf{\mathcal{M}}{'}_{\omega'}^{(\tilde{\text{m}})}(\mathbf{\mathcal{M}}{'}_{\omega'}^{(\tilde{\text{m}})} \Tilde{\textbf{P}}')^{-1}$ we obtain:

\begin{equation}
    \mathbf{\mathcal{M}}{'}_{\omega'}^{(\tilde{\text{m}})}(\mathbf{\mathcal{M}}{'}_{\omega'}^{(\tilde{\text{m}})} \Tilde{\textbf{P}}')^{-1} \left( \mathbf{\mathcal{M}}{'}_{\omega',\text{f}}^{(\text{v})} \, \textbf{e}_{\omega',\text{f}}^{\text{(out)},3} + \mathbf{\mathcal{M}}{'}_{\omega',\text{b} }^{(\text{v})} \, \textbf{e}_{\omega',\text{b}}^{\text{(in)},3}  \right) = \mathbf{\mathcal{M}}{'}_{\omega'}^{(\tilde{\text{m}})} \,  \textbf{e}_{\omega,(2)}^{(\tilde{\text{m}})} = \mathbf{\mathcal{M}}{'}_{\omega'}^{(\text{m})} \, \textbf{P}' \, \textbf{e}_{\omega,(1)}^m ,
\end{equation}
where we have used Eq. \eqref{eq:2ndint} for the second equality. Then, multiplying again the previous equation by $\mathbf{\mathcal{M}}{'}_{\omega'}^{(\text{m})}(\mathbf{\mathcal{M}}{'}_{\omega'}^{(\text{m})} \textbf{P}')^{-1}$ and using Eq. \eqref{eq:1stint} leads to:
\begin{equation}
    \mathbf{\mathcal{M}}{'}_{\omega'}^{(\text{m})}(\mathbf{\mathcal{M}}{'}_{\omega'}^{(\text{m})} \textbf{P}')^{-1} \, \mathbf{\mathcal{M}}{'}_{\omega'}^{(\tilde{\text{m}})}(\mathbf{\mathcal{M}}{'}_{\omega'}^{(\tilde{\text{m}})} \Tilde{\textbf{P}}')^{-1} \left( \mathbf{\mathcal{M}}{'}_{\omega',\text{f}}^{(\text{v})} \, \textbf{e}_{\omega',\text{f}}^{\text{(out)},3} + \mathbf{\mathcal{M}}{'}_{\omega',\text{b} }^{(\text{v})} \, \textbf{e}_{\omega',\text{b}}^{\text{(in)},3}  \right) = \mathbf{\mathcal{M}}{'}_{\omega',\text{f}}^{(\text{v})} \, \textbf{e}_{\omega',\text{f}}^{\text{(in)},1} + \mathbf{\mathcal{M}}{'}_{\omega',\text{b} }^{(\text{v})} \, \textbf{e}_{\omega',\text{b}}^{\text{(out)},1} .
\end{equation}
Finally, after rearranging and defining $\textbf{N}_{\omega'} = \mathbf{\mathcal{M}}{'}_{\omega'}^{(\text{m})}(\mathbf{\mathcal{M}}{'}_{\omega'}^{(\text{m})} \textbf{P}')^{-1} \, \mathbf{\mathcal{M}}{'}_{\omega'}^{(\tilde{\text{m}})}(\mathbf{\mathcal{M}}{'}_{\omega'}^{(\tilde{\text{m}})} \Tilde{\textbf{P}}')^{-1}$ we arrive at an expression relating the output amplitudes $\textbf{e}_{\omega',\text{f}}^{\text{(out)},3}$ and $\textbf{e}_{\omega',\text{b}}^{\text{(out)},1}$ to the input ones $\textbf{e}_{\omega',\text{f}}^{\text{(in)},1}$ and $\textbf{e}_{\omega',\text{b}}^{\text{(in)},3}$:
\begin{equation} \label{eq:pre_scat_mat}
    \textbf{N}_{\omega'} \mathbf{\mathcal{M}}{'}_{\omega',\text{f}}^{(\text{v})} \, \textbf{e}_{\omega',\text{f}}^{\text{(out)},3} - \mathbf{\mathcal{M}}{'}_{\omega',\text{b} }^{(\text{v})} \, \textbf{e}_{\omega',\text{b}}^{\text{(out)},1} = \mathbf{\mathcal{M}}{'}_{\omega',\text{f}}^{(\text{v})} \, \textbf{e}_{\omega',\text{f}}^{\text{(in)},1} - \textbf{N}_{\omega'} \mathbf{\mathcal{M}}{'}_{\omega',\text{b} }^{(\text{v})} \, \textbf{e}_{\omega',\text{b}}^{\text{(in)},3} .
\end{equation}
We can express the previous equation in vector form:
\begin{equation}
    \underbrace{\left[ \textbf{N}_{\omega'} \mathbf{\mathcal{M}}{'}_{\omega',\text{f}}^{(\text{v})} \, , - \mathbf{\mathcal{M}}{'}_{\omega',\text{b} }^{(\text{v})}  \right]}_{\textbf{A}}
    \begin{bmatrix}
        \textbf{e}_{\omega',\text{f}}^{\text{(out)},3}\\
        \textbf{e}_{\omega',\text{b}}^{\text{(out)},1}
    \end{bmatrix}
    =
    \underbrace{\left[ \mathbf{\mathcal{M}}{'}_{\omega',\text{f}}^{(\text{v})} \, , - \textbf{N}_{\omega'} \mathbf{\mathcal{M}}{'}_{\omega',\text{b} }^{(\text{v})}  \right]}_{\textbf{B}}
    \begin{bmatrix}
        \textbf{e}_{\omega',\text{f}}^{\text{(in)},1}\\
        \textbf{e}_{\omega',\text{b}}^{\text{(in)},3}
    \end{bmatrix},
\end{equation}
where we have defined the $2\cdot(2 N_{\text{F}} + 1) \times 2\cdot(2 N_{\text{F}} + 1)$ matrices $\textbf{A}$ and $\textbf{B}$. It is now easy to define our scattering matrix $\textbf{S}_{\omega'}$ as
\begin{equation}\label{eq:scatmat}
    \begin{bmatrix}
        \textbf{e}_{\omega',\text{f}}^{\text{(out)},3}\\
        \textbf{e}_{\omega',\text{b}}^{\text{(out)},1}
    \end{bmatrix}
    =
    \textbf{S}_{\omega'}
    \begin{bmatrix}
        \textbf{e}_{\omega',\text{f}}^{\text{(in)},1}\\
        \textbf{e}_{\omega',\text{b}}^{\text{(in)},3}
    \end{bmatrix},
\end{equation}
with $\textbf{S}_{\omega'} = \textbf{A}^{-1}\,\textbf{B}$.

Furthermore, since we want to see the existence of interface states between slabs, we need to extract the vector $\textbf{e}_{\omega}^{(\text{m}),i}$ at each interface $i$ in terms of the weights of the input fields. To derive their expression, we make use of the previous mode-matching equations (\ref{eq:1stint}) and substitute $\textbf{e}_{\omega',\text{b}}^{\text{(out),1}}$ expressed in terms of the incident fields: $\textbf{e}_{\omega',\text{b}}^{\text{(out),1}} = \textbf{S}_{\omega'}^{21} \, \textbf{e}_{\omega',\text{f}}^{\text{(in),1}} + \textbf{S}_{\omega'}^{22} \, \textbf{e}_{\omega',\text{b}}^{\text{(in)},3}$. We would then have:
\begin{equation}
    \mathbf{\mathcal{M}}{'}_{\omega'}^{(\text{m})} \, \textbf{e}_{\omega'}^{\text{(m)},1} = (\mathbf{\mathcal{M}}{'}_{\omega',\text{f}}^{(\text{v})} + \mathbf{\mathcal{M}}{'}_{\omega',\text{b} }^{(\text{v})} \, \textbf{S}_{\omega'}^{21}) \, \textbf{e}_{\omega',\text{f}}^{\text{(in),1}} + \mathbf{\mathcal{M}}{'}_{\omega',\text{b} }^{(\text{v})} \, \textbf{S}_{\omega'}^{22} \textbf{e}_{\omega',\text{b}}^{\text{(in)},3},
\end{equation}
\begin{equation}
    \textbf{e}_{\omega'}^{\text{(m)},1} = (\mathbf{\mathcal{M}}{'}_{\omega'}^{(\text{m})})^{-1} \, \underbrace{\left[\mathbf{\mathcal{M}}{'}_{\omega',\text{f}}^{(\text{v})} + \mathbf{\mathcal{M}}{'}_{\omega',\text{b} }^{(\text{v})} \, \textbf{S}_{\omega'}^{21} \, , \, \mathbf{\mathcal{M}}{'}_{\omega',\text{b} }^{(\text{v})} \, \textbf{S}_{\omega'}^{22}  \right]}_{\textbf{C}} \begin{bmatrix}
        \textbf{e}_{\omega',\text{f}}^{\text{(in),1}}\\
        \textbf{e}_{\omega',\text{b}}^{\text{(in)},3}
    \end{bmatrix}.
\end{equation}

Once we have the coefficients for the first slab, we can repeat the same process for the second one, taking into account that the relevant mode-matching equation is now Eq.~(\ref{eq:2ndint}), such that:
\begin{equation}
    \textbf{e}_{\omega'}^{(\tilde{\text{m}}),2} = (\mathbf{\mathcal{M}}{'}_\omega^{(\tilde{\text{m}})})^{-1} \, \mathbf{\mathcal{M}}{'}_\omega^{(\text{m})} \, {\textbf{P}'} \, (\mathbf{\mathcal{M}}{'}_\omega^{(\text{m})})^{-1} \, \textbf{C} \begin{bmatrix}
        \textbf{e}_{\omega',\text{f}}^{\text{(in)},1}\\
        \textbf{e}_{\omega',\text{b}}^{\text{(in)},3}
    \end{bmatrix}.
\end{equation}
Finally, to obtain the EM fields we only need to substitute the previous magnitudes into the following equation for the fields inside the material:
\begin{align}\label{eq:modedef}
    \begin{bmatrix}
        E'(x',t')\\
        H'(x',t')
    \end{bmatrix}^{(\text{m})} &=
    \textbf{T}(x') \sum_{\tau=1,2} \sum_{p = -N_{\text{F}}}^{N_{\text{F}}} e^{(\text{m})}_{\omega',\tau,p} \begin{bmatrix}
        E_{\omega',\tau,p}(x',t')\\
        H_{\omega',\tau,p}(x',t')
    \end{bmatrix}^{(\text{m})} = \nonumber \\
    &= \textbf{T}(x') \sum_{\tau=1,2} \sum_{p = -N_{\text{F}}}^{N_{\text{F}}} e^{(\text{m})}_{\omega',\tau,p} \sum_{n = -N_{\text{F}}}^{N_{\text{F}}} \begin{bmatrix}
        E_{\omega',\tau,p,n}\\
        H_{\omega',\tau,p,n}
    \end{bmatrix}^{(\text{m})} \, e^{i(k'_{\omega',\tau,p} + n g/\gamma)x' - i \omega' t'},
\end{align}
and into this expression for the fields in vacuum:
\begin{align}\label{eq:modedef_vac}
    \begin{bmatrix}
        E'(x',t')\\
        H'(x',t')
    \end{bmatrix}_{\text{f/b}}^{(\text{in/out})} &=
    \textbf{T}^{(\text{v})} \sum_{n = -N_{\text{F}}}^{N_{\text{F}}} e^{(\text{in/out})}_{\omega',\text{f/b},n} \begin{bmatrix}
        E_{\omega',\text{f/b},n}(x',t')\\
        H_{\omega',\text{f/b},n}(x',t')
    \end{bmatrix}^{(\text{in/out})} = \nonumber \\
    & =\textbf{T}^{(\text{v})} \sum_{n = -N_{\text{F}}}^{N_{\text{F}}} e^{(\text{in/out})}_{\omega',\text{f/b},n}  \begin{bmatrix}
        E_{\omega',\text{f/b},n}\\
        H_{\omega',\text{f/b},n}
    \end{bmatrix}^{(\text{in/out})} \, e^{i(k'_{\omega',\text{f/b},n} + n g/\gamma)x' - i \omega' t'}.
\end{align}

Lastly, we calculate the transmission spectrum in the lab-frame obtained by exciting the composite system from the left side with a plane wave of frequency $\omega_0$. To do so, we transform the incident frequency to the comoving frame, since it corresponds to the conserved quantity, and then we use the scattering matrix defined in Eq.~\eqref{eq:scatmat} to calculate the transmitted fields as seen in the lab-frame. Once the fields are obtained, we can then evaluate the time-averaged Poynting vector in the $x'$ direction, defined as:
\begin{align}
    P_{\omega',\text{f}}^{(\text{out})}(x') = - E_{\omega',\text{f}}^{(\text{out})}(x') \, \left( H_{\omega',\text{f}}^{(\text{out})}(x') \right)^*
    = - \sum_{n,n'} e_{\omega',\text{f},n}^{(\text{out})} \, e_{\omega',\text{f},n'}^{(\text{out}) \, *} \, E_{\omega',\text{f},n}^{(\text{out})} \, H_{\omega',\text{f},n'}^{(\text{out}) \,*} \; e^{i\left( k'_{\omega',\text{f},n} - k{'}^*_{\omega',\text{f},n'} + (n-n')g/\gamma \right) x'}.
\end{align}
Then, normalizing the previous expression by the Poynting vector of the incident wave, and evaluating at the third interface $x' = 2\gamma N a$, we obtain the transmittance as
\begin{equation}
    T(\omega') = \frac{P_{\omega',\text{f}}^{(\text{out})}(x' = 2\gamma N a)}{P_{\omega',\text{f}}^{(\text{in})}(x'=0)} = \frac{\sum_{n,n'} e_{\omega',\text{f},n}^{(\text{out})} \, e_{\omega',\text{f},n'}^{(\text{out}) \, *} \, E_{\omega',\text{f},n}^{(\text{out})} \, H_{\omega',\text{f},n'}^{(\text{out}) \,*} \; e^{i\left( k'_{\omega',\text{f},n} - k{'}^*_{\omega',\text{f},n'} \right) (2\gamma N a)}}{E_{\omega',\text{f},0}^{(\text{in})} H_{\omega',\text{f},0}^{(\text{in}) \,*}}.
\end{equation}

\subsection{Spatial Boundary}

The procedure to calculate the scattering matrix for two slabs truncated purely in space closely follows that described in the previous section. The main differences lie in the conserved quantity, which is now the lab-frame frequency, and in the boundary conditions enforcing the continuity of the lab-frame electromagnetic fields. These modifications can be taken into account by substituting the eigenvector matrices of both the vacuum and the modulated medium with those derived in Sec.~\ref{sec:eigen_lab}, where $\omega$ rather than $\omega'$ is well defined, as well as changing the transformation matrices in Eq.~\eqref{eq:trans_mat_exp} to the identity. In this way,  Eqs.~\eqref{eq:1stint}-\eqref{eq:3rdint} naturally enforce the continuity of $[E,H]^\text{T}$ instead of $[E',H']^\text{T}$.\\

Finally, due to the intrinsic temporal dependence of the eigenfunctions defined in Eq.~\eqref{eq:Ef_eigen_lan}-\eqref{eq:Hf_eigen_lan}, the time averaged Poynting vector still has an explicit temporal dependence
\begin{align}
    P_{\omega,\text{f}}^{(\text{out})}(x,t) &= - E_{\omega,\text{f}}^{(\text{out})}(x,t) \, \left( H_{\omega,\text{f}}^{(\text{out})}(x,t) \right)^*
    = \nonumber \\
    &= - \sum_{n,n'} e_{\omega,\text{f},n}^{(\text{out})} \, e_{\omega,\text{f},n'}^{(\text{out}) \, *} \, E_{\omega,\text{f},n,n}^{(\text{out})} \, H_{\omega,\text{f},n',n'}^{(\text{out}) \,*} e^{i\left( k_{\omega,\text{f},n} - k^*_{\omega,\text{f},n'} + (n-n')g \right) x - i (n-n')\Omega t}.
\end{align}
We next integrate the time averaged Poynting vector over one modulation period $T = 2\pi/\Omega$
\begin{equation}
    \braket{P_{\omega,\text{f}}^{(\text{out})}} = \frac{1}{T} \int_0^T P_{\omega,\text{f}}^{(\text{out})} = - \sum_{n,n'} \delta_{n,n'} \; e_{\omega,\text{f},n}^{(\text{out})} \, e_{\omega,\text{f},n'}^{(\text{out}) \, *} \, E_{\omega,\text{f},n}^{(\text{out})} \, H_{\omega,\text{f},n'}^{(\text{out}) \,*} \; e^{i\left( k_{\omega,\text{f},n} - k^*_{\omega,\text{f},n'} + (n-n')g \right) x},
\end{equation}
and finally we arrive at the transmittance for the spatial boundary case
\begin{equation}
    T(\omega) = \frac{\braket{P_{\omega,\text{f}}^{(\text{out})}(x = 2 N a)}}{\braket{P_{\omega,\text{f}}^{(\text{in})}(x=0)}} = \frac{\sum_n \, |e_{\omega,\text{f},n}^{(\text{out})}|^2 E_{\omega,\text{f},n}^{(\text{out})} \, H_{\omega,\text{f},n}^{(\text{out}) \,*}}{E_{\omega,\text{f},0}^{(\text{in})} H_{\omega,\text{f},0}^{(\text{in}) \,*}}.
\end{equation}

\section{Spatiotemporal Zak phase}

As discussed in the main text, the Hermiticity of the constitutive matrix, together with the presence of $\mathcal{P'T'}$ symmetry, allows us to define a Hermitian form of the spatiotemporal Zak phase. Under these conditions, left and right eigenfunctions can be related, which converts the biorthogonal Berry connection into a conventional one defined through a weighted inner product. To demonstrate this, we derive a general eigenvalue problem from Eq.~\eqref{eq:gen_com_Maxwell} by considering monochromatic fields such that
\begin{equation}\label{eq_pre_eigen}
    \hat{\mathbf{\mathds{L}}}' \mathbf{\Psi}'_{\text{R}} = \omega' \; \hat{\textbf{M}}' \mathbf{\Psi}'_{\text{R}},
\end{equation}
where $\hat{\mathbf{\mathds{L}}}' = i \hat{\textbf{L}}'$.
As we can see, both $\hat{\mathbf{\mathds{L}}}'$ and $\hat{\textbf{M}}'(x')$ operators are Hermitian. However, if we write Eq.~\eqref{eq_pre_eigen} as
\begin{equation}
    \hat{\textbf{H}}_{\text{R}} \mathbf{\Psi}'_{\text{R}} = \left( \hat{\textbf{M}}'\right)^{-1} \hat{\mathbf{\mathds{L}}}' \mathbf{\Psi}'_{\text{R}} = \omega' \; \mathbf{\Psi}'_{\text{R}},
\end{equation}
we clearly see that the operator $\hat{\textbf{H}}_{\text{R}}$ is not Hermitian, since the previous operators do not commute. One can prove that the adjoint operator of $\hat{\textbf{H}}_{\text{R}}$ is \cite{NonBlochTheoSTcrys} 
\begin{equation}
    \hat{\textbf{H}}_{\text{L}} = \hat{\mathbf{\mathds{L}}}' \left( \hat{\textbf{M}}'\right)^{-1} = \left( \hat{\textbf{H}}_{\text{R}} \right)^\dagger, \text{with } \; \; \; \hat{\textbf{H}}_{\text{L}} \mathbf{\Psi}'_{\text{L}} = (\omega{'})^* \, \mathbf{\Psi}'_{\text{L}} .
\end{equation}
Now, writing the adjoint operator as $\hat{\textbf{H}}_{\text{L}} = \hat{\textbf{M}}' \left( \hat{\textbf{M}}'\right)^{-1} \hat{\mathbf{\mathds{L}}}'  \left( \hat{\textbf{M}}'\right)^{-1}$, we arrive at the following expression
\begin{equation}
    \hat{\textbf{M}}' \underbrace{\left( \hat{\textbf{M}}'\right)^{-1} \hat{\mathbf{\mathds{L}}}'}_{\hat{\textbf{H}}_{\text{R}}}  \left( \hat{\textbf{M}}'\right)^{-1} \mathbf{\Psi}'_{\text{L}} = (\omega{'})^* \, \mathbf{\Psi}'_{\text{L}}  
    \longrightarrow 
    \hat{\textbf{H}}_{\text{R}} \left( \hat{\textbf{M}}'\right)^{-1} \mathbf{\Psi}'_{\text{L}} = (\omega{'})^* \, \left( \hat{\textbf{M}}'\right)^{-1} \mathbf{\Psi}'_{\text{L}}.  
\end{equation}
Considering in this section the subluminal regime, we know from the conservation of $\mathcal{P'T'}$-symmetry that the frequency spectrum is real $(\omega{'})^* = \omega'$, which means
\begin{equation}
    \mathbf{\Psi}'_{\text{R}} = \left( \hat{\textbf{M}}'\right)^{-1} \mathbf{\Psi}'_{\text{L}},
\end{equation}
such that the inner product of left and right eigenfunctions becomes
\begin{equation}
    \braket{\mathbf{\Psi}'_{\text{L}} | \mathbf{\Psi}'_{\text{R}}} = \int_{\text{UC}} dx' \left(  \mathbf{\Psi}'_{\text{L}} \right)^\dagger \, \mathbf{\Psi}'_{\text{R}} = \int_{\text{UC}} dx' \left(  \hat{\textbf{M}}' \mathbf{\Psi}'_{\text{R}} \right)^\dagger \mathbf{\Psi}'_{\text{R}} = \int_{\text{UC}} dx' \left(  \mathbf{\Psi}'_{\text{R}}\right)^\dagger  \hat{\textbf{M}}' \mathbf{\Psi}'_{\text{R}}.
\end{equation}
This relation allows us to write the Zak phase defined with a biorthogonal Berry connection as
\begin{equation}
    \theta^{\mathrm{ST}}_m =  i \int_{-g/2\gamma}^{+g/2\gamma} dk'\braket{\textbf{u}{'}_{\text{L}, k',m} | \partial_{k'} \textbf{u}'_{\text{R}, k',m}} = i \int_{-g/2\gamma}^{+g/2\gamma} dk'\braket{\textbf{u}{'}_{\text{R}, k',m} | \hat{\textbf{M}}' \partial_{k'} \textbf{u}'_{\text{R}, k',m}},
\end{equation}
which ends with the same structure as a conventional Berry connection defined with a weighted scalar product, as mentioned in the main text. 

To obtain numerically the ST Zak phase, we make use of the eigenfunctions defined in Eqs.~\eqref{eq:com_eigefun_Ek}-\eqref{eq:com_eigefun_Hk} and the discretized Zak phase definition:
\begin{equation}\label{eq:disc_Zak}
    \theta^{\mathrm{ST}}_m = -\text{Im log} \prod_i \braket{\textbf{u}'_{k'_i,m} | \textbf{u}'_{k'_i+1,m}}_{\hat{\textbf{M}}},
\end{equation}
where $\braket{\cdot | \cdot}_{\hat{\textbf{M}}}$ correspond to the weighted scalar product from Eq.~(17) of the main text, and we enforce the periodic gauge condition $\textbf{u}'_{k'=g/2\gamma ,m}(x') = e^{-ig/\gamma x'}   \textbf{u}'_{k'=-g/2\gamma ,m}(x')$.

Studying these scalar products along the comoving-frame BZ, we find that they are always real, implying that the Zak phase is determined by the overall sign of the product in Eq.~\eqref{eq:disc_Zak}. The sign changes occur when crossing the band gaps and at $k'=\omega'=0$. Importantly, the scalar products involving the $k'=0$ eigenfunctions for the first band exhibit inconsistent sign changes depending on the specific parameters of the STPhC. Although the two phases cannot be unambiguously assigned to $\alpha > 0$ or $\alpha < 0$, they remain clearly distinguished by distinct ST Zak phase values, which is enough to predict interface states. This ambiguity arises because, while the electric and magnetic eigenfunctions possess opposite parity, they become constant at $k' = \omega' = 0$, rendering their parity ill-defined and leading to inconsistent phase assignments.

\section{Band crossing position}

\begin{figure}[h!]
    \centering
    \includegraphics[width=0.45\linewidth]{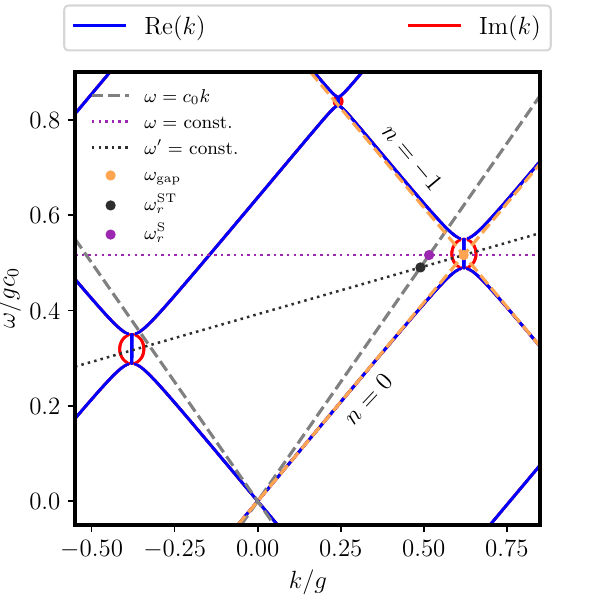}
    \caption{Representation of the band-crossing position and the resonance frequencies of the interface states for both types of boundaries. The grey and amber dashed lines indicate the free-space dispersion and the linear approximation of the STPhC band dispersion, respectively. The violet and black dotted lines represent the frequency conservation conditions in the lab and comoving frames, respectively. The band-crossing frequency $\omega_{\mathrm{gap}}$ is obtained from the intersection between the $n = 0$ forward and $n = -1$ backward modes, while the resonance frequencies are determined by applying the corresponding conservation law for each boundary type.}
    \label{fig:band-crossing}
\end{figure}

In Fig.~\ref{fig:band-crossing}, we show how to infer the resonance frequency of the interface states from the position in the band diagram where the $(0,-1 )$ Floquet bands cross, and considering the conserved quantity in the scattering process. We do this for both cases, $\omega_r^{ST}$ (black dot) and $\omega_r^S$ (purple dot).

\section{Amplification}

\begin{figure}[t]
    \centering
    \includegraphics[width=\textwidth]{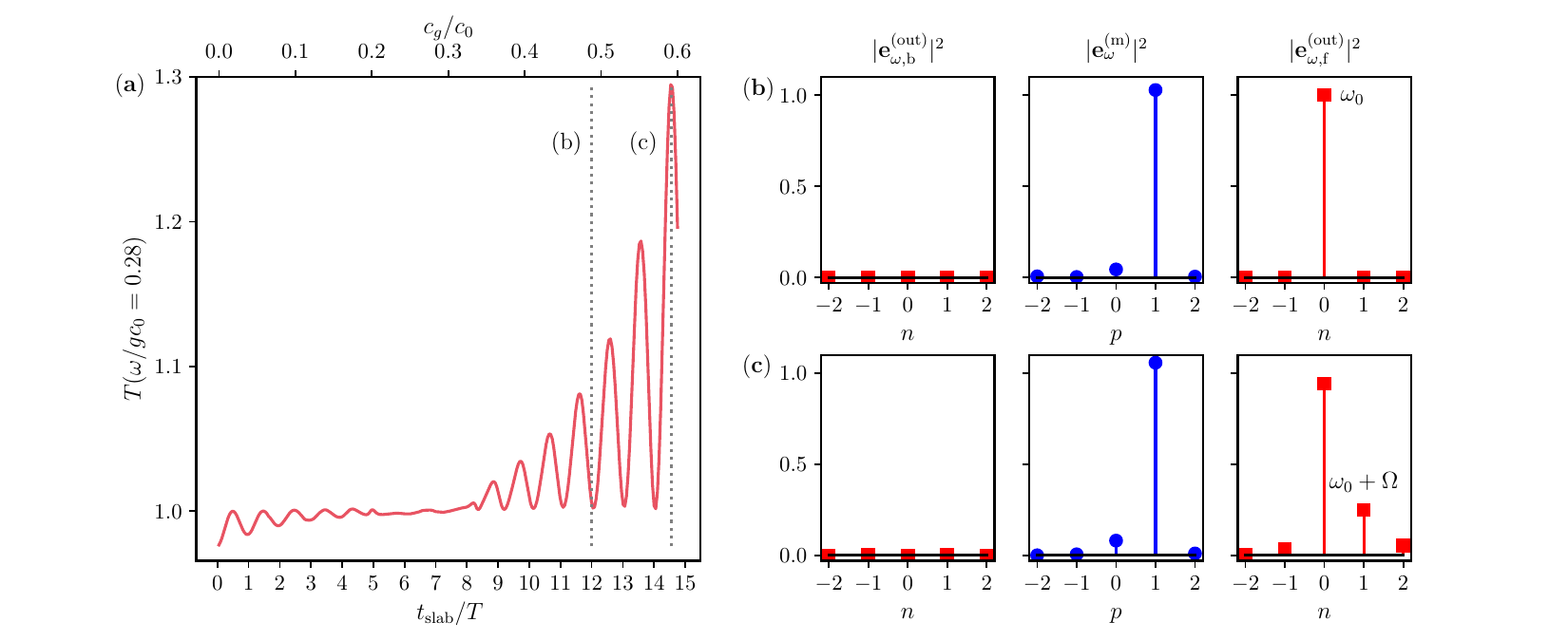}
    \caption{Amplification of the transmitted wave through higher-frequency mode generation in a spatial slab of a STPhC. (a) Transmittance of a plane wave with $\omega_0 = 0.28 g c_0$ exciting a single slab, composed of ten unit cells truncated in space, as a function of the grating velocity $c_g$ and the transit time across the slab $t_{\text{slab}}$, normalized by the modulation period $T = 2\pi/\Omega$. (b) Mode decomposition of the reflected, internal and transmitted wave for a grating velocity corresponding to a unit transmittance. (c) Same magnitudes for a $c_g$ yielding amplification.}
    \label{fig:amplification}
\end{figure}

As discussed in the main text, the spatial boundary case presents broadband amplification for certain values of the grating speed, demonstrated by the appearance of horizontal bands of transmission higher than unity in Fig.~3(g). To understand this effect, we study the transmission through a single spatially truncated slab of a STPhC excited by a plane wave of fixed frequency $\omega_0$, while varying the modulation speed $c_g$. As observed in Fig. \ref{fig:amplification}(a), the transmittance spectrum presents periodic peaks of amplification, whose magnitude increases with $c_g$ (see top $x$-axis). To understand the periodic nature of the amplification, we also plot the transmittance in terms of the time the wave takes to traverse the slab (see bottom $x$-axis),
\begin{equation}
    t_{\text{slab}} = \frac{D}{v_{\text{eff}}(c_g)},
\end{equation}
where $D$ is the size of the slab. When normalized by the modulation period $T$, we find that the wave is transmitted without amplification when it exits the slab after an integer number of modulation cycles ($t_{\text{slab}} = qT$, with $q \in \mathds{N}$), and experiences amplification when it exits halfway through the modulation ($t_{\text{slab}} = (q + 1/2)T$). This behaviour can also be interpreted in terms of the permittivity encountered at the slab boundaries. Indeed, for $t_{\text{slab}} = q T$, the wave enters and exits through the same value of the permittivity ($\epsilon^{(\text{in})} = \epsilon^{(\text{out})}$), whereas for $t_{\text{slab}} = (q + 1/2)T$, it encounters opposite permittivities, with $\epsilon^{(\text{in})} = \epsilon_0 \epsilon_m (1 \pm \alpha_e)$ and $\epsilon^{(\text{out})} = \epsilon_0 \epsilon_m (1 \mp \alpha_e)$. Therefore, amplification originates from the boundary conditions themselves, explaining both its broadband character and its independence from the slab thickness. 

Finally, using our semi-analytical approach, we analyze the mode decomposition of the reflected, internal and transmitted wave for two representative cases: the modulation is strong but the wave is only transmitted and not amplified [Fig.~\ref{fig:amplification}(b)], and the maximum amplification point [Fig.~\ref{fig:amplification}(c)]. For both cases, reflection is negligible and two internal modes are excited, with the dominant mode ($p=1$) only containing the fundamental Bloch-Floquet harmonic ($n=0$), and the $p=0$ mode including the $n=1$ harmonic.\footnote{In this section, we omit the $\tau$ index for clarity, so that all $2 \, (2 N_{\mathrm{F}} + 1)$ eigenvalues are labeled just by $p$. Since the specific value of $p$ has no physical meaning, the eigenvalues can be ordered arbitrarily, and our analysis can focus solely on the Bloch-Floquet harmonics associated with each eigenfunction.} 
Thus, higher-frequency components are present inside the slab in both regimes, though weakly. Importantly, however, the transmitted wave only presents both $n=0$ and $n=1$ harmonics in \acd{the amplifying case} Fig.~\ref{fig:amplification}(c), confirming that frequency conversion processes underlie the broadband amplification observed in this system. \acd{Indeed, while the weight of each individual harmonic mode cannot exceed unity, the presence of multiple harmonics, combined with the temporal modulation that breaks energy conservation, can result in the sum of their weights being higher than one, thus leading to amplification.}

\section{Superluminal Regime}

\subsection{Eigenvalue problem for the time-like frame}

To study the symmetries of the STPhC in the superluminal regime, we need to make a Lorentz transformation to a frame whose velocity is given by $c_f = c_0^2/c_g$ rather than $c_g$, coined the time-like frame. In this frame, the magneto-electric coupling is still present and, thus, we derive an eigenvalue problem for $\omega'$ starting from the lab-frame ansatz as was done for the subluminal case:
\begin{equation}\label{eq:mix-ans_sup}
    \begin{bmatrix}
        E (x,t)\\
        H (x,t)
    \end{bmatrix} 
    = e^{i(k x - \omega t)} \; \sum_n
    \begin{bmatrix}
        E_n \\
        H_n
    \end{bmatrix}
    \, e^{i\, n (g x - \Omega t)} = e^{i(k' x' - \omega' t')} \, \sum_n
    \begin{bmatrix}
        E_n \\
        H_n
    \end{bmatrix}
    \, e^{-i\, n \Omega/\gamma \, t'} =
    \begin{bmatrix}
        E (x',t')\\
        H (x',t')
    \end{bmatrix}.
\end{equation}
Then, using Eq.~\eqref{eq:Max} and swapping $c_g$ by $c_f$, we obtain the following eigenvalue problem:
\begin{equation}\label{eigen_com_k_sup}
\begin{pmatrix}
    \mathbf{\mathds{M}}^{\text{EE,L}}(k') & \mathbf{\mathds{M}}^{\text{EH,L}}(k') \\
    \mathbf{\mathds{M}}^{\text{HE,L}}(k') & \mathbf{\mathds{M}}^{\text{HH,L}}(k')
\end{pmatrix}
\,
\begin{bmatrix}
    \textbf{E}_{k'} \\
    \textbf{H}_{k'}
\end{bmatrix} 
=
\omega'_{k'} \,
\begin{pmatrix}
    \mathbf{\mathds{M}}^{\text{EE,R}}(k') & \mathbf{\mathds{M}}^{\text{EH,R}}(k') \\
    \mathbf{\mathds{M}}^{\text{HE,R}}(k') & \mathbf{\mathds{M}}^{\text{HH,R}}(k')
\end{pmatrix}
\,
\begin{bmatrix}
    \textbf{E}_{k'} \\
    \textbf{H}_{k'}
\end{bmatrix},
\end{equation}
where the right matrix entries read
\begin{equation}
    \mathbf{\mathds{M}}^{\text{EH,R}}_{n,n'} = \mathbf{\mathds{M}}^{\text{HE,R}}_{n,n'} = -c_f/c_0^2 \, \delta_{n,n'},
\end{equation}
\begin{equation}
    \mathbf{\mathds{M}}^{\text{EE,R}}_{n,n'} = -\epsilon_0 \epsilon_m [\delta_{n,n'} + \alpha_e (\delta_{n,n'+1} + \delta_{n,n'-1})],
\end{equation}
\begin{equation}
    \mathbf{\mathds{M}}^{\text{HH,R}}_{n,n'} = -\mu_0 \mu_m [\delta_{n,n'} + \alpha_m (\delta_{n,n'+1} + \delta_{n,n'-1})],
\end{equation}
while the left ones are
\begin{equation}
    \mathbf{\mathds{M}}^{\text{EH,L}}_{n,n'} = \mathbf{\mathds{M}}^{\text{HE,L}}_{n,n'} = (k' + n\, g/\gamma) \, \delta_{n,n'},
\end{equation}
\begin{equation}
    \mathbf{\mathds{M}}^{\text{EE,L}}_{n,n'} = \epsilon_0 \epsilon_m (c_f k' + n\, \Omega/\gamma) [\delta_{n,n'} + \alpha_e  (\delta_{n,n'+1} +\delta_{n,n'-1})],
\end{equation}
\begin{equation}
    \mathbf{\mathds{M}}^{\text{HH,L}}_{n,n'} = \mu_0 \mu_m (c_f k' + n\, \Omega/\gamma) [\delta_{n,n'} + \alpha_m (\delta_{n,n'+1} +\delta_{n,n'-1})].
\end{equation}
One can show that this eigenvalue problem is equivalent to that of the subluminal case defined in Eq.~\eqref{eigen_com_k}, upon replacing 
$c_g$ with $c_f$. This equivalence is expected, since the Lorentz transformations are applied in the same manner, with only the frame velocity being different. Once the eigenproblem is solved, we can then write the eigenfunctions for a given momentum $k'$ as 
\begin{equation}
    E_{k',\tau,p} (x',t') = e^{i( k' x' - \omega'_{k',\tau,p} t')} \sum_n E_{k',\tau,p,n} \, e^{-i\, n \Omega/\gamma \, t'}
\end{equation}
and 
\begin{equation}
    H_{k',\tau,p} (x',t') = e^{i( k' x' - \omega'_{k',\tau,p} t')} \sum_n H_{k',\tau,p,n} \, e^{-i\, n \Omega/\gamma \, t'} .
\end{equation}
However, in this frame the relevant fields are $[\textbf{D}', \textbf{B}']$. Therefore, we need to apply Lorentz transformations:
\begin{equation}
    \begin{bmatrix}
        D'(x',t')\\
        B'(x',t')
    \end{bmatrix}=
    \tilde{\textbf{T}}(t')
    \begin{bmatrix}
        E(x',t')\\
        H(x',t')
    \end{bmatrix},
\end{equation}
with
\begin{equation}\label{eq:trans_mat_sup}
    \tilde{\textbf{T}}(t') = 
    \gamma \;
    \begin{pmatrix}
        \epsilon(t') & c_f/c_0^2\\
        c_f/c_0^2  &  \mu(t')
    \end{pmatrix}
\end{equation}
the new transformation matrix. Thus, in the time-like frame the eigenfunctions are defined as
\begin{equation}\label{eq:com_eigefun_D}
    D'_{k',\tau,p}(x',t') = e^{i (k' x' - \omega'_{k',\tau,p} t')} \sum_n \gamma \, \left[ \epsilon(t') \, E_{k',\tau,p,n} + c_f/c_0^2 H_{k',\tau,p,n} \right ] e^{-i\, n \Omega/\gamma \, t'}
\end{equation}
and
\begin{equation}\label{eq:com_eigefun_B}
    B'_{k',\tau,p}(x',t') = e^{i (k' x' - \omega'_{k',\tau,p} t')} \sum_n \gamma \, \left[ \mu(t') \, H_{k',\tau,p,n} + c_f/c_0^2 E_{k',\tau,p,n} \right ] e^{-i\, n \Omega/\gamma \, t'}.
\end{equation}\\

Finally, we can derive the eigenvalues and eigenvectors in the vacuum by evaluating $\alpha_e = \alpha_m = 0$, as well as $\epsilon_m = \mu_m = 1$, in Eqs.~\eqref{eigen_com_k_sup}. We obtain
\begin{equation}
    \omega{'}_{k'}^\pm = -n \Omega/\gamma \pm c_0 k'
\end{equation}
with $\omega{'}_n^{+} \equiv \omega'_{k',\mathrm{f},n}$ and  $\omega{'}_n^{-} \equiv \omega'_{k',\mathrm{b},n}$. The eigenvectors correspond to
\begin{equation} \label{eq:com_vacummodes}
    \begin{bmatrix}
        E_{k',\mathrm{f},n} \\
        H_{k',\mathrm{f},n}
    \end{bmatrix}
    =
    \begin{bmatrix}
        1 \\
        \frac{ -k' }{\mu_0 (\omega'_{k',\mathrm{f},n} + n \Omega/\gamma)}
    \end{bmatrix} 
    =
    \frac{1}{\sqrt{1+\eta^{-2}}}
    \begin{bmatrix}
        1 \\
        -\eta^{-1}
    \end{bmatrix}
    ;
    \; \; 
    \begin{bmatrix}
        E_{k',\mathrm{b},n} \\
        H_{k',\mathrm{b},n}
    \end{bmatrix}
    =
    \begin{bmatrix}
        1 \\
        \frac{ -k' }{\mu_0 (\omega'_{k',\mathrm{b},n} + n \Omega/\gamma)}
    \end{bmatrix} 
    =
    \frac{1}{\sqrt{1+\eta^{-2}}}
    \begin{bmatrix}
        1 \\
        \eta^{-1}
    \end{bmatrix},
\end{equation}
where $\eta^2 = \mu_0/\epsilon_0$ is the wave impedance in vacuum.

\subsection{Scattering matrix and transmission spectrum}

When the modulation enters the superluminal regime, the STPhC acquires a temporal-like nature. Therefore, the two types of boundaries we consider in this regime are: spatiotemporal when the system is truncated along with the modulation, and purely temporal. As in the subluminal regime, each boundary will conserve different magnitudes: $k'$ for the ST interface and $k$ for the temporal one. Consequently, for each case we must select the eigenfunctions of the STPhC that possess a well-defined $k$ or $k'$, such that we use those obtained from Eq.~\eqref{eigen_labk} for temporal boundaries and from Eq.~\eqref{eigen_com_k_sup} for spatiotemporal ones.

\subsubsection{Spatiotemporal Boundary}

We start by defining the scattering matrix of two STPhC slabs truncated with ST interfaces. To do so, we now enforce the continuity of the time-like frame $D'_z$ and $B'_y$ fields at each interface, which are easily defined as $t' = 0, N \gamma T, 2N \gamma T$, where $N$ is the number of ST unit cells (defined as $t' \in [0, \gamma T]$) used to construct each slab. Making use of the field decomposition, we can write the matching conditions using a superposition of the eigenvectors as 
\begin{equation}
    \begin{bmatrix}
        \textbf{D}'_{k'} \\
        \textbf{B}'_{k'}
    \end{bmatrix}^{(\text{m})}
    = \tilde{\mathbf{\mathcal{M}}}{'}_{k'}^{(\text{m})} \, \textbf{e}_{k'}^{(\text{m})}; \; \; \; 
    \begin{bmatrix}
        \textbf{D}'_{k'} \\
        \textbf{B}'_{k'}
    \end{bmatrix}_{\text{f/b}}^{(\text{in/out})}
    = \tilde{\mathbf{\mathcal{M}}}{'}_{k',\text{f/b}}^{{(\text{v})}} \, \textbf{e}_{k',\text{f/b}}^{(\text{in/out})},
\end{equation}
with 
\begin{equation}
    \tilde{\mathbf{\mathcal{M}}}{'}_{k'}^{(\text{m})} = \tilde{\textbf{T}}{'}^{(\text{m})} \mathbf{\mathcal{M}}_{k'}^{(\text{m})}; \; \; \; \tilde{\mathbf{\mathcal{M}}}{'}_{k',\text{f/b}}^{(\text{v})} = \tilde{\textbf{T}}{'}^{(\text{v})} \mathbf{\mathcal{M}}_{k',\text{f/b}}^{(\text{v})},
\end{equation}
where $\mathbf{\mathcal{M}}_{k'}^{(\text{m})}$ and $\mathbf{\mathcal{M}}_{k',\text{f/b}}^{(\text{v})}$ correspond to the matrices containing the eigenvectors obtained from Eqs.~\eqref{eigen_com_k_sup} and \eqref{eigen_labk}, respectively. Finally, the $\tilde{\textbf{T}}{'}^{(\text{v})}$ and $\tilde{\textbf{T}}{'}^{(\text{m})}$ matrices correspond to the transformation matrix of Eq.~\eqref{eq:trans_mat_sup} evaluated at the interfaces for the unmodulated and modulated cases, respectively. The periodicity of the modulation ensures that at the boundaries the matrices read
\begin{equation}\label{eq:trans_mat_exp}
    \tilde{\textbf{T}}{'}^{(\text{v})} = \gamma \;
    \begin{pmatrix}
        \epsilon_0 \mathds{1} & c_f/c_0^2  \mathds{1}\\
        c_f/c_0^2 \mathds{1} & \mu_0 \mathds{1}
    \end{pmatrix}; \; \; \;
    \tilde{\textbf{T}}{'}^{(\text{m})} = \gamma \;
    \begin{pmatrix}
        \epsilon_0 \epsilon_m (1+2\alpha_e) \mathds{1} & c_f/c_0^2  \mathds{1}\\
        c_f/c_0^2  \mathds{1} & \mu_0 \mu_m (1+2\alpha_m) \mathds{1}
    \end{pmatrix}.
\end{equation}

When considering time-like interfaces, causality needs to be enforced, which consequently forbids the existence of reflected waves in the same space-time region as the incoming wave. To take this into account, we need to change the structure of Eqs.~\eqref{eq:1stint}-\eqref{eq:3rdint} in order to enforce the correct continuity equations: 
\begin{itemize}
    \item \underline{1st interface}: 
\end{itemize}
    \begin{equation}\label{eq:1stint_sup}
        \begin{bmatrix}
            \textbf{D}'_{k'} \\
            \textbf{B}'_{k'}
        \end{bmatrix}_{\text{f}}^{\text{(in)},1}
        +
        \begin{bmatrix}
            \textbf{D}'_{k'} \\
            \textbf{B}'_{k'}
        \end{bmatrix}_{\text{b}}^{\text{(in)},1}
        =
        \begin{bmatrix}
            \textbf{D}'_{k'} \\
            \textbf{B}'_{k'}
        \end{bmatrix}^{\text{(m)},1}
        \rightarrow
        \tilde{\mathbf{\mathcal{M}}}{'}_{k',\text{f}}^{(\text{v})} \, \textbf{e}_{k',\text{f}}^{\text{(in),1}} + \tilde{\mathbf{\mathcal{M}}}{'}_{k',\text{b}}^{(\text{v})} \, \textbf{e}_{k',\text{b}}^{\text{(in),1}} = \tilde{\mathbf{\mathcal{M}}}{'}_{k'}^{(\text{m})} \, \textbf{e}_{k'}^{\text{(m)},1} .
    \end{equation}
\begin{itemize}
    \item \underline{2nd interface}:
\end{itemize}
    \begin{equation}\label{eq:2ndint_sup}
        \begin{bmatrix}
            \textbf{D}'_{k'} \\
            \textbf{B}'_{k'}
        \end{bmatrix}^{(\text{m}),2}
        =
        \begin{bmatrix}
            \textbf{D}'_{k'} \\
            \textbf{B}'_{k'}
        \end{bmatrix}^{(\tilde{\text{m}}),2}
        \rightarrow
        \tilde{\mathbf{\mathcal{M}}}{'}_{k'}^{(\text{m})} \, \textbf{e}_{k'}^{\text{(m)},2} = \tilde{\mathbf{\mathcal{M}}}{'}_{k'}^{(\text{m})} \, \mathbf{\mathcal{P}}{'} \, \textbf{e}_{k'}^{\text{(m)},1} = \tilde{\mathbf{\mathcal{M}}}{'}_{k'}^{(\tilde{\text{m}})} \, \textbf{e}_{k'}^{(\tilde{\text{m}}),2} .
    \end{equation}
\begin{itemize}
    \item \underline{3rd interface}:
\end{itemize}
    \begin{equation}\label{eq:3rdint_sup}
        \begin{bmatrix}
            \textbf{D}'_{k'} \\
            \textbf{B}'_{k'}
        \end{bmatrix}^{(\tilde{\text{m}}),3}
        =
        \begin{bmatrix}
            \textbf{D}'_{k'} \\
            \textbf{B}'_{k'}
        \end{bmatrix}_{\text{f}}^{\text{(out)},3}
        +
        \begin{bmatrix}
            \textbf{D}'_{k'} \\
            \textbf{B}'_{k'}
        \end{bmatrix}_{\text{b}}^{\text{(out)},3}
        \rightarrow
        \tilde{\mathbf{\mathcal{M}}}{'}_{k'}^{(\tilde{\text{m}})} \, \textbf{e}_{k'}^{(\tilde{\text{m}}),3} = \tilde{\mathbf{\mathcal{M}}}{'}_{k'}^{(\tilde{\text{m}})} \, \Tilde{\mathbf{\mathcal{P}}}{'} \, \textbf{e}_{k'}^{(\tilde{\text{m}}),2} = \tilde{\mathbf{\mathcal{M}}}{'}_{k',\text{f}}^{(\text{v})} \, \textbf{e}_{k',\text{f}}^{\text{(out)},3} + \tilde{\mathbf{\mathcal{M}}}{'}_{k',\text{b}}^{(\text{v})} \, \textbf{e}_{k',\text{b}}^{\text{(out)},3} .
    \end{equation}
$\mathbf{\mathcal{P}}'$ corresponds to the diagonal matrix containing the phase acquired by each eigenvector between surfaces $\mathcal{P}'_{jj} = \exp(-i\, \omega'_j  D')$, where $D' = N \gamma T$ is the width of each slab and $\omega'_j$ the eigenvalues of the STPhC. As we can see, now only incoming waves are allowed before the STPhC slab, while only output waves are present afterwards. Therefore, the scattering matrix connecting input and output waves will have the structure of a transfer matrix. To derive it, we can use the same derivation as in the subluminal case to arrive at a similar expression as Eq.~\eqref{eq:pre_scat_mat}:
\begin{equation} \label{eq:pre_scat_mat_sup}
    \tilde{\textbf{N}}_{k'} \tilde{\mathbf{\mathcal{M}}}{'}_{k',\text{f}}^{(\text{v})} \, \textbf{e}_{k',\text{f}}^{\text{(out)},3} - \tilde{\mathbf{\mathcal{M}}}{'}_{k',\text{b} }^{(\text{v})} \, \textbf{e}_{k',\text{b}}^{\text{(in)},1} = \tilde{\mathbf{\mathcal{M}}}{'}_{k',\text{f}}^{(\text{v})} \, \textbf{e}_{k',\text{f}}^{\text{(in)},1} - \tilde{\textbf{N}}_{k'} \tilde{\mathbf{\mathcal{M}}}{'}_{k',\text{b} }^{(\text{v})} \, \textbf{e}_{k',\text{b}}^{\text{(out)},3} .
\end{equation}
Rearranging and writing the equation in vector form we have
\begin{equation}
    \underbrace{ \tilde{\textbf{N}}_{k'} \left[\tilde{\mathbf{\mathcal{M}}}{'}_{k',\text{f}}^{(\text{v})} \, , \tilde{\mathbf{\mathcal{M}}}{'}_{k',\text{b} }^{(\text{v})}  \right]}_{\tilde{\textbf{A}}}
    \begin{bmatrix}
        \textbf{e}_{k',\text{f}}^{\text{(out)},3}\\
        \textbf{e}_{k',\text{b}}^{\text{(out)},3}
    \end{bmatrix}
    =
    \underbrace{ \tilde{\textbf{N}}_{k'} \left[ \tilde{\mathbf{\mathcal{M}}}{'}_{k',\text{f}}^{(\text{v})} \, , \tilde{\mathbf{\mathcal{M}}}{'}_{k',\text{b} }^{(\text{v})}  \right]}_{\tilde{\textbf{B}}}
    \begin{bmatrix}
        \textbf{e}_{k',\text{f}}^{\text{(in)},1}\\
        \textbf{e}_{k',\text{b}}^{\text{(in)},1}
    \end{bmatrix},
\end{equation}
such that the scattering matrix for the superluminal scattering problem is
\begin{equation}\label{eq:scatmat_sup}
    \begin{bmatrix}
        \textbf{e}_{k',\text{f}}^{\text{(out)},3}\\
        \textbf{e}_{k',\text{b}}^{\text{(out)},3}
    \end{bmatrix}
    =
    \tilde{\textbf{S}}_{k'}
    \begin{bmatrix}
        \textbf{e}_{k',\text{f}}^{\text{(in)},1}\\
        \textbf{e}_{k',\text{b}}^{\text{(in)},1}
    \end{bmatrix},
\end{equation}
with $\tilde{\textbf{S}}_{k'} = \tilde{\textbf{A}}^{-1}\,\tilde{\textbf{B}}$.

Furthermore, we can calculate the fields inside the slabs by writing the weights $\textbf{e}_{k'}^{(\text{m}),1}$ and $\textbf{e}_{k'}^{(\tilde{\text{m}}),2}$ in terms of the input weights using Eqs.~\eqref{eq:1stint_sup}-\eqref{eq:2ndint_sup}:
\begin{equation}
    \textbf{e}_{k'}^{(\text{m}),1} = \underbrace{\left(  \tilde{\mathbf{\mathcal{M}}}{'}_{k'}^{({\text{m}})} \right)^{-1} \, \left[ \tilde{\mathbf{\mathcal{M}}}{'}_{k',\text{f}}^{(\text{v})} \, , \tilde{\mathbf{\mathcal{M}}}{'}_{k',\text{b} }^{(\text{v})}  \right]}_{\tilde{\textbf{C}}}  \begin{bmatrix}
        \textbf{e}_{k',\text{f}}^{\text{(in)},1}\\
        \textbf{e}_{k',\text{b}}^{\text{(in)},1}
    \end{bmatrix},
\end{equation}
\begin{equation}
    \textbf{e}_{k'}^{(\tilde{\text{m}}),2} = \left(  \tilde{\mathbf{\mathcal{M}}}{'}_{k'}^{(\tilde{\text{m}})} \right)^{-1} \tilde{\mathbf{\mathcal{M}}}{'}_{k'}^{({\text{m}})} \mathbf{\mathcal{P}}' \tilde{\textbf{C}} \begin{bmatrix}
        \textbf{e}_{k',\text{f}}^{\text{(in)},1}\\
        \textbf{e}_{k',\text{b}}^{\text{(in)},1}
    \end{bmatrix} .
\end{equation}

Once we have calculated the weights, the fields can be obtained using Eqs.~\eqref{eq:modedef}-\eqref{eq:modedef_vac} changing the transformation matrices to the superluminal one defined in Eq.~\eqref{eq:trans_mat_sup}, and the exponential term to $e^{ik'x' - i (\omega'_{k',\tau,p} + n\Omega/\gamma)t'}$. 

Lastly, the transmission spectrum is defined in the lab-frame by exciting the composite system with a forward propagating plane wave of momentum $k_0$. As in the subluminal case, we do so by obtaining the transmitted fields and calculating the \textit{space-averaged} Poyting vector, which is defined as
\begin{align}
    P_{k',\text{f}}^{(\text{out})}(t') = - E_{k',\text{f}}^{(\text{out})}(t') \, \left( H_{k',\text{f}}^{(\text{out})}(t') \right)^*
    = - \sum_{n,n'} e_{k',\text{f},n}^{(\text{out})} \, e_{k',\text{f},n'}^{(\text{out}) \, *} \, E_{k',\text{f},n}^{(\text{out})} \, H_{k',\text{f},n'}^{(\text{out}) \,*} \; e^{- i\left( \omega'_{k',\text{f},n} - \omega{'}^*_{k',\text{f},n'} + (n-n') \Omega/\gamma \right) t'}.
\end{align}
Then, normalizing the previous expression by the Poynting vector of the incident wave, and evaluating at the third interface $t' = 2\gamma N T$, we obtain the transmittance as
\begin{equation}
    T(k') = \frac{P_{k',\text{f}}^{(\text{out})}(t' = 2\gamma N T)}{P_{k',\text{f}}^{(\text{in})}(t'=0)} = \frac{\sum_{n,n'} e_{k',\text{f},n}^{(\text{out})} \, e_{k',\text{f},n'}^{(\text{out}) \, *} \, E_{k',\text{f},n}^{(\text{out})} \, H_{k',\text{f},n'}^{(\text{out}) \,*} \; e^{-i\left( \omega'_{k',\text{f},n} - \omega{'}^*_{k',\text{f},n'} \right) (2\gamma N T)}}{E_{k',\text{f},0}^{(\text{in})} H_{k',\text{f},0}^{(\text{in}) \,*}}.
\end{equation}

\subsubsection{Temporal Boundary}

Similarly to the subluminal case, the procedure to calculate the scattering matrix for a temporal boundary closely follows the derivation from previous section. However, we need to be mindful of two key differences: the conserved magnitude is now $k$, and the continuity conditions must be enforced on the lab-frame fields $[D,B]$. To take these differences into account, we need to change the eigenvector matrices of both vacuum and the modulated medium to those defined in Eq.~\eqref{eigen_labk} for a given $k$, as well as the transformation matrix of the STPhC. 

In the lab-frame, the transformation matrix $\tilde{\textbf{T}}^{(m)}$ simply corresponds to the constitutive matrix $\hat{\textbf{M}}(x,t)$, defined in Eq.~(3) from the main text, evaluated at the interfaces. However, evaluating at a temporal interface only gets rid of the temporal dependence, leaving 
\begin{equation}
    \tilde{\textbf{T}}^{(m)}(x) =
    \begin{pmatrix}
        \epsilon(x) & 0\\
        0 & \mu(x)
    \end{pmatrix}.
\end{equation}
In order to derive the scattering matrix for this composite system, we need to get rid of the spatial dependence too. To do so, we decompose $\epsilon(x)$ and $\mu(x)$ in their Bloch components, which leads to
\begin{equation}
    \tilde{\textbf{T}}^{(m)} = 
    \begin{pmatrix}
        \textbf{T}^{\epsilon} & \mathbb{0}_{N_\mathrm{F} \times N_\mathrm{F}}\\
        \mathbb{0}_{N_\mathrm{F} \times N_\mathrm{F}} & \textbf{T}^{\mu}
    \end{pmatrix},
\end{equation}
with
\begin{equation}
    \textbf{T}^{\epsilon}_{n,n'} = \epsilon_0 \epsilon_m [\delta_{n,n'} + \alpha_e (\delta_{n,n'+1} + \delta_{n,n'-1})]; \; \; \; \; \textbf{T}^{\mu}_{n,n'} = \mu_0 \mu_m [\delta_{n,n'} + \alpha_m (\delta_{n,n'+1} + \delta_{n,n'-1})].
\end{equation}

Finally, we obtain the transmission spectrum by calculating the spatially averaged Poyting vector
\begin{align}
    P_{k,\text{f}}^{(\text{out})}(x,t) &= - E_{k,\text{f}}^{(\text{out})}(x,t) \, \left( H_{k,\text{f}}^{(\text{out})}(x,t) \right)^*
    = \nonumber \\
    &= - \sum_{n,n'} e_{k,\text{f},n}^{(\text{out})} \, e_{k,\text{f},n'}^{(\text{out}) \, *} \, E_{k,\text{f},n,n}^{(\text{out})} \, H_{k,\text{f},n',n'}^{(\text{out}) \,*} e^{i(n-n')g x - i \left( \omega_{k,f,n} - \omega_{k,f,n'}^* + (n-n')\Omega \right) t}.
\end{align}
However, similarly to the subluminal case, the fields present an intrinsic spatial dependence due to the nature of the truncation, as we find different values of $\epsilon(x,t)$ and $\mu(x,t)$ along the temporal boundary. We integrate the spatially averaged Poyting vector over one modulation period $\braket{P_{k,\text{f}}^{(\text{out})}} = \frac{1}{a} \int_0^a P_{k,\text{f}}^{(\text{out})}$. This allows us to write the transmittance for the temporal boundary case as
\begin{equation}
    T(k) = \frac{\braket{P_{k,\text{f}}^{(\text{out})}(t = 2 N T)}}{\braket{P_{k,\text{f}}^{(\text{in})}(t=0)}} = \frac{\sum_n \, |e_{k,\text{f},n}^{(\text{out})}|^2 E_{k,\text{f},n}^{(\text{out})} \, H_{k,\text{f},n}^{(\text{out}) \,*}}{E_{k,\text{f},0}^{(\text{in})} H_{k,\text{f},0}^{(\text{in}) \,*}}.
\end{equation}

\subsection{Spatiotemporal Zak phase}

In this section we demonstrate that, as in the subluminal case, the biorthogonal Berry connection is not needed thanks to the Hermiticity of the constitutive matrix and the presence of $\mathcal{P'T'}$-symmetry. Indeed, starting from Maxwell's equations in the time-like frame for normal incidence and s-polarization
\begin{equation}
    \begin{pmatrix}
        0 & \partial_{x'}\\
        \partial_{x'} & 0
    \end{pmatrix}
    \begin{bmatrix}
        E_z'\\
        H_y'
    \end{bmatrix} =
    \partial_{t'}\left( \hat{\textbf{M}}'(t') \begin{bmatrix}
        E_z'\\
        H_y'
    \end{bmatrix} \right),
\end{equation}
we can rearrange them as 
\begin{equation}
    \partial_{x'} \left( \hat{\textbf{M}}'(t') \right)^{-1} \begin{bmatrix}
        D_z'\\
        B_y'
    \end{bmatrix} = 
    \begin{pmatrix}
        0 & \partial_{t'}\\
        \partial_{t'} & 0
    \end{pmatrix} 
    \begin{bmatrix}
        D_z'\\
        B_y'
    \end{bmatrix} .
\end{equation}
Then, we can arrive at an eigenvalue problem for the momentum $k'$ by considering $[D_z',B_y']$ as plane waves:
\begin{equation}
    \hat{\tilde{\textbf{H}}}_{\text{R}} \tilde{\mathbf{\Psi}}'_{\text{R}}= \hat{\textbf{M}}'(t') \hat{\mathds{L}}_{t'} \tilde{\mathbf{\Psi}}'_{\text{R}} = k' \tilde{\mathbf{\Psi}}'_{\text{R}}, \; \; \text{with} \; \; \hat{\mathds{L}}_{t'} = \begin{pmatrix}
        0 & -i \partial_{t'}\\
        -i \partial_{t'} & 0
    \end{pmatrix}, \; \; \text{and} \; \; \tilde{\mathbf{\Psi}}'_{\text{R}} = \begin{bmatrix}
        D_z'\\
        B_y'
    \end{bmatrix}.
\end{equation}
Again, the operator $\hat{\tilde{\textbf{H}}}_{\text{R}}$ is not Hermitian because, even though both $\hat{\textbf{M}}'(t')$ and $\hat{\mathds{L}}_{t'}$ are, they do not commute. We can obtain its adjoint operator writing
\begin{equation}
    \hat{\tilde{\textbf{H}}}_{\text{L}} = \hat{\mathds{L}}_{t'} \hat{\textbf{M}}'(t') = \left( \hat{\tilde{\textbf{H}}}_{\text{R}}\right)^{\dagger}, \; \; \text{with} \; \; \hat{\tilde{\textbf{H}}}_{\text{L}} \tilde{\mathbf{\Psi}}'_{\text{L}} = (k')^* \tilde{\mathbf{\Psi}}'_{\text{L}}.
\end{equation}
Similar to the subluminal case, we can rewrite the adjoint operator and its eigenvalue equation as
\begin{equation}
    \left( \hat{\textbf{M}}'(t') \right)^{-1} \underbrace{\hat{\textbf{M}}'(t') \hat{\mathds{L}}_{t'}}_{\hat{\tilde{\textbf{H}}}_{\text{R}}} \hat{\textbf{M}}'(t') \tilde{\mathbf{\Psi}}'_{\text{L}} = (k')^*\tilde{\mathbf{\Psi}}'_{\text{L}}  \rightarrow \hat{\tilde{\textbf{H}}}_{\text{R}} \hat{\textbf{M}}'(t') \tilde{\mathbf{\Psi}}'_{\text{L}} = (k')^* \, \hat{\textbf{M}}'(t') \tilde{\mathbf{\Psi}}'_{\text{L}}.
\end{equation}
Therefore, given that the $\mathcal{P'T'}$-broken phases are only encountered at the band gaps characterized by complex frequencies $\omega'$, we know $\mathcal{P'T'}$-symmetry will be preserved in the bands, resulting in real $k'$ eigenvalues. In turn, we can write
\begin{equation}
    \tilde{\mathbf{\Psi}}'_{\text{R}} =  \hat{\textbf{M}}'(t')  \tilde{\mathbf{\Psi}}'_{\text{L}},
\end{equation}
such that the inner product of left and right eigenfunctions becomes
\begin{equation} \label{scalar_prod_sup}
    \braket{\tilde{\mathbf{\Psi}}'_{\text{L}} | \tilde{\mathbf{\Psi}}'_{\text{R}}} = \int_{\text{UC}} dt' \left(  \tilde{\mathbf{\Psi}}'_{\text{L}} \right)^\dagger \, \tilde{\mathbf{\Psi}}'_{\text{R}} = \int_{\text{UC}} dt' \left(  \left(\hat{\textbf{M}}'(t')\right)^{-1} \tilde{\mathbf{\Psi}}'_{\text{R}} \right)^\dagger \tilde{\mathbf{\Psi}}'_{\text{R}} = \int_{\text{UC}} dt' \left(  \tilde{\mathbf{\Psi}}'_{\text{R}}\right)^\dagger  \left( \hat{\textbf{M}}'(t')\right)^{-1} \tilde{\mathbf{\Psi}}'_{\text{R}}.
\end{equation}
This relation allows us to write the Zak phase along the time-like bands defined with a biorthogonal Berry connection as
\begin{equation}
    \theta^{\mathrm{ST}}_m =  i \int_{-\Omega/2\gamma}^{+\Omega/2\gamma} d\omega'\braket{\tilde{\textbf{u}}{'}_{\text{L}, \omega',m} | \partial_{\omega'} \tilde{\textbf{u}}'_{\text{R}, \omega',m}} = i \int_{-\Omega/2\gamma}^{+\Omega/2\gamma} d\omega'\braket{\tilde{\textbf{u}}{'}_{\text{R}, \omega',m} | \left(\hat{\textbf{M}}'(t')\right)^{-1} \partial_{\omega'} \tilde{\textbf{u}}'_{\text{R}, \omega',m}},
\end{equation}
which ends with the same structure as a conventional Berry connection defined with a weighted scalar product, analogous to the subluminal case.

To obtain the ST Zak phase numerically, we note that the eigenvalue problem in the time-like frame with $k'$ as the eigenvalue can be obtained through Eqs.~\eqref{eigen_com} by changing $c_g$ and $c_f$, as was previously discussed. Then, the eigenfunctions read
\begin{equation}\label{eq:com_eigefun_D}
    D'_{\omega',m}(x',t') = e^{i (k_{\omega',m}' x' - \omega' t')} \underbrace{\sum_n \gamma \, \left[ \epsilon(t') \, E_{k',m,n} + c_f/c_0^2 H_{k',m,n} \right ] e^{-i\, n \Omega/\gamma \, t'}}_{\tilde{{u}}{'}^D_{\omega',m}(t')}
\end{equation}
and
\begin{equation}\label{eq:com_eigefun_B}
    B'_{\omega',m}(x',t') = e^{i (k_{\omega',m}' x' - \omega' t')} \underbrace{\sum_n \gamma \, \left[ \mu(t') \, H_{k',m,n} + c_f/c_0^2 E_{k',m,n} \right ] e^{-i\, n \Omega/\gamma \, t'}}_{\tilde{{u}}^B_{\omega',m}(t')},
\end{equation}
where $m$ corresponds to the band index. With the periodic part of each eigenfunction defined, we can then write the discretized ST Zak phase for superluminal modulation as
\begin{equation}\label{eq:disc_Zak_sup}
    \theta^{\mathrm{ST}}_m = -\text{Im log} \prod_i \braket{\tilde{\textbf{u}}'_{\omega'_i,m} | \textbf{u}'_{\omega'_i+1,m}}_{(\hat{\textbf{M}})^{-1}},
\end{equation}
where $\tilde{\textbf{u}}' = [\tilde{{u}}{'}^D, \tilde{{u}}{'}^B]^T$, $\braket{\cdot | \cdot}_{(\hat{\textbf{M}})^{-1}}$ corresponds to the weighted scalar product from Eq.~\eqref{scalar_prod_sup}, and we enforce the periodic gauge condition $\textbf{u}'_{\omega'=\Omega/2\gamma ,m}(t') = e^{i\Omega/\gamma t'}   \textbf{u}'_{\omega'=-\Omega/2\gamma ,m}(t')$.

\bibliography{ref}